\documentclass[11pt]{article}
\usepackage{amsmath, amssymb, amsfonts, amsthm}

\newtheorem{theorem}{Theorem}[section]
\newtheorem{proposition}[theorem]{Proposition}
\newtheorem{corollary}[theorem]{Corollary}
\newtheorem{lemma}[theorem]{Lemma}

\newcommand{\tri}{| \! | \! |}
\newcommand{\rd}{{\rm d}}
\newcommand{\be}{\begin{equation}}
\newcommand{\ee}{\end{equation}}
\newcommand{\bey}{\begin{eqnarray}}
\newcommand{\eey}{\end{eqnarray}}

\newcommand{\Htrap}{H_N^{\text{trap}}}
\newcommand{\Vtrap}{V_{\text{ext}}}
\newcommand{\psitrap}{\psi_N^{\text{trap}}}
\newcommand{\phitrap}{\phi_{\text{GP}}^{\text{trap}}}

\newcommand{\eps}{\varepsilon}

\newcommand{\bx}{{\bf x}}

\newcommand{\ph}{\varphi}

\newcommand{\la}{\langle}
\newcommand{\ra}{\rangle}

\renewcommand{\a}{\alpha}

\newcommand{\e}{\varepsilon}

\newcommand{\cU}{{\cal U}}

\newcommand{\bR}{{\mathbb R}}

\newcommand{\bN}{{\mathbb N}}
\newcommand{\bZ}{{\mathbb Z}}

\newcommand{\tr}{\mbox{Tr}}

\newcommand{\wt}{\widetilde}
\newcommand{\wh}{\widehat}
\newcommand{\ov}{\overline}

\newcommand{\const}{\mathrm{const}}

\newcommand{\cS}{{\cal S}}

\newcommand{\cE}{{\cal E}}

\newcommand{\cK}{{\cal K}}
\newcommand{\cH}{{\cal H}}
\newcommand{\cL}{{\cal L}}

\newcommand{\cJ}{{\cal J}}

\newcommand{\supp}{\operatorname{supp}}

\input epsf

\newcommand{\fh}{{\frak h}}

\newcommand{\donothing}[1]{}

\begin{document}

\title{Derivation of the Gross-Pitaevskii Equation for the \\
Dynamics of Bose-Einstein Condensate}
\author{L\'aszl\'o Erd\H os${}^2$\thanks{Partially
supported by the EU-IHP Network ``Analysis and Quantum''
HPRN-CT-2002-0027}\;, Benjamin Schlein${}^1$\thanks{Supported by NSF
postdoctoral fellowship}, \, and Horng-Tzer
Yau${}^1$\thanks{Partially supported by NSF grant
DMS-0602038} \\
\\
Department of Mathematics, Harvard University\\ Cambridge, MA 02138, USA${}^1$ \\ \\
Institute of Mathematics, University of Munich, \\
Theresienstr. 39, D-80333 Munich, Germany${}^2$
\\}

\maketitle

\begin{abstract}
Consider a system of $N$ bosons  in three dimensions interacting via
a repulsive short range pair potential $N^2V(N(x_i-x_j))$, where
$\bx=(x_1, \ldots, x_N)$ denotes the positions of the particles. Let
$H_N$ denote the Hamiltonian of the system and  let $\psi_{N,t}$ be
the solution to the Schr\"odinger equation.  Suppose that the
initial data $\psi_{N,0}$ satisfies the energy condition
\[ \langle
\psi_{N,0},  H_N^k \psi_{N,0} \rangle \leq C^k N^k \;
\]
for $k=1,2,\ldots $.
 We also assume that
the $k$-particle density matrices of the initial state are
asymptotically
 factorized as $N\to\infty$.
 We prove that
the $k$-particle density matrices of $\psi_{N,t}$ are also
asymptotically
 factorized  and  the one particle orbital wave function solves the
Gross-Pitaevskii equation, a cubic non-linear Schr\"odinger equation
with the coupling constant given by the scattering length of the
potential $V$. We also prove the same conclusion if the energy
condition holds only for $k=1$ but the factorization of $\psi_{N,0}$
is assumed in a stronger sense.

\end{abstract}

\section{Introduction}\label{sec:intro}

Bose-Einstein condensation states that at a very low temperature
Bose systems with a pair interaction exhibit a  collective mode, the
Bose-Einstein condensate.  If one neglects the interaction and
treats all bosons as independent particles, Bose-Einstein
condensation is a simple exercise \cite{Hu}. The many-body effects
were traditionally treated by the Bogoliubov approximation, which
postulates that the ratio between the
non-condensate and the condensate is small. The coupling constant
$\sigma/8 \pi$ obtained by the Bogoliubov  approximation is the
semiclassical approximation of the the scattering length $a_0$ of
the pair potential. To recover the scattering length, one needs to
perform a higher order diagrammatic re-summation,  a procedure that
yet lacks mathematical rigor for interacting  systems.

Gross \cite{G1,G2} and Pitaevskii \cite{P} proposed to model the
many-body effects by a nonlinear on-site self interaction of a
complex order parameter (the ``condensate wave function''). The
strength of the nonlinear interaction in this model is given by the
scattering length $a_0$. The {\it Gross-Pitaevskii (GP) equation} is
given by
\begin{equation}\label{nls}
i \partial_t u_t     = -\Delta u_t     +  \sigma |u_t|^2 u_t =
\frac{\delta \cE(u, \bar{u})}{\delta \bar{u}} \Big|_{u_t}, \quad
\cE(u, \bar u)= \int_{\bR^3} \Big [ \, |\nabla u|^2 + \frac {\sigma}
2 |u|^4 \Big ]\; ,
\end{equation}
where $\cE$ is the {\it Gross-Pitaevskii energy functional} and
$\sigma= 8 \pi a_0$. The Gross-Pitaevskii equation is a
phenomenological mean field type equation and its validity needs to
be established from the Schr\"odinger equation with the Hamiltonian
given by the pair interaction.

The first rigorous result concerning the many-body effects of the
Bose gas was Dyson's estimate of the ground state energy. Dyson
\cite{Dy} proved the correct leading upper bound to the energy and a
lower bound off by a factor around $10$.  Dyson's upper bound was obtained
by using trial functions with short range two-body correlations.
This short scale structure
is crucial for the emergence of the scattering length and
thus for the correct energy.
The matching lower bound
to the leading order in the low density regime  was obtained  by
Lieb and Yngvason \cite{LY1}.   Lieb and Seiringer \cite{LS} later
proved that the minimizer of the Gross-Pitaevskii energy functional
correctly describes the ground state of an
 $N$-boson system  in the limit $N\to \infty$
provided that   the  length scale of the pair potential is  of order
$1/N$. For a review on related results, see \cite{LSSY}.

The experiments on the Bose-Einstein condensation were conducted by
observing the dynamics of the condensate when the confining traps
are removed. Since the ground state of the system with traps will no
longer be  the ground state without traps, the validity of the
Gross-Pitaevskii equation for predicting the experimental outcomes
asserts that the approximation of the
many-body effects  by a nonlinear on-site self interaction of the
order parameter applies to a certain class of excited states and
their subsequent time evolution as well.

In this paper, we shall prove that the Gross-Pitaevskii equation
actually describes the dynamics of a large class of initial states.
The allowed initial states include wave functions with the
characteristic short scale two-body correlation structure of the
ground state and also wave functions of product form. Notice that
product wave functions do not have this characteristic short scale
structure,  nevertheless the GP evolution equation applies to them.
It should be noted that our theorems concern only the evolution of
the  one particle density matrix but not its energy. In fact,  for
product initial states, the GP theory is correct on the level of
density matrix, but not on the level of the energy. We shall discuss
this surprising fact in more details in Section \ref{sec:bb}.

\section{The Main Results}\label{sec:main}
\setcounter{equation}{0}

Recall that the Gross-Pitaevskii energy functional correctly
describes the energy in the large $N$ limit provided that the
scattering length is of order $1/N$ \cite{LSY1}. We thus choose the
interaction potential to be
$$
 V_N (x): = N^2 \, V \left( N x\right) = \frac{1}{N}  \;
 N^3 V \left( N x\right).
$$
This potential can also be viewed as an approximate delta function
on scale $1/N$ with a  prefactor $1/N$ which we will interpret as
the mean field average. The Hamiltonian of the Bose system is given
by
\begin{equation}\label{eq:ham1}
H_N: = - \sum_{j=1}^N \Delta_j + \sum_{j < k}^N V_N (x_j - x_k)\; ,
\quad V_N (x): = N^2 \, V \left( N x\right) .
\end{equation}
The support of the initial state will not be scaled with $N$. Thus
the density of the system is $N$ and the typical inter-particle
distance is $N^{-1/3}$, which is much bigger than the length scale
of the potential. The system is really
 a dilute gas scaled in such
a way that the size of the total system is independent of $N$.

The dynamics of the system is governed by the Schr\"odinger equation
\begin{equation}\label{eq:schr}
i\partial_t \psi_{N,t} =  H_N \psi_{N,t}
\end{equation}
for the wave function $\psi_{N,t} \in L^2_s (\bR^{3N})$, the
subspace of $L^2 (\bR^{3N})$ consisting of all functions symmetric
with respect to any permutation of the $N$ particles. We choose
$\psi_{N,t}$ to have $L^2$-norm equal to one, $\| \psi_{N,t}\| =1$.

Instead of describing the system through the wave function, we can
describe it by a density matrix $\gamma_{N} \in \cL^1 (L^2_s
(\bR^{3N}))$, where $\cL^1 (L^2_s (\bR^{3N}))$ denotes the space of
trace class operators on the Hilbert space $L^2_s (\bR^{3N})$. A
density matrix is a non-negative trace class operator with trace
equal to one. For the {\it pure} state described by the wave
function $\psi_{N}$, the density matrix $\gamma_{N} =|\psi_{N}
\rangle \langle \psi_{N}|$ is the orthogonal projection onto
$\psi_{N}$. The time evolution of a density matrix $\gamma_N$ is
determined by the Heisenberg equation
\begin{equation}\label{eq:heise}
i\partial_t \gamma_{N,t} = [ H_N , \gamma_{N,t}] \,,
\end{equation}
where $[A, B]= AB-BA$ is the commutator.

\medskip

Introduce the shorthand notation
$$
\bx:=(x_1, x_2, \ldots, x_N), \quad \bx_k:=(x_1, \ldots , x_k),
\quad \bx_{N-k}:=(x_{k+1}, \ldots , x_N)
$$
and similarly for the primed variables, $\bx'_k:=(x_1', \ldots ,
x_k')$. For $k =1,\dots ,N$, the {\it $k$-particle reduced density
matrix} (or {\it $k$-particle marginal}) associated with
$\gamma_{N,t}$ is the non-negative operator in $\cL^1 (L^2_s
(\bR^{3k}))$ defined by taking the partial trace of $\gamma_{N,t}$
over  $N-k$ variables. In other words, the kernel of
$\gamma_{N,t}^{(k)}$ is given by
\begin{equation}
\gamma_{N,t}^{(k)} (\bx_k; \bx_k'): = \int \rd \bx_{N-k}
\gamma_{N,t} \left( \bx_k, \bx_{N-k}; \bx_k', \bx_{N-k} \right) \; .
\label{eq:marginal}
\end{equation}
Our normalization implies that  \( \tr \; \gamma_{N,t}^{(k)} =1 \)
for all $k=1,\dots, N$ and for every $t \in \bR$.

We now define a topology on the density matrices. We denote by
$\cL^1_k = \cL^1 (L^2 (\bR^{3k}))$ the space of trace class
operators acting on the Hilbert space $L^2 (\bR^{3k})$. Moreover,
$\cK_k = \cK (L^2 (\bR^{3k}))$ will denote the space of compact
operators acting on $L^2 (\bR^{3k})$ equipped with the operator
norm, $\| \cdot \|_{\cK_k} := \| \cdot \|$. Since $\cL_k^1 =
\cK_k^*$, we can  define the weak* topology on
$\cL^1(L^2(\bR^{3k}))$, i.e., $\omega_{n} \to \omega$ if and only if
for every compact operator $J $ on $L^2 (\bR^{3k})$ we have
\begin{equation}
\lim_{n \to \infty}\tr \; J \, \omega_{n}  = \tr \; J \omega \, .
\end{equation}

\medskip

Throughout the paper we will assume that the unscaled interaction
potential, $V(x)$, is a nonnegative, smooth, spherically symmetric
function with a compact support in the ball of radius $R$, \be
  \mbox{supp} \; V \subset \{ x\in \bR^3 \; : \; |x|\leq R\} \;.
\label{def:R}
\ee
With the notation $r=|x|$, we will sometimes write $V(r)$ for $V(x)$.
We define the following dimensionless quantity to measure
the strength of $V$
\begin{equation}\label{eq:defrho} \rho :=
\sup_{r \geq 0} r^2 \, V(r) + \int_0^{\infty} \rd r \, r V(r) \; .
\end{equation}

Let $f$ be the zero energy scattering solution
associated with $V$ with normalization
 $\lim_{|x| \to \infty} f (x) =1$. We will  write
$f(x)=1-w_0(x)$. By definition, this function  satisfies the
equation
\begin{equation}\label{ze}
\Big [ \, -\Delta + \frac 12  \,  {V (x)}  \, \Big ] (1-w_0 (x)) = 0
\, ,
\end{equation}
and $\lim_{|x| \to \infty} w_0 (x) =0$. The scattering length $a_0$
of $V$ is defined by
 \be \label{eq:scatlength}
a_0 := \lim_{|x| \to \infty} w_0 (x) |x| \, . \ee
Since $V$ has a compact
support \eqref{def:R},
we have \be\label{fb} f(x) =  1 - \frac{a_0}{|x|} \, \qquad  |x|
\geq R \, . \ee {F}rom the zero energy equation, we also have the
identity \be\label{13} \int \rd x \; V(x) (1-w_0(x)) = 8 \pi a_0 \, .
\ee
By scaling,
the scattering length of the potential $ V_N (x)$ is $a:=
a_0/N$ and the zero energy scattering equation for the
potential $V_N$ is given by
\begin{equation}\label{eq:defw1}
\left( -\Delta + \frac{1}{2} V_N (x) \right) (1 -w(x)) = 0
\end{equation}
where $w(x) := w_0 (Nx)$. Note that $w(x) = a / |x|$, for $|x| \geq
R/N$.

\medskip

We can now state our main theorems.

\begin{theorem}\label{thm:main1}
Suppose $V \geq 0$ is a smooth, compactly supported, spherically
symmetric potential with scattering length $a_0$ and assume that
$\rho$ (defined in \eqref{eq:defrho}) is small enough. We consider a
family of systems described by initial wave functions $\psi_N \in
L_s^2 (\bR^{3N})$ such that
\begin{equation}\label{eq:thmassum1}
\langle \psi_N, H^k_N \psi_N \rangle \leq C^k N^k \,
\end{equation}
for all $k \geq 1$. We assume that the marginal densities associated
with $\psi_N$ factorize in the limit $N \to \infty$, i.e. there is
 a function $\ph \in L^2 (\bR^3)$ such that for every
$k \geq 1$,
\begin{equation}\label{eq:thmassum2}
\gamma_N^{(k)} \to |\ph \rangle \langle \ph|^{\otimes k}
\end{equation}
as $N \to \infty$ with respect to the weak* topology of $\cL^1 (L^2
(\bR^{3k}))$. Then $\ph \in H^1 (\bR^3)$, and for every fixed $k
\geq 1$ and $t \in \bR$, we have
\begin{equation}\label{eq:conv1}
\gamma_{N,t}^{(k)} \to |\ph_t \rangle \langle \ph_t|^{\otimes k}
\end{equation}
with respect to the same topology. Here $\ph_t \in H^1 (\bR^3)$ is
the solution of the nonlinear Gross-Pitaevskii equation
\begin{equation}\label{eq:GP}
i\partial_t \ph_t = -\Delta \ph_t + 8\pi a_0 |\ph_t|^2 \ph_t
\end{equation}
with initial condition $\ph_{t=0} = \ph$.
\end{theorem}

Using an approximation argument, we can relax the energy condition
(\ref{eq:thmassum1}), and only assume that $\langle \psi_N, H_N
\psi_N \rangle \leq C N$. However, in order to apply our
approximation argument, we need to assume stronger asymptotic
factorization properties on $\psi_N$.

\begin{theorem}\label{thm:main2}
Suppose $V \geq 0$ is a smooth, compactly supported, spherically
symmetric potential with scattering length $a_0$ and assume that
$\rho$ (defined in (\ref{eq:defrho})) is small enough. We consider a
family of systems described by initial wave functions $\psi_N \in
L_s^2 (\bR^{3N})$ such that
\begin{equation}\label{eq:thm2ass1}
\langle \psi_N , H_N \psi_N \rangle \leq C N \,.
\end{equation}
We assume asymptotic factorization of $\psi_N$ in the sense that
there exists $\ph \in L^2 (\bR^3)$ and, for every $N$, and every $1
\leq k \leq N$, there exists a $\xi^{(N-k)}_N \in L^2
(\bR^{3(N-k)})$ with $\| \xi_N^{(N-k)} \| =1$ such that
\begin{equation}\label{eq:thm2ass2}
\| \psi_N - \ph^{\otimes k} \otimes \xi_N^{(N-k)} \| \to 0
\end{equation}
as $N \to \infty$. This implies, in particular that, for every $k
\geq 1$,
\begin{equation}\label{eq:thm2ass3}
\gamma_N^{(k)} \to |\ph \rangle \langle \ph|^{\otimes k}
\end{equation}
as $N \to \infty$ with respect to the weak* topology of $\cL^1 (L^2
(\bR^{3k}))$. Then $\ph \in H^1 (\bR^3)$, and for every fixed $k
\geq 1$ and $t \in \bR$ we have
\begin{equation}\label{eq:convgamma}
\gamma_{N,t}^{(k)} \to |\ph_t \rangle \langle \ph_t |^{\otimes k}
\end{equation}
with respect to the same topology. Here $\ph_t \in H^1 (\bR^3)$ is
the solution of the nonlinear Gross-Pitaevskii equation
\begin{equation}
i\partial_t \ph_t = -\Delta \ph_t + 8\pi a_0 |\ph_t|^2 \ph_t
\end{equation}
with $\ph_{t=0} = \ph$.
\end{theorem}

Both theorems have analogous versions for initial data describing
mixed states (that is $\gamma_N$ is not an orthogonal projection).
For example, suppose that  $\gamma_N$ is a family of density
matrices satisfying \begin{equation}\label{eq:condomega} \tr \;
H_N^k \, \gamma_N \leq C^k N^k \qquad \text{ and } \quad
 \gamma_N^{(k)} \to \omega_0^{\otimes k} \end{equation}
where $\omega_0$ is a one-particle density matrix and
\[ \omega_0^{\otimes k} (\bx_k; \bx'_k) = \prod_{j=1}^k \omega_0
(x_j ; x'_j) .
\]
Then for every $t \in \bR$ and $k \geq 1$ we have
\begin{equation}\label{eq:thmomega}
\gamma_{N,t}^{(k)} \to \omega_t^{\otimes k}
\end{equation}
where $\omega_t$ is the solution of the nonlinear Hartree equation
\begin{equation}
i\partial_t \omega_t = [ -\Delta + 8\pi a_0 \varrho_t , \omega_t]
\qquad  \varrho_t (x) = \omega_t (x;x), \quad \omega_{t=0} =
\omega_{0}
\end{equation}
The last equation is equivalent to (\ref{eq:GP}) if $\omega_t
=|\ph_t \rangle \langle \ph_t|$.

\bigskip

Lieb and Seiringer  \cite{LS} have proved that, for pure states, the
assumption
\[ \gamma^{(1)}_{N} \to |\ph \rangle \langle \ph|
\qquad \text{as } N \to \infty
\]
implies automatically (\ref{eq:thmassum2}) for every $k \geq 1$ (see
the argument after Theorem 1 in that paper)\footnote{We thank Robert
Seiringer for pointing out this result to us.}. For mixed initial
states we still need the second condition in (\ref{eq:condomega})
for all $k \geq 1$ in order to prove (\ref{eq:thmomega}).

\bigskip

Now we  comment on the assumption of asymptotic factorization
(\ref{eq:thm2ass2}) for the initial data $\psi_N$. The most natural
example that satisfies this condition is the factorized wave
function $\psi_N (\bx) = \prod_{j=1}^N \ph (x_j)$. If, additionally,
 $\ph \in H^1 (\bR^3)$, then
(\ref{eq:thm2ass1}) is also satisfied by the Schwarz  and
Sobolev inequalities. The evolution of $\psi_N$
is therefore governed by the GP equation according to
 Theorem \ref{thm:main2}.
 This is, however, somewhat surprising because the emergence
of the scattering length in the GP equation indicates that the wave
function has a characteristic short scale correlation structure,
which is clearly absent in the factorized initial data.  We
shall discuss this issue in more details in Section \ref{sec:bb}.

\medskip

{F}rom the  physical point of view, however, the product initial
wave function is not the most relevant one. In real physical
experiments, the initial state is prepared by cooling down a trapped
Bose gas at extremely low temperatures. This state can be modelled
by the ground state $\psitrap$ of the Hamiltonian
\[ \Htrap = \sum_{j=1}^N \left(-\Delta_j + \Vtrap (x_j) \right) +
\sum_{i<j}^N V_N (x_i -x_j)\] with a trapping potential $\Vtrap (x)
\to \infty$ as $|x| \to \infty$.
In Appendix \ref{sec:trap}, we prove that
assumptions  (\ref{eq:thm2ass1}) and  (\ref{eq:thm2ass2})
are satisfied for $\psitrap$.
In other words, Theorem
\ref{thm:main2} can be used to describe the evolution of the ground
state of $\Htrap$, after the traps are removed (see Corollary
\ref{cor:main3}). This provides a mathematically rigorous analysis
of recent experiments in condensed matter physics, where the
evolution of initially trapped Bose-Einstein condensates is
observed.

\medskip

In Appendix \ref{sec:W}, we show that Theorem \ref{thm:main2} can
also be applied to a general class of initial data, which are in
some sense close to the ground state of the Hamiltonian $\Htrap$.
 The ground state of a dilute Bose system with
interaction potential $V_N$ is believed to be very close to the form
\be\label{W}
 W_N (\bx) :=  \prod_{i<j} f(N(x_i-x_j)),
\ee where $f=1-w_0$ is the zero-energy solution \eqref{ze}. We
remark that Dyson \cite{Dy} used a different function which was not
symmetric, but the short distance behavior was the same as in $W_N$.
An example of a family of initial wave functions which have local
structure given by $W$ is given by wave functions of the type
\begin{equation}\label{eq:psi=Wphi}
\psi_N (\bx) = W_N (\bx) \prod_{j=1}^N \ph (x_j)
\end{equation}
where $\ph \in H^1 (\bR^3)$. Due to the factor $W_N$, this function
carries the characteristic short scale structure of the ground
state. We will prove in Lemma \ref{lm:W} that wave functions of the
form (\ref{eq:psi=Wphi}) (with correlations cutoff at length
scales $\ell \gg  N^{-1}$) satisfies the assumptions
\eqref{eq:thm2ass1} and \eqref{eq:thm2ass2}.

\medskip

Part of Theorem \ref{thm:main2} was proved in  \cite{ESY} for
systems with the pair interaction cut off whenever three or more
particles are much closer to each other than the mean particle
distance, $N^{-1/3}$. For this model, it was proved that any
limiting point of $\gamma_N^{(k)}$ satisfies the infinite BBGKY
hierarchy (see Section \ref{sec:bb}) with coupling constant $8\pi
a_0$. The uniqueness of the solution to the hierarchy was
established in \cite{ESY2}. In the current paper we remove this
cutoff and establish  the apriori bounds needed for the uniqueness
theorem in \cite{ESY2}.
\medskip

The Hamiltonian \eqref{eq:ham1} is a special case of the Hamiltonian
\begin{equation}\label{beta}
H_{\beta, N} := -\sum_{j=1}^N \Delta_j + \frac{1}{N} \sum_{i<j}^N
N^{3\beta} V (N^{\beta} (x_i -x_j))
\end{equation}
introduced in \cite{EESY} and \cite{ESY2}. In \cite{ESY2} we have
proved a version of Theorem \ref{thm:main2} for  $0< \beta <1/2$
provided the initial data is given by a product state $\psi_{N}
(\bx) = \prod_{j=1}^k \ph (x_j)$ for some $\ph \in H^1 (\bR^3)$. In
this case the limiting macroscopic equation was given by
\[ i \partial_t \ph_t = -\Delta \ph_t + b_0 |\ph_t|^2 \ph_t \,,\]
with $b_0 = \int \rd x \, V(x)$. Note that $N^{3\beta} V(N^\beta x)$
is an approximate delta function on a scale much bigger than
$O(1/N)$, the scattering length of $\frac{1}{N} V_{N}$. This
explains why the strength of the on-site potential is given by the
semiclassical approximation $b_0$ of the $8\pi a_0$. With the
techniques used in this paper, it is straight-forward to extend the
result of \cite{ESY2}  to all $\beta <1$ with the same coefficient
$b_0$ in the limiting one-body equation provided that $\rho$ (from
\eqref{eq:defrho}) is small enough. Combining this comment with
Theorem \ref{thm:main1} and \ref{thm:main2}, we have shown that the
one particle density matrix for the $N$-body Schr\"odinger equation
with Hamiltonian given by \eqref{beta} converges to the
Gross-Pitaevskii equation with coupling constant given by
\begin{equation}
\sigma = \begin{cases} b_0,  &  \quad \text{if} \;\;  0< \beta < 1\\
8 \pi a_0,  & \quad \text{if} \;\; \beta = 1\;.
\end{cases}
\end{equation}

The case $\beta=0$ is the mean-field case and  the limiting one-body
equation is the Hartree equation: \be i\partial_t\varphi_t = -\Delta
\varphi_t + (V\ast |\varphi_t|^2)\varphi_t\; . \label{eq:hartree}
\ee This was established by Hepp \cite{H} for smooth potential.
Ginibre and Velo \cite{GV} considered singular potentials but with a
specific initial data based on second quantized formalism. Spohn
\cite{Sp} introduced a new approach to this problem using the BBGKY
hierarchy. Recent progresses on mean-field limit of quantum dynamics
have been based on the BBGKY hierarchy and we mention only a few:
the Coulomb potential case \cite{BGM, EY}, the
 pseudo-relativistic Hamiltonian with Newtonian
interaction  \cite{ES},  and the delta function interaction in one
dimension by Adami, Bardos, Golse and Teta   \cite{ABGT} \cite{AGT}.
In next section, we review the BBGKY hierarchy and the two-scale
nature of the eigenfunctions of interacting Bose systems.

\section{The BBGKY Hierarchy}\label{sec:bb}

The time evolution of the density matrices $\gamma_{N,t}^{(k)}$, for
$k=1,\dots, N$, is given by a hierarchy of $N$ equations, commonly
known as the BBGKY hierarchy:
\begin{equation}\label{eq:BBGKY}
\begin{split}
i\partial_t \gamma^{(k)}_{N,t} = \; &
\sum_{j=1}^k \left[ -\Delta_j ,
\gamma^{(k)}_{N,t} \right] + \sum_{i<j}^k \left[ V_N (x_j - x_i) ,
\gamma^{(k)}_{N,t} \right] \\ &
+ (N-k) \sum_{j=1}^k \tr_{k+1} \;
\left[ V_N (x_j -x_{k+1}), \gamma^{(k+1)}_{N,t} \right]
\end{split}
\end{equation}
for $k =1, \dots ,N$ (we use the convention that
$\gamma_{N,t}^{(k)}=0$ if $k >N$). Here $\tr_{k+1}$ denotes the
partial trace over the $(k+1)$-th particle. In particular, the
density matrix $\gamma_{N,t}^{(1)} (x_1;x'_1)$ satisfies the
equation
\begin{equation}\label{1.1}
\begin{split}
i\partial_t \gamma_{N,t}^{(1)} ({x}_1 ; {x}'_1) = \; &(-\Delta_{x_1}
+ \Delta_{x'_1}) \gamma_{N,t}^{(1)} ({x}_1; {x}'_1) \\& + (N -1)
\int \rd x_{2} \left( V_N (x_1 - x_{2}) - V_N (x'_1 - x_{2}) \right)
\gamma_{N,t}^{(2)} ({x}_1 , x_{2};{x}'_1 , x_{2}),
\end{split}
\end{equation}
To close this equation, one needs to assume some
relation between
$\gamma_{N,t}^{(2)}$ and $\gamma_{N,t}^{(1)}$. The simplest
assumption would be  the factorization property, i.e., \be
\gamma^{(2)}_{N, t} (x_1, x_2; x_1', x'_2) = \gamma^{(1)}_{N,t}
(x_1; x_1') \gamma^{(1)}_{N,t} (x_2; x'_2) \, . \ee This does not
hold for finite $N$, but it may hold for a limit point
$\gamma^{(k)}_{t}$ of $\gamma^{(k)}_{N, t}$ as $N\to \infty$, i.e.,
\be\label{g2} \gamma^{(2)}_{t} (x_1, x_2; x_1', x'_2) =
\gamma^{(1)}_{t} (x_1; x_1') \gamma^{(1)}_{t} (x_2; x'_2) \, . \ee
Under this assumption, $\gamma_{t}^{(1)}$ satisfies the limiting
equation
\begin{equation}\label{1.2}
i\partial_t \gamma_{t}^{(1)} ({x}_1; {x}'_1) = \; (-\Delta_{x_1} +
\Delta_{x'_1}) \gamma_{t}^{(1)} ({x}_1; {x}'_1) +  \left( Q_t(x_1)
-Q_t(x'_1)\right) \gamma_{t}^{(1)} ({x}_1; {x}'_1)
\end{equation}
where \be\label{b0} Q_t(x) := \lim_{N \to \infty} N \int \rd y
V_N(x-y) \rho_t(y), \quad \rho_t(x) = \gamma_{t}^{(1)} ({x}; {x})\,
. \ee If $\rho_t(x)$ is continuous, then $Q_t$ is given by
$$
Q_t(x) = b_0 \rho_t(x) .
$$
Thus (\ref{1.2})
 gives the GP equation with  a coupling constant
$\sigma = b_0$ instead of $\sigma = 8\pi a_0$. This explains the
case if $\beta< 1$. For $\beta=1$, we note that  $b_0/8 \pi $ is the
first Born approximation to the scattering length $a_0$  and the
following inequality holds: \be\label{slbound} a_0 \le \frac {b_0}{8
\pi}= \frac 1 {4 \pi} \int_{\bR^3} \,  \frac  1 2 \, V (x) \, dx \,
. \ee Recall that the ground state of a dilute Bose system with
interaction potential $V_N$ is believed  to be very close to $
 W (\bx)$ (see \eqref{W}).
We assume, for the moment, that the ansatz,  $\psi_t (\bx) = W(\bx)
\phi_t (\bx)$ with $\phi_t$ a product function, holds  for all time.
The reduced density matrices for $\psi_t (\bx) $ satisfy
\be\label{g3}
\gamma^{(2)}_t (x_1, x_2; x_1', x_2')\sim
f(N(x_1-x_2))f(N(x'_1-x_2')) \gamma_{t}^{(1)} ({x}_1;
{x}'_1)\gamma_t^{(1)}(x_2; x_2')\,.
\ee
Together with \eqref{13} and
the assumption that $\rho_t$ is smooth on scale $1/N$, we have
\be\label{g4}
\lim_{N \to \infty}  N \int \rd x_{2} V_N (x_1 - x_{2})
\gamma_{N,t}^{(2)} ({x}_1 , x_{2} ; {x}'_1 , x_{2}) = 8 \pi a_0
\gamma_{t}^{(1)} ({x}_1; {x}'_1) \rho_t(x_1) \, .
\ee
This formula
is valid for $|x_1-x'_1|\gg 1/N$.  We have used that $\lim_{|x| \to
\infty} f (x) =1$. For pure states,
this gives the GP equation with the correct
dependence on the scattering length.

Notice that the correlation in $\gamma^{(2)}$ occurs at the scale
$1/N$, which vanishes  in a weak limit and the product relation
\eqref{g2} will hold. However, this short distance correlation shows
up in the GP equation due to the singular potential $N V_N (x_1 -
x_{2})$.  This phenomena occurs for the ground state as proved in
\cite{LSY1}. Our task is to characterize wave functions with this
short scale structure and establish it for the time evolved states.
The key observation is the following Proposition. Recall the
assumptions on $V$ from Section \ref{sec:main} and
 that $1-w(x)$ denotes the zero energy solution to
 $-\Delta + \frac{1}{2}V_{N}$  (\ref{eq:defw1}).
 We will use the short notation $w_{ij}:= w(x_i-x_j)$, $\nabla
w_{ij} = (\nabla w) (x_i - x_j)$ (note that $\nabla w_{ij} = -
\nabla w_{ji}$).

\begin{proposition}[$H_N^2$-energy estimate]\label{prop:H2}
Suppose that $\rho$ (defined in (\ref{eq:defrho})) is small enough.
Then, there exists a universal constant $c>0$ such that, for every
$\psi \in L^2_s (\bR^N)$, and for every fixed indices $i \neq j$,
$i,j =1,\dots ,N$, we have
\begin{equation}\label{eq:H2}
\langle \psi, H_N^2 \psi \rangle \geq (1- c\rho) N(N-1) \int \,
\left( 1- w_{ij}\right)^2 \, |\nabla_i \nabla_j \, \phi_{ij}|^2
\end{equation}
where $\phi_{ij}$ defined by $\psi = (1- w_{ij}) \phi_{ij}$.
\end{proposition}

If $\phi_{ij}$ is singular when $x_i$ approaches $x_j$, then
$\nabla_i \nabla_j  \phi_{ij}$ cannot be $L^2$-integrable. This
Proposition thus shows that  the short distance behavior of any
function $\psi$ with $\langle \psi, H_N^2 \psi \rangle \le C N^{2}$
is given by $(1-w(x_i -x_j))$  when $x_{i}$ is near $x_{j}$.

\medskip

We emphasized the importance of the local structure $(1-w(x_i
-x_j))$ for obtaining the scattering length $a_0$. While Theorem
\ref{thm:main2} concerns only the one particle density matrix in the
weak limit and no statement on the local structure is made at all,
the validity of the GP equation does suggest the existence of this
structure. For the initial data \eqref{eq:psi=Wphi} beginning with
this local structure, it simply means its preservation by the
dynamics. This is indeed the case if the local structure of the
initial data $\psi$ is precise enough so that $\langle \psi, H_N^2
\psi \rangle \le C N^2$, see Proposition \ref{prop:H2}.

For the product initial state, there is no such structure to begin
with. Theorem \ref{thm:main2} thus indicates  that on some short
length scale a local structure similar to $(1-w(x_i -x_j))$ forms in
a very short time which approaches zero in the limit $N\to \infty$.
Heuristically, notice that the two particle dynamics is described by
the operator
$$
i\partial_t -\Delta_{x_1}-\Delta_{x_2}
- V_N(x_1-x_2) = N^2 \left [ i\partial_T
-\Delta_{X_1}-\Delta_{X_2}- V(X_1-X_2) \right ]
$$
where $X_i = Nx_i$ and $T= N^2 t$ are the microscopic coordinates.
The small positive time behavior of the original wave function on
the short length scale is the same as the long time behavior
in the microscopic coordinates. Clearly, we expect the long time
dynamics to be characterized by the relaxation to the zero energy
solution. This picture, however, is far from rigorous as the true
$N$-body dynamics develops higher order correlations as well.

On the other hand, the local structure $(1-w(x_i -x_j))$ cannot be
the only singular piece of the wave function in positive time for
product initial states. A simple calculation shows that the energy
per particle
of a product initial state $\psi_N (\bx)= \prod_{j=1}^N \ph
(x_j)$ is given by
\begin{equation}
\lim_{{N\to \infty}}N^{-1}\langle \psi_{N}, H_N \psi_{N} \rangle =
\int_{\bR^3} \rd x \; |\nabla \ph (x)|^{2} + \frac { b_0} 2 \int_{\bR^3}
\rd x \;
|\ph(x)|^{4}
\end{equation}
where  $b_0= \int V$. This is different from the GP energy
functional \eqref{nls} due to the coupling constant. Since the energy is a
constant of the motion, this implies that the GP theory does not
predict the evolution of the energy. If we grant that the local
structure $(1-w(x_i -x_j))$ does form for positive time $t>0$, the
discrepancy in energy suggests that there is some energy on
intermediate length scales of order $N^{-\alpha}$,
$0<\alpha <1$  which is not captured by the GP
theory.  This excess energy apparently does not
participate in the evolution of the density matrix on
length scale of order one which is the only scale that is
visible by our weak limit.
We do not know if such a picture can be established
rigorously.

\medskip

{\it Notation.} We will denote  an arbitrary constant by $C$. In
general $C$ can depend on the choice of the unscaled potential $V$.
Universal constants, {\it independent of $V$}, will be denoted by
$c$. We write $f(N) = o(N^{\alpha})$ if there is $\delta>0$ such
that $N^{-\alpha +\delta} f(N) \to 0$ as $N \to \infty$ (unless
stated otherwise, this convergence does not need to be uniform in
the other relevant parameters). We also write $f(N) \ll g(N)$ if
$f(N)/g(N) = o(1)$. Integrations without specified domains are
always understood on the whole space ($\bR^3$, $\bR^{3k}$ or
$\bR^{3N}$ according to the integrand) with the Lebesgue measure.

\section{Proof of Theorem \ref{thm:main1} and Theorem \ref{thm:main2}}
\label{sec:outline}
\setcounter{equation}{0}

In this section we present the main steps of the proofs and we
reduce the argument to a sequence of key theorems and propositions.
These will be proven in the rest of the paper.

\medskip

We start with defining the space of  density matrices
that depend continuously on the time parameter with respect to the
weak* topology. To use Arzela-Ascoli compactness argument, we will need
to establish the concept of uniform continuity in this space,
thus we have to metrize the weak* topology.

Since $\cK_k$ is separable, we can fix a dense countable subset of
the unit ball of $\cK_k$: we denote it by $\{J^{(k)}_i\}_{i \ge 1}
\in \cK_k$, with $\| J^{(k)}_i \|_{\cK_k} \leq 1$ for all $i \ge 1$.
Using the operators $J^{(k)}_i$ we define the following metric on
$\cL^1_k$: for $\gamma^{(k)}, \bar \gamma^{(k)} \in \cL^1_k$ we set
\begin{equation}\label{eq:etak}
\eta_k (\gamma^{(k)}, \bar \gamma^{(k)}) : = \sum_{i=1}^\infty
2^{-i} \left| \tr \; J^{(k)}_i \left( \gamma^{(k)} - \bar
\gamma^{(k)} \right) \right| \, .
\end{equation}
Then the topology induced by the metric $\eta_k$ and
the weak* topology are equivalent on the unit ball of $\cL^1_k$
 (see \cite{Ru}, Theorem 3.16) and hence on
any ball of finite radius as well. In other words, a uniformly
bounded sequence $\gamma_N^{(k)} \in \cL^1_k$ converges to
$\gamma^{(k)} \in \cL^1_k$ with respect to the weak* topology, if
and only if $\eta_k (\gamma^{(k)}_N , \gamma^{(k)}) \to 0$ as $N \to
\infty$.

For a fixed $T > 0$, let $C ([0,T], \cL^1_k)$ be the space of
functions of $t \in [0,T]$ with values in $\cL^1_k$ which are
continuous with respect to the metric $\eta_k$. On $C ([0,T],
\cL^1_k)$ we define the metric
\begin{equation}\label{eq:whetak}
\widehat \eta_k (\gamma^{(k)} (\cdot ) , \bar \gamma^{(k)} (\cdot ))
:= \sup_{t \in [0,T]} \eta_k (\gamma^{(k)} (t) , \bar \gamma^{(k)}
(t))\,.
\end{equation}
Finally, we denote by $\tau_{\text{prod}}$ the topology on the
space $\bigoplus_{k \geq 1} C([0,T], \cL^1_k)$ given by the product
of the topologies generated by the metrics $\wh \eta_k$ on $C([0,T],
\cL^1_k)$.

\medskip
\begin{proof}[Proof of Theorem \ref{thm:main1}] The proof is divided
in several steps.

\medskip

{\it Step 1. Compactness of $\Gamma_{N,t}=\{ \gamma^{(k)}_{N,t}
\}_{k\geq 1}$.} We set $T>0$ and work on the interval $t\in [0, T]$.
Negative times can be handled analogously. We will prove in Theorem
\ref{thm:compactness} that the sequence $\Gamma_{N,t}^{(k)} = \{
\gamma^{(k)}_{N,t} \}_{k \geq 1} \in \bigoplus_{k \geq 1} C([0,T],
\cL^1_k)$ is compact with respect to the product topology
$\tau_{\text{prod}}$ defined above (we use the convention that
$\gamma^{(k)}_{N,t} =0$ if $k >N$). Moreover, we also prove in Theorem
\ref{thm:compactness}, that any limit point $\Gamma_{\infty,t} =
\{\gamma_{\infty,t}^{(k)} \}_{k\geq 1} \in \bigoplus_{k \geq 1} C (
[0,T], \cL^1_k)$ is such that, for every $k \geq 1$,
$\gamma_{\infty,t}^{(k)} \geq 0$, and $\gamma_{\infty,t}^{(k)}$ is
symmetric w.r.t. permutations. In Proposition \ref{prop:apriorik} we
also show that
\begin{equation}\label{eq:apri}
\tr \; (1-\Delta_1) \dots (1-\Delta_k) \, \gamma^{(k)}_{\infty,t}
\leq C^k
\end{equation}
for every $t \in [0,T]$ and every $k \geq 1$. Note that, for finite
$N$, the densities $\gamma^{(k)}_{N,t}$ do not satisfy estimates
such as (\ref{eq:apri}) (at least not uniformly in $N$), because
they contain a short scale structure. Only after taking the weak
limit, we can prove (\ref{eq:apri}).

\medskip

{\it Step 2. Convergence to the infinite hierarchy.} In Theorem
\ref{thm:convergence} we prove that any limit point $
\Gamma_{\infty,t} = \{ \gamma_{\infty,t}^{(k)} \}_{k\geq 1} \in
\bigoplus_{k\geq 1} C([0,T], \cL^1_k)$ of $\Gamma_{N,t} = \{
\gamma_{N,t}^{(k)} \}_{k\ge1}$ with respect to the product topology
$\tau_{\text{prod}}$ is a solution of the infinite hierarchy of
integral equations ($k=1,2, \ldots$)
\begin{equation}\label{eq:BBGKYinf}
\gamma_{\infty,t}^{(k)}  = \; \cU^{(k)} (t) \gamma_{\infty,0}^{(k)}
- 8\pi i a_0 \sum_{j=1}^k \int_0^t \rd s \, \cU^{(k)} (t-s)
\tr_{k+1} \left[ \delta (x_j -x_{k+1}),
\gamma_{\infty,s}^{(k+1)}\right]\,
\end{equation}
with initial data $\gamma_{\infty,0}^{(k)} = |\ph\rangle
\langle\ph|^{\otimes k}$. Here $\tr_{k+1}$ denotes the partial trace
over the $(k+1)$-th particle, and  $\cU^{(k)} (t)$ is the free
evolution, whose action on $k$-particle density matrices is given by
\[\cU^{(k)} (t) \gamma^{(k)} := e^{it\sum_{j=1}^k \Delta_j}
\gamma^{(k)} e^{-it\sum_{j=1}^k \Delta_j}\,. \] Note that
\eqref{eq:BBGKYinf} is the (formal) limit of the $N$-particle BBGKY
hierarchy  \eqref{eq:BBGKY} (written in integral form) if we replace
the limit of $NV_N(x)$ with $8\pi a_0\delta(x)$ (see \eqref{g4}).

The one-particle wave function $\ph$ was introduced in
(\ref{eq:thmassum2}). {F}rom (\ref{eq:thmassum1}) and the positivity
of the potential we note that
\begin{equation}\label{eq:ph1}
C N \geq \langle \psi_N, \left(H_N +N \right) \psi_N \rangle \geq N
\tr \left(1 - \Delta \right) \gamma^{(1)}_{N}\,.
\end{equation}
Since by (\ref{eq:thmassum2}), $\gamma^{(1)}_N \to |\ph\rangle
\langle \ph|$ as $N \to \infty$, w.r.t. the weak * topology of
$\cL^1 (L^2 (\bR^{3}))$, it follows from (\ref{eq:ph1}) that $\tr \,
(1-\Delta) |\ph \rangle \langle \ph | \leq C$, and therefore that
$\ph \in H^1 (\bR^3)$.

We remark here that
the family of factorized densities,
\begin{equation}
\label{eq:factsol} \gamma^{(k)}_t = |\ph_t \rangle \langle
\ph_t|^{\otimes k},
\end{equation}
is a solution of the infinite hierarchy (\ref{eq:BBGKYinf}) if
$\ph_t$ is the solution of the nonlinear Gross-Pitaevskii equation
(\ref{eq:GP}) with initial data $\ph_{t=0}= \ph$. The nonlinear
Schr\"odinger equation (\ref{eq:GP}) is well posed in $H^1 (\bR^3)$
and it conserves the energy, $\cE (\ph) := \frac{1}{2} \int |\nabla
\ph|^2 + 4 \pi a_0 \int |\ph|^4$.
{F}rom $\ph \in H^1 (\bR^3)$, we thus
obtain that $\ph_t \in H^1 (\bR^3)$ for every $t \in \bR$, with a
uniformly bounded $H^1$-norm. Therefore
\begin{equation}\label{eq:phik}
\tr \; (1-\Delta_1) \dots (1-\Delta_k) |\ph_t \rangle \langle
\ph_t|^{\otimes k} \leq \| \ph_t \|_{H^1}^k \leq C^k
\end{equation}
for all $t \in \bR$, and a constant $C$ only depending on the
$H^1$-norm of $\ph$.

\medskip

{\it Step 3. Uniqueness of the solution to the infinite hierarchy.}
In Section 9 of \cite{ESY2} we proved the following theorem, which
states the uniqueness of solution to the infinite hierarchy
(\ref{eq:BBGKYinf}) in the space of densities satisfying the a
priori bound (\ref{eq:apri}). The proof of this theorem is based on
a diagrammatic expansion of the solution of (\ref{eq:BBGKYinf}).
\begin{theorem}\label{thm:uniqueness}[Theorem 9.1 of \cite{ESY2}]
Suppose $\Gamma = \{ \gamma^{(k)} \}_{k \geq 1} \in \bigoplus_{k
\geq 1} \cL^1_k$ is such that
\begin{equation}
\tr\; (1-\Delta_1) \dots (1-\Delta_k) \gamma^{(k)} \leq C^k \, .
\end{equation}
Then, for any fixed $T >0$, there exists at most one solution
$\Gamma_t = \{ \gamma^{(k)}_t \}_{k \geq 1} \in \bigoplus_{k \geq 1}
C([0,T], \cL^1_k)$ of (\ref{eq:BBGKYinf}) such that
\begin{equation}\label{eq:bougam}
\tr\; (1-\Delta_1) \dots (1-\Delta_k) \gamma^{(k)}_t \leq C^k
\end{equation}
for all $t \in [0,T]$ and for all $k \geq 1$.
\end{theorem}

\medskip

{\it Step 4. Conclusion of the proof.} {F}rom Step 2 and Step 3 it
follows that the sequence $\Gamma_{N,t} = \{ \gamma^{(k)}_{N,t}
\}_{k \geq 1} \in \bigoplus_{k \geq 1} C([0,T],\cL_k^1)$ is
convergent with respect to the product topology
$\tau_{\text{prod}}$; in fact a compact sequence with only one limit
point is always convergent. Since  the family of densities
$\Gamma_t = \{ \gamma^{(k)}_t \}_{k \geq 1}$ defined in
(\ref{eq:factsol}) satisfies (\ref{eq:phik}) and it
is a solution of (\ref{eq:BBGKYinf}),
 it follows that $\Gamma_{N,t} \to
\Gamma_t$ w.r.t. the topology $\tau_{\text{prod}}$.
The estimates are  uniform in $t\in [0,T]$,
thus we can also conclude that $\wh\eta_k(\gamma^{(k)}_{N,t}, \gamma_t^{(k)})
\to 0$.
In particular
this implies that, for every fixed $k \geq 1$, and $t \in [0,T]$,
$\gamma_{N,t}^{(k)} \to \gamma_t^{(k)}$ with respect to the weak*
topology of $\cL^1_k$. This completes the proof of Theorem
\ref{thm:main1}. Actually, the estimates
are uniform in $t\in [0,T]$, and
thus we can also conclude that $\wh\eta_k(\gamma^{(k)}_{N,t}, \gamma_t^{(k)})
\to 0$.
\end{proof}

\medskip

Next we prove Theorem \ref{thm:main2}; to this end we regularize the
initial wave function, and then we apply the same arguments as in
the proof of Theorem \ref{thm:main1}.

\begin{proof}[Proof of Theorem \ref{thm:main2}]

Fix $\kappa >0$ and $\chi \in C_0^{\infty} (\bR)$, with $0\leq
\chi\leq 1$, $\chi (s) = 1$, for $0 \leq s \leq 1$, and $\chi (s)
=0$ if $s \geq 2$. We define the regularized initial wave function
\[
\wt \psi_N := \frac{ \chi (\kappa H_N /N) \psi_N }{ \| \chi (\kappa
H_N /N) \psi_N \|} ,
\]
and  we denote by $\wt\psi_{N,t}$ the solution of the Schr\"odinger
equation (\ref{eq:schr}) with initial data $\wt \psi_N$. Denote by
$\wt \Gamma_{N,t} = \{ \wt \gamma_{N,t}^{(k)} \}_{k=1}^\infty$ the
family of marginal densities associated with $\wt \psi_{N,t}$. By
convention, we set $\wt\gamma^{(k)}_{N,t}:=0$ if $k >N$. The tilde
in the notation indicates the dependence on the cutoff parameter
$\kappa$. In Proposition \ref{prop:initialdata}, part i), we prove
that
\begin{equation}\label{eq:thm2-1}
\langle \wt \psi_{N,t} , H_N^k \wt\psi_{N,t} \rangle \leq {\wt C}^k
N^k \end{equation} if $\kappa >0$ is sufficiently small (the
constant $\wt C$ depends on $\kappa$). Moreover, using the strong
asymptotic factorization assumption (\ref{eq:thm2ass2}), we prove in
part iii) of Proposition \ref{prop:initialdata} that for every
$J^{(k)} \in \cK_k$,
\begin{equation}\label{eq:thm2-2}
\tr \; J^{(k)} \left( \wt \gamma_N^{(k)} - |\ph \rangle \langle
\ph|^{\otimes k} \right) \to 0
\end{equation}
as $N \to \infty$. {F}rom (\ref{eq:thm2-1}) and (\ref{eq:thm2-2}),
we observe that the assumptions (\ref{eq:thmassum1}) and
(\ref{eq:thmassum2}) of Theorem \ref{thm:main1} are satisfied by the
regularized wave function $\wt \psi_N$ and by the regularized
marginal densities $\wt \gamma^{(k)}_{N,t}$. Therefore, applying
Theorem \ref{thm:main1}, we obtain that, for every $t \in \bR$ and
$k \geq 1$,
\begin{equation}\label{eq:conve} \wt \gamma_{N,t}^{(k)} \to |\ph_t \rangle \langle
\ph_t|^{\otimes k} \end{equation} where $\ph_t$ is the solution of
(\ref{eq:GP}).

\medskip

It remains to prove that the densities $\gamma^{(k)}_{N,t}$
associated with the original wave function $\psi_{N,t}$ (without
cutoff $\kappa$) converge and have the same limit as the regularized
densities $\wt \gamma^{(k)}_{N,t}$. This follows from Proposition
\ref{prop:initialdata}, part ii), where we prove that
\[
\| \psi_{N,t} - \wt\psi_{N,t} \| = \| \psi_{N} -
\wt\psi_{N} \| \leq C \kappa^{1/2} \; ,
\]
 where the constant
$C$ is independent of $N$ and $\kappa$. This implies that, for every
$J^{(k)} \in \cK_k$, we have
\begin{equation}\label{eq:remove}
\Big| \tr \; J^{(k)} \left( \gamma^{(k)}_{N,t} - \wt
\gamma^{(k)}_{N,t} \right) \Big| \leq C \kappa^{1/2}
\end{equation}
where the constant $C$ depends on $J^{(k)}$, but is independent of
$N$,  $k$ or $\kappa$. Therefore, for fixed $k \geq 1$, $t \in \bR$,
$J^{(k)} \in \cK_k$,  we have
\begin{equation}\label{eq:lastproof}
\begin{split}
\Big| \tr \; J^{(k)} \left( \gamma^{(k)}_{N,t} - |\ph_t \rangle
\langle \ph_t |^{\otimes k} \right) \Big| \leq &\; \Big| \tr \;
J^{(k)} \left( \gamma^{(k)}_{N,t} - \wt \gamma^{(k)}_{N,t} \right)
\Big| + \Big| \tr \; J^{(k)} \left( \wt \gamma^{(k)}_{N,t} - |\ph_t
\rangle \langle \ph_t |^{\otimes k} \right) \Big| \\
\leq & \; C \kappa^{1/2} + \Big| \tr \; J^{(k)} \left( \wt
\gamma^{(k)}_{N,t} - |\ph_t \rangle \langle \ph_t |^{\otimes k}
\right) \Big| \, .
\end{split}
\end{equation}
Since $\kappa >0$ was arbitrary, it follows from (\ref{eq:conve})
that the l.h.s. of (\ref{eq:lastproof}) converges to zero as $N \to
\infty$. This completes the proof of Theorem \ref{thm:main2}.
\end{proof}

\section{Energy Estimates}\label{sec:energy}
\setcounter{equation}{0}

In this section we prove two energy estimates that are the most
important new tools used in the proof of the main theorem. Both
estimates concern the smoothness of the solution $\psi_{N,t}(\bx)$
of the Schr\"odinger equation (\ref{eq:schr}),
 uniformly in $N$ (for $N$
large enough) and in $t \in \bR$. However, due to the short
scale structure of the interaction, $V_N$, uniform smoothness, say
in the $x_1$ variable, cannot be expected near the collision points
$|x_1-x_j|\sim 1/N$, $j=2, 3,  \ldots , N$.
 The key observation is that $x_1\to \psi_{N,t}(\bx)$ will
nevertheless be smooth away from these regimes, whose total volume
is negligible. For technical reasons, the excluded regime will be
somewhat larger, $|x_1-x_j|\ge \ell$, but still with $N\ell^3\ll 1$.
The same statement holds for the smoothness in an arbitrary but
fixed number of variables, $x_1, \ldots, x_k$. This is the content
of our second energy estimate Proposition \ref{prop:Hk}.

Our first energy estimate, Proposition \ref{prop:H2}, controls only
two derivatives, but it is more refined: it establishes smoothness
of $\psi_{N,t}(\bx)$ in the $x_i$ and $x_j$ variables (for any fixed
pair $i,j$) after removing the explicit short scale factor
$(1-w(x_i-x_j))$. This factor represents the short scale effect of
the two body interaction $V_N(x_i-x_j)$ on the wave function and it
is responsible for the emergence of the scattering length
\eqref{eq:scatlength}.

\subsection{$H_N^2$ Energy Estimate}

In this section, we shall prove Proposition \ref{prop:H2}. We first
collect some important properties of $w(x)$ \eqref{eq:defw1} in the
following lemma. This lemma is an improved version of Lemma A.2 from
\cite{ESY}. By defining $\rho$ somewhat differently (see
\eqref{eq:defrho}), we also correct a minor error in (A.6) and
(A.19) of \cite{ESY}.

\begin{lemma}\label{lm:w}
Suppose $V \geq 0$ is smooth, spherical symmetric, compactly
supported and with scattering length $a_0$. Let
\begin{equation}\label{eq:defrho1}
\rho = \sup_{r \geq 0} r^2 V(r) + \int_0^{\infty} \rd r \, r \, V(r)
\end{equation} and let $a=a_0/N$ be the scattering length of the rescaled
potential $V_N$. Then the following hold with constants uniform in
$N$.
\begin{itemize}
\item[i)] There exists a constant $C_0>0$, which depends on the unscaled potential
$V$, such that \begin{equation}\label{eq:w>01} C_0\leq 1-w(x) \leq 1
\qquad \text{for all } x \in \bR^3.\end{equation} Moreover, there
exists a universal constant $c$ such that
\begin{equation}\label{eq:w>0} 1-c\rho \leq 1 - w(x) \leq 1 \qquad
\text{for all } x \in \bR^3 \; .
\end{equation}
\item[ii)] Let $R$ be such that $\supp V \subset \{ x \in \bR^3 : |x| \leq R \}$. Then
\[ w(x) = \frac{a}{|x|} \qquad \text{for all } x \quad \text{with }
|x| > R/N \; .
\]
\item[iii)] There exist constants $C_1$, $C_2$, depending on $V$, such that
\begin{equation}\label{eq:w-1}
|\nabla w (x) | \leq C_1 N, \qquad  |\nabla^2 w(x)| \leq C_2 N^2\, ,
\qquad \text{for all } x \in \bR^3 \, .
\end{equation}
Moreover, there exists a universal constant $c$ such
that
\begin{equation}\label{eq:w-2}
|\nabla w(x)| \leq c \frac{a}{|x|^2}, \qquad |\nabla w(x)|  \leq c
\, \frac{\rho}{|x|}, \qquad |\nabla^2 w(x)| \leq c \,
\frac{\rho}{|x|^2}\, \qquad \text{for all } x \in \bR^3.
\end{equation}
\item[iv)] We have
\begin{equation}\label{eq:a_0}
8\pi a = \int \rd x \, V_N(x) (1-w(x))\,.
\end{equation}
\end{itemize}
\end{lemma}

\begin{proof}
We prove part i) and iii) in Appendix \ref{app:w}. Part ii) follows
trivially by the definition of the scattering length $a$ and by the
fact that the potential has compact support. As for part iv), note
that, due to the spherical symmetry of $V_N$ and $w(x)$, with the
notation $r =|x|$, the function $g(r):= r f(r) = r(1-w(r))$
satisfies
\[
- g'' (r) +
\frac{1}{2} V_N (r) g(r) = 0 \; .
\]
By ii) of this lemma, $g(r) = r-a$ for $r >Ra$.  We thus
obtain
\begin{equation}
\begin{split} \int \rd x \, V_N (x) (1-w(x)) &= 4\pi \int_0^{\infty}
\rd r \, r^2 V_N (r) (1-w(r)) = 8\pi \int_0^{\infty} \rd r \, r \,
g'' (r) \\ &= 8\pi \lim_{ \varrho \to \infty} \left( r \, g' (r) -
g(r) \right) |_0^{\varrho} = 8 \pi a \,.\end{split}
\end{equation}
\end{proof}

\medskip

\begin{proof}[Proof of Proposition \ref{prop:H2}]
For $j=1,\dots ,N$, we define \begin{equation}\label{eq:frakh}
 \fh_j
:= -\Delta_j + \frac{1}{2} \sum_{i \neq j} V_N (x_j -x_i)
.\end{equation} Then we clearly have
\[ H_N = \sum_{j=1}^N \fh_j . \]
Since $\psi$ is symmetric with respect to permutations, we have
\begin{equation}\label{eq:H2-1} \langle \psi, H_N^2 \psi \rangle = \sum_{i,j}^N \langle \psi,
\fh_i \fh_j \psi \rangle = N(N-1) \langle \psi, \fh_1 \fh_2 \psi
\rangle + N \langle \psi, \fh_1^2 \psi \rangle \geq N (N-1) \langle
\psi, \fh_1 \fh_2 \psi \rangle \, .\end{equation} Of course, instead
of the indices $1,2$ we could have chosen any $i \neq j$.

\medskip

We have
\begin{equation}
\fh_1 \psi = - \Delta_1 \psi + \frac{1}{2} V_N (x_1 -x_2) \psi +
\frac{1}{2} \sum_{j \geq 3} V_N (x_1 -x_j) \psi
\end{equation}
Next we write $\psi = (1- w_{12}) \phi_{12}$ and we observe that
\begin{equation}
-\Delta_1 [(1-w_{12}) \phi_{12}] = (1- w_{12}) (-\Delta_1 \phi_{12}) +
2 \nabla w_{12} \, \nabla_1 \phi_{12}  + \Delta w_{12} \, \phi_{12}.
\end{equation}
Hence
\begin{equation}
\begin{split}
(1-w_{12})^{-1} \fh_1\big[ (1-w_{12}) \phi_{12}\big] = &-\Delta_1 \phi_{12} +
2 \frac{\nabla w_{12}}{1-w_{12}} \nabla_1 \phi_{12} \\ &+
\frac{\left(-\Delta_1 +(1/2) V_N (x_1 -x_2) \right)
(1-w_{12})}{1-w_{12}} \phi_{12} \\ &+ \frac{1}{2} \sum_{j \geq 3}
V_{N} (x_1 -x_j) \phi_{12}.
\end{split}
\end{equation}
Using the definition of $w (x)$ (see (\ref{eq:defw1})), we obtain
\begin{equation}\label{eq:h1}
(1-w_{12})^{-1} \fh_1 \big[ (1-w_{12}) \phi_{12}\big] = L_1 \phi_{12} +
\frac{1}{2} \sum_{j \geq 3} V_{N} (x_1 - x_j) \phi_{12}
\end{equation}
where we defined  \[ L_1 := -\Delta_1 + 2 \frac{\nabla
w_{12}}{1-w_{12}} \nabla_1 \; .
\]
Note that this operator is symmetric with respect to the measure
$(1-w_{12})^2 \rd \bx$, i.e.
\begin{equation}
\int \, (1- w_{12})^2 \ov{\phi} \, ( L_1 \chi) = \int  \, (1- w_{12})^2
\, (L_1 \ov{\phi}) \, \chi = \int  \, (1- w_{12})^2 \, \nabla_1\ov{\phi}
\nabla_1 \chi \, .
\end{equation}
Analogously to (\ref{eq:h1}), we have
\begin{equation}
(1-w_{12})^{-1} \fh_2 \big[ (1-w_{12}) \phi_{12} \big]= L_2 \phi_{12} +
\frac{1}{2} \sum_{j \geq 3} V_N (x_2 - x_j) \phi_{12}
\end{equation}
with
\[ L_2= -\Delta_2 + 2 \frac{\nabla w_{21}}{1-w_{12}} \nabla_2 \, .\]
Therefore, from (\ref{eq:H2-1}) we find
\begin{equation}\label{eq:H2-2}
\begin{split}
\langle \psi, H_N^2 \psi \rangle \geq \; &N (N-1) \int \,
(1-w_{12})^2 \, \left( L_1 + \frac{1}{2} \sum_{j \geq 3} V_N (x_1 -
x_j) \right) \overline{\phi}_{12} \left( L_2 + \frac{1}{2} \sum_{j
\geq 3} V_N (x_2 - x_j) \right) \phi_{12} \\ = \; &N (N-1) \int \,
(1-w_{12})^2 \, L_1 \overline{\phi}_{12} \, L_2 \phi_{12} \\ &+
\frac{N (N-1)}{2} \sum_{j \geq 3} \int \, (1-w_{12})^2 \, \left\{
V_N (x_2 - x_j) \; L_1 \overline{\phi}_{12} \phi_{12} + V_N (x_1 -
x_j) \overline{\phi}_{12} L_2 \phi_{12} \right\}
\\&+ \frac{N (N-1)}{4} \sum_{i,j \geq 3} \int \, (1-w_{12})^2 \, V_N (x_1 -
x_j) V_N (x_2 -x_i) |\phi_{12}|^2 \\ = \; &N (N-1) \int \,
(1-w_{12})^2 \, L_1 \overline{\phi}_{12} \, L_2 \phi_{12} \\ &+
\frac{N (N-1)}{2} \sum_{j \geq 3} \int \, (1-w_{12})^2 \, \left\{
V_N (x_1 - x_j) |\nabla_2 \phi_{12}|^2 + V_N (x_2 - x_j) |\nabla_1
\phi_{12}|^2 \right\}
\\ &+ \frac{N (N-1)}{4} \sum_{i,j \geq 3} \int \, (1-w_{12})^2 \, V_N (x_1 -
x_j) V_N (x_2 -x_i) |\phi_{12}|^2
\\ \geq \; &N (N-1)\int \, (1-w_{12})^2 \, L_1 \overline{\phi}_{12} \, L_2
\phi_{12}.
\end{split}
\end{equation}
Here we used that the potential is positive and that the sum
$\sum_{j \geq 3} V_N (x_1 - x_j)$ is independent of $x_2$ (and
analogously $\sum_{j \geq 3} V_N (x_2 - x_j)$ is independent of
$x_1$).

\medskip

{F}rom (\ref{eq:H2-2}) we find
\begin{equation}\label{eq:H2-3}
\begin{split}
\langle \psi, H_N^2 \psi \rangle \geq \, &N (N-1) \int \,
(1-w_{12})^2 \, \nabla_1 \overline{\phi}_{12} \nabla_1 \, L_2
\phi_{12} \\ = \, &N(N-1) \int \, (1-w_{12})^2 \, |\nabla_1 \nabla_2
\, \phi_{12}|^2 + N (N-1) \int \, (1- w_{12})^2 \, \nabla_1 \,
\overline{\phi}_{12} [\nabla_1, L_2] \phi_{12} \, .
\end{split}
\end{equation}
To control the last term, we note that
\begin{equation*}
\begin{split}\left| \left[
\nabla_1, \frac{\nabla w_{21}}{1-w_{21}} \right] \right| \leq \;
&\frac{\left|\nabla^2 w_{21}\right|}{1-w_{12}} + \left(\frac{\nabla
w_{12}}{1-w_{12}} \right)^2  \leq  c \rho \frac{1}{|x_1 - x_2|^2}\,
\end{split}
\end{equation*}
by \eqref{eq:w>0} and \eqref{eq:w-2}, for $\rho$ small enough.
Therefore we have
\begin{equation}\label{eq:H2-3b}
\begin{split}
\left|\int \, (1- w_{12})^2 \, \nabla_1 \phi_{12} [\nabla_1, L_2]
\phi_{12} \right| &\leq c \rho \, \int  \, (1-w_{12})^2 \,
\frac{1}{|x_1 - x_2|^2} |\nabla_1 \phi_{12}| |\nabla_2 \phi_{12}|
\\ & \leq c \rho  \, \int  \, \frac{1}{|x_1 -
x_2|^2} |\nabla_1 \phi_{12}|^2
\\ &\leq c \rho \int \, |\nabla_1 \nabla_2 \, \phi_{12}|^2
\\ &\leq c \rho \int \, (1- w_{12})^2 \, |\nabla_1 \nabla_2 \phi_{12}|^2
\end{split}
\end{equation}
where we used (\ref{eq:w>0}) to remove and then reinsert the factor
$(1-w_{12})^2$ (assuming $\rho$ is small enough), and where we used
the Hardy inequality to control the $1/|x|^2$ singularity. {F}rom
(\ref{eq:H2-3}) we have
\begin{equation}
\langle \psi, H_N^2 \psi \rangle \geq (1-c\rho) N (N-1) \int \,
(1-w_{12})^2 \, |\nabla_1 \nabla_2 \phi_{12}|^2.
\end{equation}
This completes the proof of the Proposition \ref{prop:H2}.
\end{proof}

For fixed  $2\leq k \leq N$ and $i,j \leq k$, with $i \neq j$, we
define the densities $\gamma^{(k)}_{N,i,j,t}$ by
\begin{equation}\label{eq:gammaij}
\gamma^{(k)}_{N,i,j,t} := (1-w_{ij})^{-1} \gamma^{(k)}_{N,t}
(1-w_{ij})^{-1}\; ,
\end{equation}
where $(1-w_{ij})^{-1} = (1-w (x_i -x_j))^{-1}$ is viewed as a
multiplication operator. The kernel of $\gamma^{(k)}_{N,i,j,t}$ is
given by
\begin{equation}\label{eq:gammaij-ker}
\gamma^{(k)}_{N,i,j,t} (\bx_k ;\bx'_k) = (1-w(x_i-x_j))^{-1} \,
(1-w(x'_i-x'_j))^{-1} \gamma^{(k)}_{N,t} (\bx_k;\bx'_k)\; .
\end{equation}
Then, for every $k$, and every $i,j \leq k$, with $i\neq j$,
$\gamma_{N,i,j,t}^{(k)}$ is a positive operator, with $\tr \,
\gamma^{(k)}_{N,i,j,t} \leq C$, uniformly in $N,t$.

\begin{proposition}[A-priori bounds for $
\gamma^{(k)}_{N,i,j,t}$]\label{prop:apriori-phiij} For any
sufficiently small $\rho$, there exists a constant $C > 0$, such
that
\begin{equation}
\tr \; (1-\Delta_i) (1-\Delta_j) \, \gamma_{N,i,j,t}^{(k)} \leq C
\end{equation}
for all $t \in \bR$, $2 \leq k \leq N$, $i,j \leq k$, $i \neq j$,
and for all $N$ large enough.
\end{proposition}
\begin{proof}
For fixed $i\neq j$ we define the function
$\phi_{i,j,t}$ by $\psi_{N,t}
 = (1-w_{ij}) \phi_{i,j,t} $ (the $N$ dependence of
$\phi_{i,j,t}$ is omitted in the notation). Then we observe that
\begin{equation}\label{eq:apriori-phiij-1}
\tr \; (1-\Delta_i) (1-\Delta_j) \, \gamma_{N,i,j,t}^{(k)} = \| S_i
S_j \, \phi_{i,j,t} \|^2 = \| \phi_{i,j,t} \|^2 + 2 \| \nabla_i
\phi_{i,j,t} \|^2 + \| \nabla_i \nabla_j \phi_{i,j,t}\|^2
\end{equation}
with $S_n := (1-\Delta_n)^{1/2}$. Next we note that, by
(\ref{eq:w>0}),
\begin{equation}\label{eq:apriori-phiij-2}
\| \phi_{i,j,t} \|^2 = \int \rd \bx \, |\phi_{i,j,t} (\bx) |^2 \leq
C \int \rd \bx \, |\psi_{N,t} (\bx)|^2 \leq C
\end{equation}
uniformly in $N$ and $t$. Moreover
\begin{equation}\label{eq:nablapsi}
\begin{split}
\| \nabla_i \phi_{i,j,t} \|^2 = \; &\int \rd \bx \, \left| \nabla_i
\frac{\psi_{N,t} (\bx)}{1-w(x_i-x_j)} \right|^2 \\ \leq \; & \int
\rd \bx \, \frac{1}{(1-w(x_i-x_j))^2} |\nabla_i \psi_{N,t} (\bx)|^2
+ \int \rd \bx \, \left|\frac{\nabla_i w(x_i
-x_j)}{(1-w(x_i-x_j))^2}\right|^2 | \psi_{N,t} (\bx)|^2 \\ \leq \;
&C \int \rd \bx \, |\nabla_i  \psi_{N,t} (\bx)|^2 + C \int \rd \bx
\, \frac{1}{|x_i -x_j|^2} | \psi_{N,t} (\bx)|^2
\\ \leq \; &C \int \, \rd \bx \, |\nabla_i \psi_{N,t}(\bx) |^2 \,.
\end{split}
\end{equation}
where we used (\ref{eq:w>01}), (\ref{eq:w-2}) and Hardy inequality.
Next we note that, for every $i=1,\dots,N$,
\begin{equation}
\langle \psi_{N,t}, H_N \,   \psi_{N,t} \rangle \geq N \langle
\psi_{N,t}, \Delta_i  \psi_{N,t} \rangle = N \int  \,
|\nabla_i \psi_{N,t}|^2 \, .
\end{equation}
Therefore, from (\ref{eq:nablapsi}),
\begin{equation}\label{eq:apriori-phiij-3}
\begin{split}
\| \nabla_i \phi_{i,j,t} \|^2 \leq \; C N^{-1} \langle \psi_{N,t} ,
H_N \, \psi_{N,t} \rangle = C N^{-1} \langle \psi_N, H_N \, \psi_N
\rangle \leq C
\end{split}
\end{equation}
by (\ref{eq:thmassum1}) and by conservation of energy.
 Finally, to bound the last term on the
r.h.s. of (\ref{eq:apriori-phiij-1}), we note that, for a
sufficiently small $\rho$,
\begin{equation}\label{eq:apriori-phiij-4}
\begin{split}
\| \nabla_i \nabla_j \phi_{i,j,t}\|^2 \leq \; &C \int \rd \bx\;
(1-w(x_i -x_j))^2 \, |\nabla_i \nabla_j \phi_{i,j,t} (\bx)|^2 \\
\leq \; &\frac{C}{N(N-1)} \langle \psi_{N,t} , H_N^2  \psi_{N,t}
\rangle  \\ = \;& \frac{C}{N(N-1)} \langle \psi_{N} , H_N^2 \psi_{N}
\rangle \leq C
\end{split}
\end{equation}
for all $N$ large enough. Here we used (\ref{eq:w>01}) in the first
line, Proposition \ref{prop:H2} in the second line, the conservation
of $H_N^2$ in the third line, and the assumption
(\ref{eq:thmassum1}) in the last inequality. Proposition
\ref{prop:apriori-phiij} now follows from
(\ref{eq:apriori-phiij-1}), (\ref{eq:apriori-phiij-2}),
(\ref{eq:apriori-phiij-3}), and (\ref{eq:apriori-phiij-4}).
\end{proof}

\subsection{Higher Order Energy Estimates}

We will choose a cutoff length scale $\ell$. For technical reasons,
we will have to work with exponentially decaying cutoff functions,
so we set
\begin{equation}\label{eq:hdef}
h (x): = e^{-\frac{\sqrt{x^2 + \ell^2}}{\ell}}.
\end{equation}
Note that $h \simeq 0$ if $|x| \gg \ell$, and $h \simeq e^{-1}$ if
$|x| \ll \ell$. For $i=1,\dots ,N$ we define the cutoff function
\begin{equation}
\theta_i (\bx) := \exp \left(-\frac{1}{\ell^{\eps}} \sum_{j\neq i} h
(x_i -x_j)\right)
\end{equation}
for some $\eps >0$. Note that  $\theta_i(\bx)$
is exponentially small if
there is at least one other particle at distance of order $\ell$
from $x_i$, while $\theta_i(\bx)$ is exponentially close to 1
 if there is no other particle
near $x_i$ (on the length scale $\ell$).

As for the choice of $\ell$, to make sure that the presence of
particles at distances smaller than $\ell$ from $x_i$ is a rare
event, we will need to assume $N \ell^3 \ll 1$. This condition is
not used in Proposition \ref{prop:Hk} below, but if $N\ell^3 \gg 1$,
then our estimates were empty in the limit $N \to \infty$ as the
r.h.s. of the estimate (\ref{eq:Hk})
below  tended to zero. On the other hand, choosing
$\ell$ too small makes the price to pay for localizing the kinetic
energy on the length scale $\ell$ too high. In Proposition
\ref{prop:Hk} we will actually have to assume $N \ell^2 \gg1$.

 Next we define
\begin{equation}
\theta_i^{(n)}(\bx) := \theta_i (\bx)^{2^n} = \exp \left( -\frac{2^n}{\ell^{\eps}}
\sum_{j \neq i} h (x_i -x_j) \right) \,
\end{equation}
and their cumulative versions, for $n, k\in \bN$,
\be\label{eq:thetan}
  \Theta_k^{(n)} (\bx) := \theta_1^{(n)} (\bx)
\dots \theta_k^{(n)} (\bx) = \exp \left( -\frac{2^n}{\ell^{\eps}}
\sum_{i\leq k}\sum_{j \neq i} h (x_i -x_j) \right) \; . \ee To cover
all cases in one formula, we introduce the notation
$\Theta_k^{(n)}=1$ for any $k\leq 0$, $n\in \bZ$. We will need to
use the functions $\theta_i^{(n)}$ (instead of $\theta_i (\bx)$) to
take into account the deterioration of the kinetic energy
localization estimates. For example the bound \( |\nabla_j \theta_i
(\bx)| \leq C \ell^{-1} \theta_i (\bx)\) is wrong, while
\[ |\nabla_j \theta_i^{(n)} (\bx) | \leq C \ell^{-1} \theta_i^{(n-1)} (\bx)\] is
correct and similar bounds hold for $\Theta_k^{(n)}$.
 This, and other important properties of the function
$\Theta_k^{(n)}$, used throughout the proof of Proposition
\ref{prop:Hk} are collected in Lemma \ref{lm:theta} of the Appendix.

\medskip

\begin{proposition}[$H^k$ energy estimates]\label{prop:Hk}
Suppose $\ell \gg N^{-1/2}$ and that $\rho$ (from \eqref{eq:defrho})
is small enough. Then for $C_0
> 0$ sufficiently small (depending on the constant $(1-c\rho)$ in
Proposition \ref{prop:H2}) and for every integer $k\geq 1$ there
exists $N_0 = N_0 (k, C_0)$ such that
\begin{equation}\label{eq:Hk}
\begin{split}
\langle \psi, \left(H_N + N \right)^k \psi \rangle \geq \; &C_0^k
N^k \, \int \Theta^{(k)}_{k-1} \;
|\nabla_1 \dots \nabla_k \psi|^2 \\ &+C_0^k N^{k-1} \int
\Theta_{k-1}^{(k)} \; |\nabla_1^2 \nabla_2 \dots \nabla_{k-1} \psi |^2 \\
&+ C_0^k N^{k+1} \int \Theta_{k-1}^{(k)}
(\bx)  \; V_N (x_{k} -x_{k+1}) \, |\nabla_1 \dots \nabla_{k-1} \psi
(\bx)|^2\rd\bx
\end{split}
\end{equation} for every wave function $\psi \in L^2_s (\bR^{3N})$
and for every $N \geq N_0$.
\end{proposition}

In order to keep the exposition of the main ideas as clear as
possible, we defer the proof of this proposition, which is quite
long and technical, to  Section
\ref{sec:proofHk}, at the end of the paper.

\section{Compactness of the Marginal
 Densities}
\setcounter{equation}{0}

In this section we prove the compactness of the sequence $
\Gamma_{N,t} = \{ \gamma^{(k)}_{N,t} \}_{k\ge1}$ w.r.t. the topology
$\tau_{\text{prod}}$. (See Section \ref{sec:outline} for the
definition of $\tau_{\text{prod}}$ and recall the convention that
$\gamma^{(k)}_{N,t}=0$ if $k > N$.)
 Moreover, in Proposition
\ref{prop:apriorik}, we prove important a-priori bounds on any limit
point $\Gamma_{\infty,t}$ of the sequence $\Gamma_{N,t}$.

\begin{theorem}\label{thm:compactness}
Assume that $\rho$ is small enough and fix an arbitrary $T>0$.
Suppose that $\Gamma_{N,t} = \{\gamma^{(k)}_{N,t} \}_{k\geq 1}$ is
the family of marginal density associated with the solution
$\psi_{N,t}$ of the Schr\"odinger equation (\ref{eq:schr}),  and
that (\ref{eq:thmassum1}) is satisfied. Then $\Gamma_{N,t} \in
\bigoplus_{k \geq 1} C([0,T], \cL_k^1)$ Moreover the sequence
$\Gamma_{N,t} \in \bigoplus_{k \geq 1} C([0,T], \cL_k^1)$ is compact
with respect to the product topology $\tau_{\text{prod}}$ generated
by the metrics $\wh \eta_k$ (defined in Section \ref{sec:outline}).
For any limit point $ \Gamma_{\infty,t} = \{ \gamma_{\infty,t}^{(k)}
\}_{k \geq 1}$, $ \gamma^{(k)}_{\infty,t}$ is symmetric w.r.t.
permutations, $ \gamma^{(k)}_{\infty,t} \geq 0$, and
\begin{equation}\label{eq:bou} \tr \; \gamma^{(k)}_{\infty,t} \leq 1
\,\end{equation} for every $k \geq 1$.
\end{theorem}

\begin{proof}
By a standard ``choice of the diagonal subsequence''-argument it is
enough to prove the compactness of $ \gamma_{N,t}^{(k)}$, for fixed
$k \geq 1$, with respect to the metric $\wh \eta_k$. In order to
prove the compactness of $ \gamma_{N,t}^{(k)}$ with respect to the
metric $\wh \eta_k$, we show the equicontinuity of
$\gamma_{N,t}^{(k)}$ with respect to the metric $\eta_k$.
The following lemma gives a useful criterium to prove the
equicontinuity of a sequence in $ C([0,T],
\cL^1_k)$. Its proof is very similar to the proof of Lemma 9.2 in
\cite{ESY}; the only difference is that here we keep $k$ fixed and
we consider sequences in $\cL^1_k$, while in \cite{ESY} we
considered equicontinuity in the direct sum $C([0,T], \cH) =
\oplus_{k \geq 1} C([0,T], \cH_k)$ over all $k \geq 1$, for some
Sobolev space $\cH_k$.

\begin{lemma}\label{lm:equi}  Fix $k \in \bN$ and $T > 0$.
A sequence $\gamma_{N, t}^{(k)} \in \cL^1_k$, $N=k,k+1, \ldots$,
with $\gamma_{N,t}^{(k)} \geq 0$ and $\tr \; \gamma_{N,t}^{(k)} = 1$
for all $t \in [0,T]$ and $N \geq k$, is equicontinuous in $C([0,T],
\cL^1_k)$ with respect to the metric $\eta_k$, if and only if there
exists a dense subset $\cJ_k$ of $\cK_k$ such that for any
$J^{(k)}\in \cJ_k$ and for every $\eps
>0$ there exists a $\delta > 0$ such that
\begin{equation}\label{eq:equi02}
\sup_{N\ge 1}\Big| \tr \;  J^{(k)} \left( \gamma_{N,t}^{(k)} -
\gamma_{N,s}^{(k)} \right) \Big| \leq \eps
\end{equation}
for all $t,s \in [0,T]$ with $|t -s| \leq \delta$. $\;\;\Box$
\end{lemma}

\medskip

For the proof of the equicontinuity of
$\gamma_{N,t}^{(k)}$ with respect to the metric $\eta_k$, we will
choose the set $\cJ_k$ in Lemma \ref{lm:equi} to
consist of all $J^{(k)} \in \cK_k$ such that $S_i S_j
J^{(k)} S_i S_j$ is bounded, for all $i \neq j$, and $i,j \leq k$.
We recall the notation
 $S_n = (1-\Delta_n)^{1/2}$.

Rewriting the BBGKY hierarchy
(\ref{eq:BBGKY}) in integral form we obtain for any $s\leq t$
\begin{equation}
\begin{split}
\gamma_{N,t}^{(k)} = \; &\gamma_{N,s}^{(k)} -i \sum_{j=1}^k \int_s^t
\rd r \, [ -\Delta_j , \gamma_{N,r}^{(k)} ] - i
\sum_{i<j}^k \int_s^t  \rd r \, [ V_N (x_i -x_j), \gamma_{N,r}^{(k)} ] \\
& -i (N-k) \sum_{j=1}^k \int_s^t \rd r \, \tr_{k+1} [ V_N (x_j
-x_{k+1}) , \gamma_{N,r}^{(k+1)} ] \,.
\end{split}
\end{equation}
Multiplying the last equation with $J^{(k)} \in \cJ_k$ and taking
the trace we get the bound (recall the definition
(\ref{eq:gammaij}) of the densities $\gamma_{N,i,j,t}^{(k)}$)
\begin{equation}\label{eq:equi-1}
\begin{split}
\Big| \tr \, J^{(k)} &\left(  \gamma_{N,t}^{(k)} -
\gamma_{N,s}^{(k)} \right) \Big| \leq \sum_{j=1}^k \int_s^t \rd r \,
\Big| \tr \; \left( S_j^{-1} J^{(k)} S_j - S_j J^{(k)} S_j^{-1}
\right) \, S_j  \gamma_{N,r}^{(k)} S_j \Big| \\ &+ \sum_{i<j}^k
\int_s^t \rd r \, \Big| \tr \; \left(S_i S_j J^{(k)} S_i S_j \right)
\left( S_i^{-1} S_j^{-1} V_N (x_i -x_j) (1-w_{ij}) S_i^{-1}
S_j^{-1}\right)
\\ &\hspace{1.5cm} \times \left(S_i S_j  \gamma^{(k)}_{N,i,j,r} S_i S_j
\right) \left(S_i^{-1} S_j^{-1} (1-w_{ij}) S_i^{-1} S_j^{-1}\right)\Big| \\
&+ \sum_{i<j}^k \int_s^t \rd r \, \Big| \tr \; \left(S_i S_j J^{(k)}
S_i S_j\right)  \left(S_i^{-1} S_j^{-1} (1-w_{ij}) S_i^{-1}
S_j^{-1}\right)
\\ &\hspace{1.5cm} \times \left(S_i S_j  \gamma^{(k)}_{N,i,j,r} S_i S_j
\right) \left( S_i^{-1} S_j^{-1} V_N (x_i -x_j) (1-w_{ij}) S_i^{-1}
S_j^{-1}\right) \Big|\\
&+\left(1-\frac{k}{N} \right) \sum_{j=1}^k \int_s^t \rd r \, \Big|
\tr \; \left(S_j J^{(k)} S_j \right)\; \left(S_j^{-1} S_{k+1}^{-1} N
V_N
(x_j - x_{k+1}) (1- w_{j,k+1}) S_{k+1}^{-1} S_j^{-1} \right) \\
&\hspace{1.5cm} \times \left( S_{k+1} S_j \gamma^{(k+1)}_{N,j,k+1,r}
S_j S_{k+1}\right) \left(S_j^{-1} S_{k+1}^{-1} (1-w_{j,k+1})
S_{k+1} S_j^{-1}\right)\Big| \\
&+\left(1-\frac{k}{N} \right) \sum_{j=1}^k \int_s^t \rd r \, \Big|
\tr \; \left(S_j J^{(k)} S_j\right) \left(S_j^{-1} S_{k+1}
(1-w_{j,k+1}) S_{k+1}^{-1} S_j^{-1}\right) \\ &\hspace{1.5cm} \times
\left(S_{k+1} S_j  \gamma^{(k+1)}_{N,j,k+1,r} S_j S_{k+1}\right) \;
\left(S_j^{-1} S_{k+1}^{-1} N V_N (x_j - x_{k+1}) (1- w_{j,k+1})
S_{k+1}^{-1} S_j^{-1}\right)\Big|\; .
\end{split}
\end{equation}
Here we used that $S_{k+1}$
commutes with $J^{(k)}$. Next we observe that (see Lemma
\ref{lm:sobolev} below),
\begin{equation}
\| S_i^{-1} S_j^{-1} N V_N (x_i - x_j) (1-w_{ij}) S_i^{-1} S_j^{-1}
\| \leq C N \int  V_N \, (1-w) \leq C \,,
\end{equation}
by part iv) of Lemma \ref{lm:w}. Moreover
\begin{equation}
\| S_i^{-1} S_j^{-1} (1- w_{ij}) S_i^{-1} S_j^{-1} \| \leq C
\end{equation}
and \be\label{eq:equi-2}
\begin{split}
\| S_j^{-1} S_{k+1}^{-1} (1- w_{j,k+1}) S_{k+1} S_j^{-1} \|
\leq & \;  \Big\| S_j^{-1} S_{k+1}^{-1} (1- w_{j,k+1}) S_{k+1}^2 S_j^{-2}
 (1- w_{j,k+1}) S_{k+1}^{-1} S_j^{-1}\Big\|^{\frac{1}{2}} \\
\leq & \; C + \Big\| S_j^{-1} S_{k+1}^{-1} \nabla_{k+1}
  (1- w_{j,k+1})  S_j^{-2}
 (1- w_{j,k+1})  \nabla_{k+1} S_{k+1}^{-1} S_j^{-1}\Big\|^{\frac{1}{2}} \\
& +  \Big\| S_j^{-1} S_{k+1}^{-1} (\nabla_{k+1} w_{j,k+1}) S_j^{-2}
 (\nabla_{k+1} w_{j,k+1}) S_{k+1}^{-1} S_j^{-1}\Big\|^{\frac{1}{2}} \\
\leq & \; C + \Big\| S_j^{-1} S_{k+1}^{-1} (\nabla w_{j,k+1})^2
  S_{k+1}^{-1} S_j^{-1}\Big\|^{\frac{1}{2}} \leq C
\end{split}
\ee In the last step we used  the second bound in
\eqref{eq:w-2}. Since $J^{(k)} \in \cJ_k$ is such that $\| S_i S_j
J^{(k)} S_i S_j \| \leq C$ for all $i,j=1, \dots ,k$, it follows
from (\ref{eq:equi-1})--\eqref{eq:equi-2} that
\begin{equation}
\Big| \tr \, J^{(k)} \left(  \gamma_{N,t}^{(k)} - \gamma_{N,s}^{(k)}
\right) \Big| \leq C_k (t-s) \max_{n=k,k+1} \max_{i\neq j, i,j \leq
n} \sup_{r \in [s,t]} \tr \; \left| S_i S_j \,
\gamma_{N,i,j,r}^{(n)} S_i S_j \right|
\end{equation}
for a constant $C_k$ depending on $k$ and on $J^{(k)}$, but
independent of $t,s,N$. {F}rom Proposition \ref{prop:apriori-phiij},
and from the fact that
 the subset $\cJ^{(k)}$ is dense in $\cK_k$, it follows
that the sequence $ \gamma_{N,t}^{(k)} \in C([0,T], \cL^1_k)$ is
equicontinuous. Since, moreover, $\tr \;  \gamma^{(k)}_{N,t} =1$
uniformly in $t \in [0,T]$ and $N$, the compactness of the sequence
$ \gamma^{(k)}_{N,t}$ w.r.t. the metric $\wh \eta_k$ follows from
the Arzela-Ascoli theorem.  This proves the compactness of $
\Gamma_{N,t} = \{  \gamma^{(k)}_{N,t} \}_{k \geq 1} \in \bigoplus_{k
\geq 1} C([0,T], \cL^1_k)$ with respect to the product topology
$\tau_{\text{prod}}$.

\medskip

 Now suppose that $
\Gamma_{\infty,t} = \{  \gamma^{(k)}_{\infty,t} \}_{k \geq 1}
\in\bigoplus_{k\geq 1} C([0,T], \cL^1_k)$ is a limit point of $
\Gamma_{N,t}$ with respect to $\tau_{\text{prod}}$. Then, for any
$k\geq 1$, $ \gamma_{\infty,t}^{(k)} \in C([0,T], \cL^1_k)$ is a
limit point of $ \gamma_{N,t}^{(k)}$. The bound
\[ \tr \; \Big|  \gamma_{\infty,t}^{(k)}\Big| \leq 1 \] follows
because the norm can only drop in the weak limit.

 To prove that $ \gamma^{(k)}_{\infty,t}$
is non-negative, we observe that, for an arbitrary $\ph \in L^2
(\bR^{3k})$ with $\| \ph \| =1$, the orthogonal projection $P_{\ph}
= |\ph \rangle\langle \ph |$ is in $\cK_k$ and therefore we have
\begin{equation}
\begin{split}
\langle \ph ,  \gamma^{(k)}_{\infty,t} \ph \rangle = \tr \; P_{\ph}
 \gamma^{(k)}_{\infty,t} = \lim_{j \to \infty} \tr \; P_{\ph}
\gamma^{(k)}_{N_j,t} = \lim_{j \to \infty} \langle \ph,
\gamma^{(k)}_{N_j ,t} \ph \rangle \geq 0 \, ,
\end{split}
\end{equation}
for an appropriate subsequence $N_j$ with $N_j \to \infty$ as $j \to
\infty$.

 Similarly, the symmetry of $ \gamma^{(k)}_{\infty,t}$
w.r.t. permutations is inherited from the symmetry of $
\gamma^{(k)}_{N,t}$ for finite $N$. For a permutation $\pi \in
\cS_k$, we denote by $\Xi_{\pi}$ the operator on $L^2 (\bR^{3k})$
defined by
\[ \Xi_{\pi} \ph (x_1, \dots x_k) = \ph (x_{\pi 1}, \dots , x_{\pi
k}) \,. \] Then the permutation symmetry of $
\gamma_{\infty,t}^{(k)}$ is defined by
\begin{equation}\label{eq:permu}
\Xi_{\pi}  \gamma^{(k)}_{\infty,t} \Xi_{\pi}^{-1} =
\gamma_{\infty,t}^{(k)}
\end{equation}
for every $\pi \in \cS_k$. To prove (\ref{eq:permu}), we note that,
for an arbitrary $J^{(k)} \in \cK_k$ and a permutation $\pi \in
\cS_k$, we have, for an appropriate subsequence $N_j \to \infty$, as
$j \to \infty$,
\begin{equation}
\begin{split}
\tr \; J^{(k)}  \gamma^{(k)}_{\infty,t} &= \lim_{j \to \infty} \;
J^{(k)}  \gamma^{(k)}_{N_j ,t} = \lim_{j \to \infty} \tr \; J^{(k)}
\Xi_{\pi}  \gamma^{(k)}_{N_j ,t} \Xi_{\pi}^{-1} = \lim_{j \to
\infty} \tr \; \Xi_{\pi}^{-1} J^{(k)} \Xi_{\pi}  \gamma_{N_j
,t}^{(k)} \\ &= \tr \; \Xi_{\pi}^{-1} J^{(k)} \Xi_{\pi}
\gamma_{\infty,t}^{(k)} = \tr \; J^{(k)} \Xi_{\pi}
\gamma_{\infty,t}^{(k)} \Xi_{\pi}^{-1} \; ,
\end{split}
\end{equation}
where we used that, since $J^{(k)} \in \cK_k$, also $\Xi_{\pi}^{-1}
J^{(k)} \Xi_{\pi} \in \cK_k$.
\end{proof}

In the next proposition we prove important a-priori bounds on the
limit points $ \Gamma_{\infty,t}$. These bounds are essential in the
proof of the uniqueness of the solution to the infinite hierarchy
(\ref{eq:BBGKYinf}), in Theorem \ref{thm:uniqueness}.

\begin{proposition}\label{prop:apriorik}
Suppose that $\rho$  is small enough, and assume that
(\ref{eq:thmassum1}) is satisfied. Let $\Gamma_{\infty,t} = \{
\gamma_{\infty,t}^{(k)}\}_{k \geq 1} \in \bigoplus_{k \geq 1}
C([0,T], \cL^1_k)$ is a limit point of the sequence $\Gamma_{N,t} =
\{  \gamma_{N,t}^{(k)} \}_{k=1}^N$ w.r.t. the product topology
$\tau_{\text{prod}}$. Then $ \gamma_{\infty,t}^{(k)}$ (has a version
which) satisfies
\begin{equation}\label{eq:apriorik}
\tr \; (1-\Delta_1) \dots (1-\Delta_k) \gamma_{\infty,t}^{(k)} \leq
C_1^k
\end{equation}
for a constant $C_1$ independent of $t \in [0,T]$ and $k \geq 1$.
\end{proposition}

\begin{proof}
We fix $\ell$ as a function of $N$, such that $N\ell^2 \gg 1$, and
$N\ell^3 \ll 1$. Moreover we fix $\e
>0$ so small that $N \ell^{3-\e} \ll 1$. With this choice of $\ell$
and $\eps$, we construct, for integer $n,k$ the cutoff functions
$\Theta^{(n)}_k (\bx)$ as in (\ref{eq:thetan}). For $k\in \bN$, we
will use the notation
\[ D_k := \nabla_1 \dots \nabla_k, \qquad D'_k:
= \nabla'_1 \dots \nabla'_k, \qquad \mbox{with} \quad \nabla'_j = \nabla_{x'_j}\; .
\]
We also set $D_k=I$ for $k\leq 0$ to cover all cases in a single
formula. {F}rom Proposition \ref{prop:Hk}, it follows that, for any
fixed $k \geq 1$,
\begin{equation}\label{eq:aprioripsi}
\begin{split}
\int  \, \Theta^{(k)}_{k-1} \, |D_k  \psi_{N,t}|^2
 &\leq \frac{1}{C_0^k N^k} \langle  \psi_{N,t} , (H_N + N)^k
 \psi_{N,t} \rangle \\ &= \frac{1}{C_0^k N^k} \langle
\psi_{N,0} , (H_N + N)^k  \psi_{N,0} \rangle \leq C_2^k
\end{split}
\end{equation}
for any $N$ large enough (depending only on $k$). In the last
inequality we applied the assumption (\ref{eq:thmassum1}).

\medskip

For $k=1, \dots,N$, we define the densities $U^{(k)}_{N,t}$ by their
kernels
\begin{equation}\label{eq:apriorik-0}
\begin{split}
U^{(k)}_{N,t} (\bx_k; \bx'_k) := \int \rd \bx_{N-k} \;
\Theta_k^{(k)} (\bx_k, \bx_{N-k}) \Theta_k^{(k)} (\bx'_k, \bx_{N-k})
D_k \psi_{N,t} (\bx_k,\bx_{N-k}) \; D'_k \overline{ \psi}_{N,t}
(\bx_k,\bx_{N-k}).
\end{split}
\end{equation}
Note that the operator $U^{(k)}_{N,t}$ is the $k$-particle marginal
density associated with the $N$-body wave function $\Theta_k^{(k)}
(\bx) D_k  \psi_{N,t} (\bx)$. Therefore $U^{(k)}_{N,t} \geq 0$.
Moreover, it follows from (\ref{eq:aprioripsi}) that, for $N$ large
enough,
\begin{equation}\label{eq:apriorik-1}
\begin{split}
\tr \; U^{(k)}_{N,t} = \;& \int  \, \big[ \Theta^{(k)}_k \big]^2
 \;
|D_k \,  \psi_{N,t}|^2 \leq  \int  \,
\Theta^{(k)}_{k-1} \; |D_k  \psi_{N,t} |^2  \leq C_2^k\,.
\end{split}
\end{equation}
It follows from (\ref{eq:apriorik-1}) that for every fixed integer
$k \geq 1$, and for every $t \in [0,T]$, the sequence
$U^{(k)}_{N,t}$ is compact w.r.t. the weak* topology of $\cL^1_k$.
Moreover, if $U^{(k)}_{\infty,t}$ denotes an arbitrary limit point
of $U^{(k)}_{N,t}$, then
\begin{equation}\label{eq:apriorik-1b}
 \tr \; U^{(k)}_{\infty,t} \leq C_2^k \,.
 \end{equation}

Next we assume that $ \gamma_{\infty,t}^{(k)} \in C([0,T], \cL^1_k)$
is a limit point of $ \gamma_{N,t}^{(k)}$ w.r.t. to the topology
$\wh\eta_k$. It follows that for any fixed $t \in [0,T]$, $
\gamma_{\infty,t}^{(k)}$ is a limit point of $ \gamma_{N,t}^{(k)}$
w.r.t. the weak* topology of $\cL_k^1$. Because of the compactness
of the sequence $U^{(k)}_{N,t}$ w.r.t. the weak * topology of
$\cL^1_k$, we can assume, by passing to a common subsequence $N_i$,
that there exists a limit point $U^{(k)}_{\infty,t} \in \cL^1_k$ of
$U^{(k)}_{N,t}$ such that,
\begin{equation}\label{eq:apriorik-2}
\tr \; J^{(k)} \;  \gamma_{N_i,t}^{(k)} \to \tr \; J^{(k)} \;
\gamma_{\infty,t}^{(k)}
\end{equation}
and
\begin{equation}\label{eq:apriorik-3}
\tr \; J^{(k)} \; U_{N_i,t}^{(k)} \to \tr \; J^{(k)} \;
U_{\infty,t}^{(k)}
\end{equation}
for every $J^{(k)} \in \cK_k$. For notational simplicity, we will
drop the index $i$, but keep in mind that the limits hold only along
a subsequence.

Next we fix $J^{(k)} \in \cK_k$ such that $\nabla_1 \dots \nabla_k
J^{(k)} \nabla^*_k \dots \nabla^*_1$ is compact and such that
\begin{equation}\label{eq:assump}
\begin{split}
\sup_{\bx_k} \int \rd \bx'_k \; \sum_{b=0}^4|\nabla^b_{x'_n} \,
\nabla_{i_1} \dots \nabla_{i_j} \nabla'_{r_1} \dots \nabla'_{r_m}
J^{(k)} (\bx_k ; \bx'_k) | < \infty \\
\sup_{\bx'_k} \int \rd \bx_k \; \sum_{b=0}^4 |\nabla^b_{x_n} \,
\nabla_{i_1} \dots \nabla_{i_j} \nabla'_{r_1} \dots \nabla'_{r_m}
J^{(k)} (\bx_k ; \bx'_k) | < \infty
\end{split}
\end{equation}
for every $j,m,n \leq k$, and $(i_1, \dots i_j), (r_1,\dots,r_m)
\subset \{ 1,2, \dots,k \}$. Then we have, applying
(\ref{eq:apriorik-2}) to the derivatives of $J^{(k)}$,
\begin{equation}\label{eq:apriorik-4}
\tr \; \nabla_1 \dots \nabla_k J^{(k)} \nabla^*_k \dots \nabla^*_1
\,  \gamma_{N_i ,t}^{(k)} \to \tr \; \nabla_1 \dots \nabla_k J^{(k)}
\nabla^*_k \dots \nabla^*_1 \,  \gamma_{\infty,t}^{(k)}
\end{equation}
as $N_i \to \infty$. For such observable $J^{(k)}$ we rewrite the
l.h.s. of (\ref{eq:apriorik-3}), using (\ref{eq:apriorik-0}), as
\begin{equation}\label{eq:apriorik-6}
\begin{split}
\tr \; J^{(k)} \; U_{N,t}^{(k)} = \; \int \rd \bx_k \rd \bx'_k \rd
\bx_{N-k} \; J^{(k)} (\bx_k ; \bx'_k) \, &\Theta_k^{(k)}
(\bx_k,\bx_{N-k})  \Theta_k^{(k)} (\bx'_k , \bx_{N-k})  \\ &\times
D_k \psi_{N,t} (\bx_k, \bx_{N-k}) D'_k \overline{ \psi}_{N,t}
(\bx'_k, \bx_{N-k})\,.
\end{split}
\end{equation}
{F}rom (\ref{eq:apriorik-6}), we will show later that
\begin{equation}\label{eq:apriorik-7}
\begin{split}
\tr \; J^{(k)} \; U_{N,t}^{(k)} = \; \int \rd \bx_k \rd \bx'_k \rd
\bx_{N-k} \; \left(D_k \, D'_k \, J^{(k)}\right) (\bx_k ; \bx'_k) \,
 \psi_{N,t} (\bx_k, \bx_{N-k}) \overline{ \psi}_{N,t} (\bx'_k,
\bx_{N-k}) + o(1)
\end{split}
\end{equation}
as $N \to \infty$.

Before proving (\ref{eq:apriorik-7}), let us show how Proposition
\ref{prop:apriorik} follows from it. Equation (\ref{eq:apriorik-7})
implies that
\begin{equation}
\begin{split}
\tr \; J^{(k)} \; U_{N,t}^{(k)} = \; & \tr \; \nabla_1 \dots
\nabla_k J^{(k)} \nabla^*_k \dots \nabla^*_1  \gamma_{N,t}^{(k)} + o(1) \\
\to \; & \tr \; \nabla_1 \dots \nabla_k J^{(k)} \nabla^*_k \dots
\nabla^*_1  \gamma_{\infty,t}^{(k)}
\end{split}
\end{equation}
as $N \to \infty$ (using (\ref{eq:apriorik-4})). Comparing with
(\ref{eq:apriorik-3}), we obtain that
\begin{equation}
\tr \; J^{(k)} \; U^{(k)}_{\infty,t} = \tr \; \nabla_1 \dots
\nabla_k J^{(k)} \nabla^*_k \dots \nabla^*_1 \,
\gamma_{\infty,t}^{(k)} \, .
\end{equation}
Since the set of all
 $J^{(k)} \in \cK_k$ with the property that $\nabla_1 \dots
\nabla_k J^{(k)} \nabla^*_1 \dots \nabla^*_k \in \cK_k$ and
such that (\ref{eq:assump}) is satisfied is a dense subset of
$\cK_k$, it follows that
\begin{equation}
\nabla_1 \dots \nabla_k  \gamma_{\infty,t}^{(k)} \nabla^*_k \dots
\nabla^*_1 = U^{(k)}_{\infty,t}\,.
\end{equation}
{F}rom (\ref{eq:apriorik-1b}), we find
\begin{equation}\label{eq:apriorik-8}
\tr \; (-\Delta_1) \dots (-\Delta_k)  \gamma^{(k)}_{\infty,t} \leq
C_2^k \,.
\end{equation}

Now suppose that $ \Gamma_{\infty,t} = \{ \gamma_{\infty,t}^{(k)}
\}_{k \geq 1} \in C([0,T], \cL^1_k)$ is a limit point of the
sequence $ \Gamma_{N,t}$. Then, for every fixed $k \geq 1$ and $t
\in [0,T]$, $ \gamma_{\infty,t}^{(k)}$ is a limit point of $
\gamma_{N,t}^{(k)}$ and thus satisfies (\ref{eq:apriorik-8}), for a
constant $C_2$ independent of $t$ and $k$. Moreover, for any $m \leq
k$ we also have
\begin{equation}\label{eq:apriorik-8b}
\tr \; (-\Delta_1) \dots (-\Delta_m)  \gamma^{(k)}_{\infty,t} \leq
\; C_2^m \,.
\end{equation}
To prove the last equation, we repeat the same argument leading from
(\ref{eq:apriorik-0}) to (\ref{eq:apriorik-8}), but with the
densities $U^{(k)}_{N,t}$ replaced by
\begin{equation}
U^{(k)}_{m,N,t} (\bx_k;\bx'_k) = \int \rd \bx_{N-k} \;
\Theta_k^{(k)} (\bx_k, \bx_{N-k}) \Theta_k^{(k)} (\bx'_k, \bx_{N-k})
D_m  \psi_{N,t} (\bx_k,\bx_{N-k}) \; D'_m \overline{ \psi}_{N,t}
(\bx_k,\bx_{N-k}) \,.
\end{equation}
{F}rom (\ref{eq:apriorik-8}), (\ref{eq:apriorik-8b}), and from the
permutation symmetry of $ \gamma_{\infty,t}^{(k)}$, we find
\begin{equation}
\tr \; (1-\Delta_1) \dots (1-\Delta_k)  \gamma_{\infty,t}^{(k)} =
\sum_{m=0}^k {k \choose m} \tr \; (-\Delta_1) \dots (-\Delta_m)
\gamma_{\infty,t}^{(m)} \leq (C_2 + 1)^k
\end{equation}
which completes the proof of Proposition \ref{prop:apriorik}.

\bigskip

It remains to prove (\ref{eq:apriorik-7}). To this end, we rewrite
the r.h.s. of (\ref{eq:apriorik-6}) by using $\Theta_k^{(k)} =
\theta_k^{(k)}\Theta_{k-1}^k$ as follows:
\begin{equation}\label{eq:apriorik-9}
\tr \; J^{(k)} \; U_{N,t}^{(k)} =  (I) - (II) \ee with
\be\label{eq:apriorik-9.5}
\begin{split}
(I): = \; &\int \rd \bx_k \rd \bx'_k \rd \bx_{N-k} J^{(k)} (\bx_k;
\bx'_k) \, \Theta_{k-1}^{(k)} (\bx_k,\bx_{N-k}) \Theta_k^{(k)}
(\bx'_k,\bx_{N-k}) \\ &\hspace{3cm} \times D_k  \psi_{N,t}
(\bx_k, \bx_{N-k}) \, D'_k \overline{ \psi}_{N,t} (\bx'_k,\bx_{N-k})\\
(II):= \; &\int \rd \bx_k \rd \bx'_k \rd \bx_{N-k} J^{(k)} (\bx_k;
\bx'_k) (1-\theta_k^{(k)} (\bx_k, \bx_{N-k})) \Theta_{k-1}^{(k)}
(\bx_k, \bx_{N-k}) \Theta_k^{(k)} (\bx'_k,\bx_{N-k}) \\
&\hspace{3cm} \times D_k  \psi_{N,t} (\bx_k, \bx_{N-k}) \, D'_k
\overline{ \psi}_{N,t} (\bx'_k,\bx_{N-k}) \,.
\end{split}
\end{equation}
By integration by parts
\begin{equation}\label{eq:apriorik-10}
(I) = (Ia) + (Ib)
\ee
with
\be
\begin{split}
(Ia):= \; & - \int \rd \bx_k \rd \bx'_k \rd \bx_{N-k}
 \nabla_k J^{(k)} (\bx_k; \bx'_k) \,
\Theta_{k-1}^{(k)} (\bx_k,\bx_{N-k}) \, \Theta_k^{(k)}
(\bx'_k,\bx_{N-k}) \\ &\hspace{3cm} \times  D_{k-1}  \psi_{N,t}
(\bx_k, \bx_{N-k}) \, D'_k \overline{\psi}_{N,t} (\bx'_k,\bx_{N-k})
\\
(Ib):= &- \int \rd \bx_k \rd \bx'_k \rd \bx_{N-k} J^{(k)} (\bx_k;
\bx'_k) \, \nabla_k \Theta_{k-1}^{(k)} (\bx_k, \bx_{N-k})
\Theta_k^{(k)} (\bx'_k,\bx_{N-k}) \\ &\hspace{3cm} \times  D_{k-1}
\psi_{N,t} (\bx_k, \bx_{N-k}) \, D'_k \overline{\psi}_{N,t}
(\bx'_k,\bx_{N-k})
\end{split}
\end{equation}
The main term is (Ia). To bound the term (Ib), we use Schwarz
inequality with some $\a>0$:
\begin{equation}\label{eq:apriorik-11}
\begin{split}
|(Ib)|\leq \;& \a \int \rd \bx_k \rd \bx'_k \rd \bx_{N-k} |J^{(k)} (\bx_k;
\bx'_k)| \, \Big|\nabla_k \Theta_{k-1}^{(k)} (\bx_k, \bx_{N-k})
\Big|^2 \, |D_{k-1}  \psi_{N,t} (\bx_k, \bx_{N-k})|^2\\
&+ \a^{-1} \int \rd \bx_k \rd \bx'_k \rd \bx_{N-k} |J^{(k)} (\bx_k;
\bx'_k)| \, \Theta_k^{(k+1)} (\bx'_k,\bx_{N-k}) |D'_k  \psi_{N,t}
(\bx'_k,\bx_{N-k})|^2 \\
\leq \;& \a \left( \sup_{\bx_k} \int \rd \bx'_k |J^{(k)} (\bx_k;
\bx'_k)| \right) \; \int \rd \bx \; \Big|\nabla_k \Theta_{k-1}^{(k)}
(\bx)
\Big|^2 \, |D_{k-1}  \psi_{N,t} (\bx)|^2\\
&+ \a^{-1} \left( \sup_{\bx'_k} \int \rd \bx_k |J^{(k)}
(\bx_k;\bx'_k)| \right) \; \int \rd \bx'_k \rd \bx_{N-k} \;
\Theta_{k-1}^{(k)} (\bx'_k,\bx_{N-k}) |D'_k  \psi_{N,t}
(\bx'_k,\bx_{N-k})|^2.
\end{split}
\end{equation}
Using that \begin{equation} \Big|\nabla_k \Theta_{k-1}^{(k)}(\bx)
\Big|^2 \leq C \ell^{-2} \left( \frac{2^k}{\ell^{\eps}} \sum_{m=2}^k
h (x_1 -x_m) \right)^2 \, \Theta_{k-1}^{(k+1)} (\bx)
\end{equation}
we obtain that
\begin{equation}\label{eq:apriorik-12}
\begin{split}
\int \rd \bx  \, \Big|\nabla_k &\Theta_{k-1}^{(k)} (\bx) \Big|^2 \,
|D_{k-1}  \psi_{N,t} (\bx)|^2
\\ \leq \; & C \ell^{-2} \int \rd \bx \; \left( \frac{2^k}{\ell^{\eps}} \sum_{m=2}^k h
(x_1 -x_m) \right)^2 \, \Theta_{k-1}^{(k+1)} (\bx) |D_{k-1}
\psi_{N,t} (\bx)|^2\\
\leq \; & C (N-k)^{-1} \ell^{-2} \sum_{i \geq k} \int \rd \bx \;
\left( \frac{2^k}{\ell^{\eps}} \sum_{m=2}^k h (x_i -x_m) \right)^2
\, \Theta_{k-1}^{(k+1)} (\bx) |D_{k-1}  \psi_{N,t} (\bx)|^2,
\end{split}
\end{equation}
where we used the symmetry of the $D_{k-1}  \psi_{N,t}$ w.r.t.
permutations of the last $N-k$ variables. Since
\begin{equation}
\begin{split}
\sum_{i \geq k} \left( \frac{2^k}{\ell^{\eps}} \sum_{m=2}^k h (x_i
-x_m) \right)^2 \Theta_{k-1}^{(k+1)} (\bx) &\leq \left(
\frac{2^k}{\ell^{\eps}} \sum_{i \geq k} \sum_{m=2}^k h (x_i -x_m)
\right)^2 \Theta_{k-1}^{(k+1)} (\bx)  \leq C \Theta_{k-1}^{(k)}
(\bx)
\end{split}
\end{equation}
(see part ii) of Lemma \ref{lm:theta}), it follows from
(\ref{eq:apriorik-12}) that
\begin{equation}
\begin{split}
\int   \, \Big|\nabla_k \Theta_{k-1}^{(k)}  \Big|^2 \,
|D_{k-1}  \psi_{N,t} |^2  &\leq C \ell^{-2} (N-k)^{-1} \, \int
 \;
\Theta_{k-1}^{(k)}  \; | D_{k-1}  \psi_{N,t} |^2 \\
&\leq C \ell^{-2} (N-k)^{-1} \int  \; \Theta_{k-2}^{(k-1)}
 \; | D_{k-1}  \psi_{N,t} |^2 \\ &\leq C_k \; \ell^{-2}
(N-k)^{-1}
\end{split}
\end{equation}
by (\ref{eq:aprioripsi}) (here the constant $C_k$ depends on $k$ and
on the observable $J^{(k)}$). {F}rom (\ref{eq:apriorik-11}), from
the assumptions (\ref{eq:assump}), and again using
(\ref{eq:aprioripsi}), it follows that
\begin{equation}\label{eq:apriorik-13}
|(Ib)| \leq \; C_k \left( \a (N-k)^{-1} \ell^{-2} + \a^{-1} \right)
= o(1)
\end{equation}
because $N \ell^2 \gg 1$.

Next we consider the term (II) in
(\ref{eq:apriorik-9.5}). By  Schwarz inequality, we have
\begin{equation}\label{eq:apriorik-14}
\begin{split}
|(II)|\leq \; & \a \int \rd \bx_k \rd \bx'_k \rd \bx_{N-k} \;
|J^{(k)} (\bx_k; \bx'_k)| \Theta_{k-1}^{(k+1)} (\bx_k, \bx_{N-k})
|D_k \psi_{N,t} (\bx_k, \bx_{N-k})|^2 \\ &+ \a^{-1} \int \rd \bx_k
\rd \bx'_k \rd \bx_{N-k} \; |J^{(k)} (\bx_k; \bx'_k)|
(1-\theta_k^{(k)}
(\bx_k, \bx_{N-k})) \Theta_{k}^{(k+1)} (\bx'_k, \bx_{N-k}) \\
&\hspace{3cm} \times |D'_k
 \psi_{N,t} (\bx'_k, \bx_{N-k})|^2 \\
\leq \; & \a \left(\sup_{\bx_k} \int \rd \bx'_k \, |J^{(k)} (\bx_k;
\bx'_k)| \right) \int \rd \bx \; \Theta_{k-1}^{(k+1)} (\bx) |D_k
\psi_{N,t} (\bx)|^2 \\ &+ \a^{-1} \left( \sup_{\bx'_k, \bx_{N-k}}
\int \rd \bx_k \, |J^{(k)} (\bx_k; \bx'_k)| (1-\theta_k^{(k)}
(\bx_k, \bx_{N-k})) \right) \\ &\hspace{3cm} \times  \int \rd \bx'_k
\rd \bx_{N-k} \; \Theta_{k}^{(k+1)} (\bx'_k, \bx_{N-k}) |D'_k
\psi_{N,t} (\bx'_k, \bx_{N-k})|^2 \\ \leq \; & C_k \left( \a +
\a^{-1}\sup_{\bx'_k, \bx_{N-k}} \int \rd \bx_k \, |J^{(k)} (\bx_k;
\bx'_k)| (1-\theta_k^{(k)} (\bx_k, \bx_{N-k})) \right) \; ,
\end{split}
\end{equation}
where we used (\ref{eq:aprioripsi}). Next we note that
\begin{equation}\label{eq:apriorik-15}
\begin{split}
\int \rd \bx_k \, |J^{(k)} (\bx_k; \bx'_k)| (1-\theta_k^{(k)}
(\bx_k, \bx_{N-k})) &\leq \frac{2^k}{\ell^{\eps}} \sum_{m \neq k}
\int \rd \bx_k \, |J^{(k)} (\bx_k; \bx'_k)| h (x_k -x_m) \\ & \leq
C^k N \ell^{3-\eps} \int \rd \bx_k \, |\nabla^4_k J^{(k)} (\bx_k ;
\bx'_k)| + |J^{(k)} (\bx_k;\bx'_k)| \,
\end{split}
\end{equation}
because, with $h(x) = \exp (- (x^2 + \ell^2)^{1/2}/\ell)$, we have, by the
Sobolev inequality,
\begin{equation}
\int \rd x \, h (x) |f(x)| \leq \| h\|_1 \| f\|_\infty
\leq C \ell^{3} \int \sum_{b=0}^4 |\nabla^b f| \;.
\end{equation}
{F}rom (\ref{eq:apriorik-14}), (\ref{eq:apriorik-15}), and from the
assumptions (\ref{eq:assump}) we find
\begin{equation}
|(II)| \leq C_k \, \left( \a + \a^{-1} N\ell^{3-\eps} \right) \to 0
\end{equation}
as $N \to \infty$, because $N\ell^{3-\eps}  \ll 1$.

\medskip

{F}rom (\ref{eq:apriorik-9}), (\ref{eq:apriorik-13}) and last
equation we find
\begin{equation}
\begin{split}
\tr \; J^{(k)} \; U_{N,t}^{(k)}
 = \; & \int \rd \bx_k \rd \bx'_k \rd \bx_{N-k} \nabla_k J^{(k)} (\bx_k; \bx'_k) \,
\Theta_{k-1}^{(k)} (\bx_k,\bx_{N-k}) \, \Theta_k^{(k)}
(\bx'_k,\bx_{N-k}) \\ &\hspace{3cm} \times  D_{k-1}  \psi_{N,t}
(\bx_k, \bx_{N-k}) \, D'_k \overline{\psi}_{N,t} (\bx'_k,\bx_{N-k})
+ o(1)
\end{split}
\end{equation}
Repeating the same arguments to move the derivative $\nabla'_k$ from
$ \psi_{N,t}$ to $J^{(k)}$, we obtain
\begin{equation}
\begin{split}
\tr \; J^{(k)} \; U_{N,t}^{(k)} = \; & \int \rd \bx_k \rd \bx'_k \rd
\bx_{N-k} \nabla_k \nabla'_k J^{(k)} (\bx_k; \bx'_k) \,
\Theta_{k-1}^{(k)} (\bx_k,\bx_{N-k}) \, \Theta_{k-1}^{(k)}
(\bx'_k,\bx_{N-k}) \\ &\hspace{3cm} \times D_{k-1}  \psi_{N,t}
(\bx_k, \bx_{N-k}) \, D'_{k-1} \overline{\psi}_{N,t}
(\bx'_k,\bx_{N-k}) + o(1)
\end{split}
\end{equation}
Iterating this argument $k-1$ more times to move all derivatives to
the observable, we prove (\ref{eq:apriorik-7}).
\end{proof}

The following lemma was used in the proof of Theorem
\ref{thm:compactness}, and will also be used in the next sections,
in order to bound potentials by the action of derivatives.

\begin{lemma}\label{lm:sobolev}
\begin{itemize}
\item[i)] Suppose $V \in L^{3/2} (\bR^3)$. Then
\begin{equation}\label{3/2}
\int \rd x \, V (x) |\ph (x)|^2 \leq C \| V \|_{L^{3/2}} \int \rd x
\left( |\nabla \ph (x)|^2 + |\ph (x)|^2 \right)
\end{equation}
\item[ii)] Suppose $V  \in L^1 (\bR^3)$. Then the operator $V (x_1 -x_2)$,
viewed as a multiplication operator on $L^2 (\bR^3\times \bR^3, \rd
x_1 \, \rd x_2)$, satisfies the  following operator inequalities
\begin{equation}\label{eq:sobolev2}
V (x_1 -x_2) \leq C \| V \|_{L^1} \, (1-\Delta_1) (1-\Delta_2),
\quad \text{and}  \quad V (x_1 -x_2) \leq C \| V \|_{L^1}
(1-\Delta_1)^2.
\end{equation}
\end{itemize}
\end{lemma}
The proof of \eqref{3/2} is given in Lemma 5.2 of \cite{ESY},
the proof of the first inequality of \eqref{eq:sobolev2}
is found in Lemma 5.3 of \cite{EY}. The last inequality follows
from the usual Sobolev imbedding. $\;\;\;\Box$

\section{Convergence to the infinite hierarchy}
\setcounter{equation}{0}

The aim of this section is to prove that any limit point $
\Gamma_{\infty,t} \in \bigoplus_{k \geq 1} C([0,T], \cL^1_k)$ of the
sequence $ \Gamma_{N,t}$ satisfies the infinite hierarchy
(\ref{eq:BBGKYinf}).

\begin{theorem}\label{thm:convergence} Suppose
the assumptions of Theorem \ref{thm:main1} are satisfied and
fix $T>0$. Suppose $
\Gamma_{\infty,t} = \{ \gamma_{\infty,t}^{(k)} \}_{k\geq 1} \in
\bigoplus_{k \geq 1} C([0,T], \cL^1_k)$ is a limit point of $
\Gamma_{N,t} = \{ \gamma_{N,t}^{(k)} \}_{k=1}^N$ with respect to the
topology $\tau_{\text{prod}}$. Then $ \Gamma_{\infty,t}$ is a
solution of the infinite BBGKY hierarchy
\begin{equation}\label{infbbgky}
\begin{split}
 \gamma^{(k)}_{\infty,t} = \cU^{(k)} (t) \,
\gamma^{(k)}_{\infty,0} - 8\pi i a_0 \sum_{j=1}^k \int_0^t \rd s \,
\cU^{(k)} (t-s) \tr_{k+1}\, \left[\delta (x_j -x_{k+1}),
\gamma^{(k+1)}_{\infty,s} \right]
\end{split}
\end{equation}
with initial data $ \gamma^{(k)}_{\infty,0} = |\ph
\rangle\langle\ph|^{\otimes k}$.
\end{theorem}

{\it Remark.} Note that in terms of kernels
\[ \left( \tr_{k+1} \, \delta (x_j -x_{k+1})
\gamma^{(k+1)}_{\infty,s} \right) (\bx_k;\bx'_k) =
\gamma^{(k+1)}_{\infty,s} (\bx_k,x_j; \bx'_k,x_j) \,. \] To define
this kernel properly,
we  choose a function
$g\in C_0^{\infty} (\bR^3)$, $g \geq 0$, $\int g =1$, and we let
$g_r(x) = r^{-3} g (x/r)$.
 Then the definition is given by the limit
\begin{equation}\label{eq:limrr'}
\begin{split}
\lim_{r,r' \to 0} \int \rd x'_{k+1} \rd x_{k+1} \, g_r (x'_{k+1} -
x_{k+1}) g_{r'} &(x_{k+1} - x_j)
 \gamma^{(k+1)}_{\infty,s} (\bx_k , x_{k+1} ; \bx'_k , x'_{k+1}) \\
&= :  \gamma^{(k+1)}_{\infty,s} (\bx_k, x_j ; \bx'_k , x_j).
\end{split}
\end{equation}
The existence of this limit in a weak sense (tested against a
sufficiently smooth observable) follows from the apriori estimate
(\ref{eq:apriorik}) and from the following lemma (whose proof was
given in Lemma 8.2 in~\cite{ESY2}).

\begin{lemma}\label{lm:sobsob}   Suppose that
$\delta_\alpha(x)$ is  a function satisfying $0 \leq \delta_{\alpha}
(x) \leq C \alpha^{-3} {\bf 1} (|x| \leq \alpha)$ and $\int
\delta_{\alpha} (x) \rd x=1$ (for example $\delta_{\alpha} (x) =
\alpha^{-3} g (x/\alpha)$, for a bounded probability density $g(x)$
supported in $\{ x : |x| \leq 1\}$). Moreover, for $J^{(k)} \in
\cK_k$, and for $j=1,\dots ,k$, we define the norm
\be\label{eq:Jnorm} \tri J^{(k)} \tri_{j} := \sup_{\bx_k, \bx'_k}
\la x_1 \ra^4 \dots \la x_k \ra^4 \la x'_1 \ra^4 \dots \la x'_k
\ra^4  \left( |J^{(k)} (\bx_k ; \bx'_k)| + |\nabla_{x_j} J^{(k)}
(\bx_k;\bx'_k)| + |\nabla_{x'_j} J^{(k)} (\bx_k;\bx'_k)| \right) \,
\ee for any $j\leq k$ and for any function $J^{(k)}(\bx_k ;
\bx'_k)$ (here $\la x \ra^2 := 1+ x^2$).
 Then if $\gamma^{(k+1)} (\bx_{k+1};\bx'_{k+1})$ is the
kernel of a density matrix on $L^2 (\bR^{3(k+1)})$, we have, for any
$j\leq k$,
\begin{multline}\label{eq:gammaintbound}
\Big| \int \rd \bx_{k+1} \rd \bx'_{k+1} \, J^{(k)} (\bx_k ; \bx'_k)
\left(\delta_{\alpha_1} (x_{k+1} - x'_{k+1}) \delta_{\alpha_2} (x_j
-x_{k+1}) - \delta (x_{k+1} -x'_{k+1}) \delta (x_j -
x_{k+1})\right)\\ \times \gamma^{(k+1)} (\bx_{k+1} ; \bx'_{k+1})
\Big| \\ \leq (\const.)^k \, \tri J^{(k)} \tri_j \left( \alpha_1 +
\sqrt{\alpha_2}\right) \, \tr \, | S_j S_{k+1} \gamma^{(k+1)} S_j
S_{k+1}|\;.
\end{multline}
Recall  that $S_{\ell} = (1-\Delta_{x_{\ell}})^{1/2}$.
The same bound holds if $x_j$ is replaced with $x_j'$ in
(\ref{eq:gammaintbound}) by symmetry.
\end{lemma}

\begin{proof}[Proof of Theorem \ref{thm:convergence}]
For every integer $k
\geq 1$, and every $J^{(k)} \in \cK_k$, we have
\begin{equation}\label{eq:conv-0}
\sup_{t \in [0,T]} \, \tr \; J^{(k)} \, \left( \gamma_{N_i,t}^{(k)} -
 \gamma_{\infty,t}^{(k)} \right) \to 0
\end{equation}
along a subsequence $N_i\to \infty$.
 For an arbitrary integer $k \geq 1$, we define
$$
   \Omega_k : =  \prod_{j=1}^k \left( \la
x_j \ra + S_j \right)   \; .
$$
In the following we assume that the observable $J^{(k)} \in \cK_k$
is such that
\begin{equation}\label{eq:assJ} \Big\| \Omega_k^7
 J^{(k)}\Omega_k^7 \Big\|_{\text{HS}} < \infty ,
\end{equation}
where  $\| A \|_{\text{HS}}$ denotes the Hilbert-Schmidt norm of the
operator $A$, that is $\| A \|^2_{\text{HS}} = \tr A^* A$.
 Note that the set of observables $J^{(k)}$
satisfying the condition (\ref{eq:assJ}) is a dense subset of
$\cK_k$.

It is straightforward to check that
\begin{equation}\label{eq:assJsimple}
\| S_1 \dots S_k \, J^{(k)} S_1 \dots S_k \| \leq \Big\| \Omega_k^7
 J^{(k)}\Omega_k^7 \Big\|_{\text{HS}} .
\end{equation}
Moreover,  for any $j\leq k$ \be \label{HStoj} \tri J^{(k)}\tri_j
\leq (\mbox{const.})^k \Big\| \Omega_k^7 J^{(k)}\Omega_k^7
\Big\|_{\text{HS}} , \ee
where the norm $\tri . \tri_j$ is defined
in (\ref{eq:Jnorm}). This follows from the standard Sobolev
inequality $\| f\|_\infty \leq (\const.) \,  \| f \|_{W^{2,2}}$ in three
dimensions applied to each variable separately in the form
\begin{equation*}
\begin{split}
\Big(\sup_{x,x'} \; \langle x \rangle^4 \langle x'\rangle^4
|\nabla_x J(x,x')|\Big)^2 &\leq (\const.) \int \rd x \rd x' \Big|
  (1-\Delta_x) \Big[\langle x \rangle^4
 \big(\nabla_x J(x,x') \big)\langle x'\rangle^4\Big]\Big|^2
\\ &\leq (\const.) \; \tr \; (1-\Delta)  \langle x \rangle^4 \nabla \; J
 \, \langle x \rangle^8 \, J^* \;  \nabla^*  \;
 \langle x \rangle^4 \, (1-\Delta)
\\& \leq (\const.) \; \tr \; \Omega^7 J\Omega^{14} J^* \Omega^7
\end{split}
\end{equation*}
with $\Omega = \langle x \rangle + (1-\Delta)^{1/2}$. Similar
estimates are valid for each term in the definition of $\tri \cdot
\tri_j$, for $j \leq k$. Here we  commuted derivatives and the
weights $\langle x \rangle$; the commutators can be estimated using
Schwarz inequalities.

\medskip

For $J^{(k)} \in \cK_k$ satisfying (\ref{eq:assJ}), we prove that
\begin{equation}\label{eq:conv-1}
\tr \, J^{(k)}  \gamma_{\infty,0}^{(k)} = \tr \, J^{(k)} |\ph
\rangle \langle \ph|^{\otimes k}
\end{equation}
and that, for $t \in [0,T]$,
\begin{equation}\label{eq:conv-2}
\begin{split}
\tr \; J^{(k)}  \gamma_{\infty,t}^{(k)} = \tr \; J^{(k)} \cU^{(k)}
(t)  \gamma_{\infty,0}^{(k)} -8\pi a_0 i \sum_{j=1}^k \int_0^t \rd s
\tr \, J^{(k)} \cU^{(k)} (t-s) \left[ \delta (x_j -x_{k+1}),
\gamma^{(k+1)}_{\infty,s} \right]\, .
\end{split}
\end{equation}
Note that the trace in the last term of (\ref{eq:conv-2}) is over
$k+1$ variables. The theorem then follows from (\ref{eq:conv-1}) and
(\ref{eq:conv-2}), because the set of $J^{(k)} \in \cK_k$ satisfying
(\ref{eq:assJ}) is dense in $\cK_k$.

The relation (\ref{eq:conv-1}) follows from the assumption
(\ref{eq:thmassum2}) and (\ref{eq:conv-0}).

In order to prove (\ref{eq:conv-2}), we fix $t \in [0,T]$, we
rewrite the BBGKY hierarchy (\ref{eq:BBGKY}) in integral form and we
test it against the observable $J^{(k)}$. We obtain
\begin{equation}\label{eq:conv-4}
\begin{split}
\tr \; J^{(k)} \, \gamma_{N,t}^{(k)} = \; & \tr \; J^{(k)} \,
\cU^{(k)} (t) \gamma_{N,0}^{(k)} - i \sum_{i<j}^k \int_0^t  \rd s \,
\tr \; J^{(k)} \, \cU^{(k)}
(t-s) [ V_N (x_i -x_j), \gamma_{N,s}^{(k)} ] \\
& -i (N-k) \sum_{j=1}^k \int_0^t \rd s \, \tr J^{(k)} \cU^{(k)}
(t-s) [ V_N (x_j -x_{k+1}) , \gamma_{N,s}^{(k+1)} ] \,.
\end{split}
\end{equation}
{F}rom (\ref{eq:conv-0}) it follows immediately that
\begin{equation}\label{eq:conv-4-1}
\tr \; J^{(k)} \, \gamma_{N,t}^{(k)} \to \tr \; J^{(k)}
\gamma_{\infty,t}^{(k)}
\end{equation}
and also that
\begin{multline}\label{eq:conv-4-2}
\tr \; J^{(k)} \,\cU^{(k)} (t)  \gamma_{N,0}^{(k)} =  \tr \;
\left(\cU^{(k)} (-t) J^{(k)}\right) \,  \gamma_{N,0}^{(k)} \\ \to
\tr\;\left(\cU^{(k)} (-t) J^{(k)}\right)  \gamma_{\infty,0}^{(k)}
=\tr \; J^{(k)} \,\cU^{(k)} (t)  \gamma_{\infty,0}^{(k)}
\end{multline}
as $N \to \infty$. Here we used that, if $J^{(k)} \in \cK_k$, then
also $\cU^{(k)} (-t) J^{(k)} \in \cK_k$.

Next we consider the second term on the r.h.s. of (\ref{eq:conv-4})
and we prove that it converges to zero, as $N \to \infty$. To this
end, we recall the definition (\ref{eq:gammaij-ker}) \[
\gamma_{N,i,j,t}^{(k)} (\bx_k;\bx'_k) = (1-w(x_i -x_j))^{-1}
(1-w(x'_i -x'_j))^{-1}  \gamma_{N,t}^{(k)} (\bx_k;\bx'_k)\] for
every $i\neq j$, $i,j \leq k$. Then we obtain
\begin{equation}\label{eq:long}
\begin{split}
\left| \tr \; J^{(k)} \right. & \, \cU^{(k)} (t-s) [ V_N (x_i
-x_j),\left. \gamma_{N,s}^{(k)} ] \right| \\ \leq \; & \Big| \tr \;
\left(S_i S_j (\cU^{(k)} (s-t) J^{(k)}) S_i S_j \right) \left(
S_i^{-1} S_j^{-1} V_N (x_i -x_j) (1-w_{ij}) S_i^{-1} S_j^{-1}\right)
\\ &\hspace{2cm} \times \left(S_i S_j  \gamma^{(k)}_{N,i,j,s} S_i S_j
\right) \left(S_i^{-1} S_j^{-1} (1-w_{ij}) S_i^{-1} S_j^{-1}\right)\Big| \\
&+  \Big| \tr \; \left(S_i S_j (\cU^{(k)} (s-t) J^{(k)}) S_i
S_j\right)  \left(S_i^{-1} S_j^{-1} (1-w_{ij}) S_i^{-1}
S_j^{-1}\right)
\\ &\hspace{2cm} \times \left(S_i S_j  \gamma^{(k)}_{N,i,j,s} S_i S_j
\right) \left( S_i^{-1} S_j^{-1} V_N (x_i -x_j) (1-w_{ij}) S_i^{-1}
S_j^{-1}\right) \Big| \,.
\end{split}
\end{equation}
Since, by part iv) of Lemma \ref{lm:w},
\begin{equation}\label{eq:conv-4-2b}
\| S_i^{-1} S_j^{-1} V_N (x_i -x_j) (1-w_{ij}) S_i^{-1}
S_j^{-1} \| \leq C \int \rd x V_N (x) (1-w(x)) \leq C N^{-1}
\end{equation}
and
\begin{equation}
\| S_i^{-1} S_j^{-1} (1-w(x_i -x_j)) S_i^{-1} S_j^{-1} \| \leq 1
\end{equation}
we find
\begin{equation*}
\left| \tr \; J^{(k)} \right. \, \cU^{(k)} (t-s) [ V_N (x_i
-x_j),\left. \gamma_{N,s}^{(k)} ] \right| \leq \;   C N^{-1} \| S_i
S_j \left( \cU^{(k)} (s-t) J^{(k)} \right) S_i S_j \| \; \tr \;
S_i^2 S_j^2  \gamma_{N,i,j,s}^{(k)}\,.
\end{equation*}
{F}rom $\|S_i S_j \left( \cU^{(k)} (s-t) J^{(k)} \right) S_i S_j \|
= \| S_i S_j J^{(k)} S_i S_j \| <\infty$, and from Proposition
\ref{prop:apriori-phiij} it follows immediately that, for any $t \in
[0,T]$,
\begin{equation}\label{eq:conv-5a}
\sum_{i<j}^k \int_0^t  \rd s \, \tr \; J^{(k)} \, \cU^{(k)} (t-s) [
V_N (x_i -x_j), \gamma_{N,s}^{(k)} ] \to 0
\end{equation}
as $N \to \infty$ (the convergence is not uniform in $k$).

Finally we consider the last term on the r.h.s. of
(\ref{eq:conv-4}). First of all, we note that
\begin{equation}\label{eq:conv-5b}
k \sum_{j=1}^k \int_0^t \rd s \, \tr \, J^{(k)} \cU^{(k)} (t-s) [
V_N (x_j -x_{k+1}) , \gamma_{N,s}^{(k+1)} ] \to 0
\end{equation}
as $N \to \infty$. In fact
\begin{equation}
\begin{split}
\Big| \tr \, J^{(k)} &\cU^{(k)} (t-s) [ V_N (x_j -x_{k+1}) ,
\gamma_{N,s}^{(k+1)} ] \Big| \\ \leq \, &\Big| \tr \; \left(S_j
\cU^{(k)} (s-t) J^{(k)} S_j \right)\; \left(S_j^{-1} S_{k+1}^{-1}
V_N
(x_j - x_{k+1}) (1- w_{j,k+1}) S_{k+1}^{-1} S_j^{-1} \right) \\
&\hspace{1cm} \times \left( S_{k+1} S_j \gamma^{(k+1)}_{N,j,k+1,s}
S_j S_{k+1}\right) \left(S_j^{-1} S_{k+1}^{-1} (1-w_{j,k+1})
S_{k+1} S_j^{-1}\right)\Big| \\
&+ \Big| \tr \; \left(S_j J^{(k)} S_j\right)
\left(S_j^{-1} S_{k+1} (1-w_{j,k+1}) S_{k+1}^{-1} S_j^{-1}\right) \\
&\hspace{1cm} \times \left(S_{k+1} S_j \gamma^{(k+1)}_{N,j,k+1,s}
S_j S_{k+1}\right) \; \left(S_j^{-1} S_{k+1}^{-1} N V_N (x_j -
x_{k+1}) (1- w_{j,k+1}) S_{k+1}^{-1} S_j^{-1}\right)\Big|\,.
\end{split}
\end{equation}
As in (\ref{eq:conv-4-2b}) we have $\| S_j^{-1} S_{k+1}^{-1} V_N
(x_j -x_{k+1}) (1-w_{j,k+1}) S_j^{-1} S_{k+1}^{-1} \| \leq C
N^{-1}$. Moreover (see (\ref{eq:equi-2})),
\begin{equation}
\| S_j^{-1} S_{k+1} (1-w_{j,k+1}) S_{k+1}^{-1} S_j^{-1} \| \leq C.
\end{equation}
By an argument very similar to \eqref{eq:long}--\eqref{eq:conv-5a}
and by Proposition \ref{prop:apriori-phiij} we obtain
(\ref{eq:conv-5b}).

\medskip

It remains to consider
\begin{equation}\label{eq:conv-6}
\begin{split}
N \sum_{j=1}^k \int_0^t \rd s &\, \tr \, J^{(k)} \cU^{(k)} (t-s) [
V_N (x_j -x_{k+1}) , \gamma_{N,s}^{(k+1)} ]
\\ = \; &\sum_{j=1}^k \int_0^t \rd s \, \tr \, \left(\cU^{(k)} (s-t)
J^{(k)}\right) \left[ N V_N (x_j -x_{k+1}) (1-w_{j,k+1}),
\gamma_{N,j,k+1,s}^{(k+1)} \right] \\ &- \sum_{j=1}^k \int_0^t \rd s
\, \tr \, \left(\cU^{(k)} (s-t) J^{(k)}\right) N V_N (x_j -x_{k+1})
(1-w_{j,k+1}) \gamma_{N,j,k+1,s}^{(k+1)} w_{j,k+1}
\\ &+\sum_{j=1}^k \int_0^t \rd s \, \tr \, \left(\cU^{(k)} (s-t)
J^{(k)}\right) w_{j,k+1} \gamma_{N,j,k+1,s}^{(k+1)} (1-w_{j,k+1}) N
V_N (x_j -x_{k+1})\,.
\end{split}
\end{equation}
The terms on the third and fourth lines converge to zero, as $N \to
\infty$. For example, the contributions on the third line can be
bounded by
\begin{equation}
\begin{split}
 \Big| \tr \, &\left(\cU^{(k)} (s-t) J^{(k)}\right)  N V_N (x_j -x_{k+1})
(1-w_{j,k+1}) \gamma_{N,j,k+1,s}^{(k+1)} w_{j,k+1} \Big| \\ \leq \;
& \| S_j \left(\cU^{(k)} (s-t) J^{(k)}\right) S_j \| \| S_j^{-1}
S_{k+1}^{-1} \left(N V_N (x_j -x_{k+1}) (1-w_{j,k+1})\right)
S_j^{-1} S_{k+1}^{-1} \|  \\ &\times \| S_j^{-1} S_{k+1} w_{j,k+1}
S_j^{-1} S_{k+1}^{-1} \| \; \tr \; S_j^2 S_{k+1}^2 \,
\gamma_{N,j,k+1,s}^{(k+1)}\,.
\end{split}
\end{equation}
Then we use
\begin{equation}
\| S_j^{-1} S_{k+1}^{-1} NV_N (x_j -x_{k+1}) (1-w_{j,k+1}) S_j^{-1}
S_{k+1}^{-1} \| \leq C \end{equation}
and
\begin{equation}\label{eq:SwS}
\begin{split}
\|S_j^{-1} S_{k+1} w_{j,k+1} S_j^{-1} S_{k+1}^{-1} \|
\leq  & \; \Big\| S_{k+1}^{-1} S_j^{-1} w_{j,k+1} S_{k+1}^2 w_{j,k+1}
S_j^{-1} S_{k+1}^{-1} \Big\|^{1/2} \\
\leq & \;\| S_j^{-1} w_{j,k+1}^2 S_j^{-1} \|^{1/2}
  + \|  S_{k+1}^{-1} S_j^{-1} (\nabla w_{j,k+1})^2
 S_j^{-1}  S_{k+1}^{-1}\|^{1/2} \\
\leq & \;  C N^{-1} + CN^{-1/4}\,.
\end{split}
\end{equation}
 To prove (\ref{eq:SwS}), we applied Lemma \ref{lm:sobolev} and
the fact that, by Lemma \ref{lm:w}, with $R$ such that $\supp V
\subset \{ x \in \bR^3 : |x| \leq R\}$,
\[ w(x) \leq C \chi (|x| <R/N) + a \frac{\chi (|x|> R/N)}{|x|} \leq
\frac{C}{N|x|},\] and \[ |\nabla w(x)|^2 \leq C \frac{a}{|x|^2} \,
|\nabla w(x)| \leq C \frac{1}{N^{1/2} |x|^{5/2}} \] (the last bound
is obtained interpolating the first bound in (\ref{eq:w-1}) and the
second bound in (\ref{eq:w-2})). It follows that
\begin{equation}
\begin{split}
\Big| \sum_{j=1}^k \int_0^t \rd s \, \tr \, \left(\cU^{(k)} (s-t)
J^{(k)}\right) &N V_N (x_j -x_{k+1}) (1-w_{j,k+1})
\gamma_{N,j,k+1,s}^{(k+1)} w_{j,k+1} \Big|\\ \leq \; & C t k
N^{-1/4} \max_{j\leq k} \, \sup_{s \in [0,t]} \tr\; S_j S_{k+1}
\gamma_{N,j,k+1,s}^{(k+1)} S_j S_{k+1}
\end{split}
\end{equation}
which converges to zero, as $N \to \infty$, by using Proposition
\ref{prop:apriori-phiij}. The fourth line of (\ref{eq:conv-6}) can be
handled analogously. Hence, from (\ref{eq:conv-6}),
\begin{equation}\label{eq:conv-6b}
\begin{split}
N \sum_{j=1}^k \int_0^t \rd s &\, \tr \, J^{(k)} \cU^{(k)} (t-s) [
V_N (x_j -x_{k+1}) , \gamma_{N,s}^{(k+1)} ]
\\ = \; &\sum_{j=1}^k \int_0^t \rd s \, \tr \, \left(\cU^{(k)} (s-t)
J^{(k)}\right) \left[ N V_N (x_j -x_{k+1}) (1-w_{j,k+1}),
\gamma_{N,j,k+1,s}^{(k+1)} \right] \\ &+ C_{k,T} \; o_N(1)
\end{split}
\end{equation}
where $o_N (1) \to 0$ as $N\to \infty$ and $C_{k,T}$ is a constant
depending on $k$ and on $T$.

To handle the r.h.s. of (\ref{eq:conv-6b}), we choose a compactly
supported positive function $h \in C_0^{\infty} (\bR^3)$ with $\int
\rd x \, h(x) =1$. For $\beta >0$, we define $\delta_{\beta} (x) =
\beta^{-3} h(x/\beta)$, i.e. $\delta_{\beta}$ is an approximate
delta-function on the scale $\beta$. Then we have
\begin{equation}\label{eq:conv-7}
\begin{split}
\sum_{j=1}^k &\int_0^t \rd s \, \tr \, \left(\cU^{(k)} (s-t)
J^{(k)}\right) \left[ N V_N (x_j -x_{k+1}) (1-w_{j,k+1}),
\gamma_{N,j,k+1,s}^{(k+1)} \right] \\ = \; &\sum_{j=1}^k \int_0^t
\rd s \, \tr \, \left(\cU^{(k)} (s-t) J^{(k)}\right) \left[ N V_N
(x_j -x_{k+1}) (1-w_{j,k+1}) - 8\pi a_0 \delta_{\beta} (x_j
-x_{k+1}), \gamma_{N,j,k+1,s}^{(k+1)} \right] \\ &+\sum_{j=1}^k
\int_0^t \rd s \, \tr \, \left(\cU^{(k)} (s-t) J^{(k)}\right) \left[
8\pi a_0 \delta_{\beta} (x_j -x_{k+1}), \gamma_{N,j,k+1,s}^{(k+1)}
\right]
\\ = \; & \sum_{j=1}^k
\int_0^t \rd s \, \tr \, \left(\cU^{(k)} (s-t) J^{(k)}\right) \left[
8\pi a_0 \delta_{\beta} (x_j -x_{k+1}), \gamma_{N,j,k+1,s}^{(k+1)}
\right] \\ &+C_{k,T} \left( O(N^{-1/2}) + O(\beta^{1/2}) \right)
\end{split}
\end{equation}
for some constant $C_{k,T}$ which depends on $k\geq 1$, on $T$, and
on $J^{(k)}$ ($O(\beta^{1/2})$ is independent of $N$). Here we used
that, by (\ref{eq:a_0}),
\begin{equation}
\int \rd x N V_N(x) (1-w(x)) = 8\pi a_0\,,
\end{equation}
and we applied Lemma \ref{lm:sobsob}.  To apply Lemma
\ref{lm:sobsob}, we used Proposition \ref{prop:apriori-phiij} and
that, by (\ref{HStoj}),
$$
 \tri \cU^{(k)}_0 (s-t) J^{(k)} \tri_j
\leq C \; \big\| \Omega_k^7 \; \cU^{(k)}_0 (s-t) J^{(k)} \;
\Omega_k^7 \big\|_{\text{HS}}
$$
with a $k$-dependent constant $C$. Since $e^{-i(s-t)\Delta_j} \la
x_j \ra^{m} e^{i(s-t) \Delta_j} = \la x_j +2 (s-t) p_j \ra^{m}$, for
any $j=1,\dots,k$, $m\in \bN$, we obtain that
$$
 \tri \cU^{(k)}_0 (s-t) J^{(k)} \tri_j
\leq C (1 + |t-s|^7)\big\| \Omega_k^7  \; J^{(k)} \;  \Omega_k^7
\big\|_{\text{HS}} \; .
$$

\medskip

To control the first term on the r.h.s. of (\ref{eq:conv-7}) we go
back to $ \gamma_{N,t}^{(k+1)}$. We write
\begin{equation}\label{eq:ga}
\gamma^{(k+1)}_{N,j,k+1,s} = \gamma^{(k+1)}_{N,s} +
\Big(\frac{1}{1-w_{j,k+1}}-1\Big) \gamma^{(k+1)}_{N,s} +
\frac{1}{1-w_{j,k+1}} \gamma^{(k+1)}_{N,s}
\Big(\frac{1}{1-w_{j,k+1}}-1\Big)\,.
\end{equation}
When we insert (\ref{eq:ga}) in the r.h.s. of (\ref{eq:conv-7}), the
contributions arising from the last two terms in (\ref{eq:ga})
converge to zero, as $N \to \infty$, for any fixed $\beta >0$. For
example, to bound the contribution of the second term on the r.h.s.
of (\ref{eq:ga}), we use that
\begin{equation}\label{eq:conv-9}
\begin{split}
\Big| \tr \,& \left(\cU^{(k)} (s-t) J^{(k)}\right) \left[ 8\pi a_0
\delta_{\beta} (x_j -x_{k+1}), \left( \frac{1}{1-w_{j,k+1}}-1\right)
\gamma_{N,s}^{(k+1)} \right] \Big| \\ \leq \; &C \Big| \tr \;
\left(\cU^{(k)} (s-t) J^{(k)}\right) \left( S_{k+1}^{-1}
\delta_{\beta} (x_j -x_{k+1}) S_{k+1} \right) \left( S_{k+1}^{-1}
\frac{w_{j,k+1}}{1-w_{j,k+1}} S_{k+1}^{-1}\right) \left(
S_{k+1} \gamma_{N,s}^{(k+1)}S_{k+1}\right) \Big|\\
&+ C \Big| \tr \; \left(\cU^{(k)} (s-t) J^{(k)}\right)
\left(S_{k+1}^{-1} \frac{w_{j,k+1}}{1-w_{j,k+1}} S_{k+1}^{-1}\right)
\left( S_{k+1} \gamma_{N,s}^{(k+1)}S_{k+1}\right) \left(
S_{k+1}^{-1} \delta_{\beta} (x_j -x_{k+1}) S_{k+1}\right)  \Big|
\\ \leq \; &C \| S_{k+1}^{-1} \delta_{\beta} (x_j -x_{k+1}) S_{k+1} \|
\; \Big\| S_{k+1}^{-1}\frac{w_{j,k+1}}{1-w_{j,k+1}} S_{k+1}^{-1}\Big\| \;
\tr \; S_{k+1}^2  \gamma_{N,s}^{(k+1)}\,.
\end{split}
\end{equation}
Now we have
\begin{equation}
\left\| S_{k+1}^{-1} \frac{w_{j,k+1}}{1-w_{j,k+1}} S_{k+1}^{-1}
\right\| \leq C N^{-1}
\end{equation}
because $w(x) \leq C a |x|^{-1}$ and thus, as an operator
inequality, $w_{j,k+1} \leq Ca S_{k+1}^2$ (and $a \simeq N^{-1}$).
Moreover
\begin{equation}\label{eq:conv-9b}
\begin{split}
\tr \; S_{k+1}^2  \gamma_{N,s}^{(k)} &=  \langle  \psi_{N,s},
(1-\Delta_{k+1})  \psi_{N,s} \rangle \\ &\leq  N^{-1} \langle
\psi_{N,s}, (H_N +N)  \psi_{N,s} \rangle \\ &= N^{-1} \langle
\psi_{N}, (H_N +N)  \psi_{N} \rangle \leq C
\end{split}
\end{equation}
by the assumption (\ref{eq:thmassum1}). It is also easy to see that
\begin{equation}
\| S_{k+1}^{-1} \delta_{\beta} (x_j -x_{k+1}) S_{k+1} \| \leq C
\beta^{-4}
\end{equation}
for $\beta <1$. The contribution arising from the last term on the
r.h.s. of (\ref{eq:ga}) can also be controlled similarly. Therefore,
it follows from (\ref{eq:conv-6b}), (\ref{eq:conv-7}),
(\ref{eq:ga}), and (\ref{eq:conv-9}) that
\begin{equation}\label{eq:conv-10}
\begin{split}
N \sum_{j=1}^k \int_0^t \rd s &\, \tr \, J^{(k)} \cU^{(k)} (t-s)
\left[ V_N (x_j -x_{k+1}) , \gamma_{N,s}^{(k+1)} \right]
\\
= \; & 8\pi a_0  \sum_{j=1}^k \int_0^t \rd s \, \tr \, \left(\cU^{(k)}
(s-t) J^{(k)}\right) \left[ \delta_{\beta} (x_j -x_{k+1}),
 \gamma_{\infty,s}^{(k+1)} \right] \\
&+8\pi a_0 \sum_{j=1}^k \int_0^t \rd s \, \tr \,
\left(\cU^{(k)} (s-t) J^{(k)}\right) \left[ \delta_{\beta} (x_j
-x_{k+1}),
\gamma_{N,s}^{(k+1)} - \gamma_{\infty,s}^{(k+1)} \right] \\
&+C_{k,T} \left(  O(\beta^{1/2}) + o_N (1) \right)\,,
\end{split}
\end{equation}
where $o_N(1) \to 0$ as $N \to \infty$ (for any fixed $\beta >0$).
The first term is the main term.
To control the second term, we rewrite it, for $\e
>0$, as
\begin{equation}\label{eq:conv-11}
\begin{split}
\sum_{j=1}^k \int_0^t \rd s \, \tr \, &\left(\cU^{(k)} (s-t)
J^{(k)}\right) \left[ \delta_{\beta} (x_j -x_{k+1}),
\gamma_{N,s}^{(k+1)} - \gamma_{\infty,s}^{(k+1)} \right] \\ = \;
&\sum_{j=1}^k \int_0^t \rd s \, \tr\,\left(\cU^{(k)} (s-t)
J^{(k)}\right) \delta_{\beta} (x_j -x_{k+1}) \frac{1}{1+\e S_{k+1}}
\left(\gamma_{N,s}^{(k+1)} -  \gamma_{\infty,s}^{(k+1)} \right)
\\&+ \sum_{j=1}^k \int_0^t \rd s \, \tr\,\left(\cU^{(k)} (s-t)
J^{(k)}\right) \delta_{\beta} (x_j -x_{k+1}) \left( 1-\frac{1}{1+\e
S_{k+1}}\right) \left(\gamma_{N,s}^{(k+1)} -
\gamma_{\infty,s}^{(k+1)} \right)\\ &- \sum_{j=1}^k \int_0^t \rd s\,
\tr \, \delta_{\beta} (x_j -x_{k+1}) \frac{1}{1+\e S_{k+1}}
\left(\cU^{(k)} (s-t) J^{(k)} \right) \left(\gamma_{N,s}^{(k+1)} -
\gamma_{\infty,s}^{(k+1)} \right)
\\ &- \sum_{j=1}^k \int_0^t \rd s \, \tr\,\delta_{\beta} (x_j
-x_{k+1}) \left( 1-\frac{1}{1+\e S_{k+1}}\right) \left(\cU^{(k)}
(s-t) J^{(k)}\right) \left(\gamma_{N,s}^{(k+1)} -
\gamma_{\infty,s}^{(k+1)} \right) \,.
\end{split}
\end{equation}
The second term on the r.h.s. of (\ref{eq:conv-11}) can be bounded
by using that
\begin{equation}
\begin{split}
\Big| \tr \, \left(\cU^{(k)} (s-t) J^{(k)}\right) &\delta_{\beta}
(x_j -x_{k+1}) \left( 1-\frac{1}{1+\e S_{k+1}}\right)
\left(\gamma_{N,s}^{(k+1)} - \gamma_{\infty,s}^{(k+1)} \right)\Big|
\\ \leq \; &\e \Big\| \left(\cU^{(k)} (s-t) J^{(k)}\right)
\delta_{\beta} (x_j -x_{k+1}) \Big\|  \, \left(\tr\; S_{k+1}
\gamma_{N,s}^{(k+1)}S_{k+1} + \tr S_{k+1} \gamma_{\infty,s}^{(k+1)}
S_{k+1} \right) \\ \leq \; &C \beta^{-3} \e \, \left( \tr \;
S_{k+1}^2  \gamma_{N,s}^{(k+1)} + \tr S_{k+1}^2
\gamma_{\infty,s}^{(k+1)} \right)
\\ \leq \; &C \beta^{-3} \e
\end{split}
\end{equation}
where we used (\ref{eq:conv-9b}) and Proposition
\ref{prop:apriorik}. Also the fourth term on the r.h.s. of
(\ref{eq:conv-11}) can be controlled analogously. As for the first
and third term on the r.h.s. of (\ref{eq:conv-11}), we note that for
every fixed $\e >0$, $\beta >0$ and $s \in [0,t]$, the integrand
converges to zero, as $N \to \infty$, by (\ref{eq:conv-0}), and
because \begin{equation}\label{eq:cUs} \left(\cU^{(k)} (s-t)
J^{(k)}\right) \delta_{\beta} (x_j -x_{k+1}) \frac{1}{1+\e S_{k+1}},
\quad \delta_{\beta} (x_j -x_{k+1}) \frac{1}{1+\e S_{k+1}}
\left(\cU^{(k)} (s-t) J^{(k)}\right) \in \cK_{k+1} \, .
\end{equation} Since, moreover, the integrand is bounded uniformly
in $s \in [0,t]$ (because for fixed $\e, \beta >0$ the norm of the
operators (\ref{eq:cUs}) is bounded uniformly in $s$), it follows
from Lebesgue dominated convergence theorem and from
(\ref{eq:conv-10}) that
\begin{equation}\label{eq:conv-12}
\begin{split}
N \sum_{j=1}^k \int_0^t \rd s &\, \tr \, J^{(k)} \cU^{(k)} (t-s)
\left[ V_N (x_j -x_{k+1}) , \gamma_{N,s}^{(k+1)} \right]
 \\
= &\; \sum_{j=1}^k \int_0^t \rd s \, \tr \, \left(\cU^{(k)} (s-t)
J^{(k)}\right) \left[ 8\pi a_0 \delta_{\beta} (x_j -x_{k+1}),
 \gamma_{\infty,s}^{(k+1)} \right]\\
&+C_{k,T} \left(O(\beta^{1/2}) + \beta^{-3} O(\e) + o_N (1) \right)
\end{split}
\end{equation}
where the convergence $o_N (1) \to 0$ as $N \to \infty$ depends on
$\eps$ and $\beta$. By applying
 Lemma \ref{lm:sobsob} again and by using that,
 by
Proposition \ref{prop:apriorik}, \[ \max_{j=1,\dots, k} \, \sup_{t
\in [0,T]} \tr (1-\Delta_j) (1-\Delta_{k+1}) \,
\gamma_{\infty,t}^{(k+1)} \leq C \,.\]
we can replace $\delta_\beta(x_j-x_{k+1})$ with $\delta(x_j-x_{k+1})$
in \eqref{eq:conv-12}
at the expense of an error $O(\beta^{-1/2})$.

\medskip

{F}rom (\ref{eq:conv-4}), (\ref{eq:conv-4-1}), (\ref{eq:conv-4-2}),
(\ref{eq:conv-5a}), (\ref{eq:conv-5b}), and (\ref{eq:conv-12})
with $\delta(x_j-x_{k+1})$ it
follows, letting $N \to \infty$ with fixed $\beta >0$ and $\e > 0$,
that
\begin{equation}
\begin{split}
\tr \; J^{(k)}  \gamma^{(k)}_{\infty,t}
= \; &\tr \; J^{(k)} \cU^{(k)} (t) \,  \gamma_{\infty,0}^{(k)} \\
&-i \sum_{j=1}^k \int_0^t \rd s \, \tr \, \left(\cU^{(k)} (s-t)
J^{(k)}\right) \left[ 8\pi a_0 \delta (x_j -x_{k+1}),
\gamma_{\infty,s}^{(k+1)} \right] + O(\beta^{1/2})+ \beta^{-4} O(\e)\; .
\nonumber
\end{split}
\end{equation}
Eq. (\ref{eq:conv-2}) now
follows from the last equation letting first $\e \to 0$ and then
$\beta\to 0$.
\end{proof}

\section{Regularization of the Initial Wave Function}
\setcounter{equation}{0}

In this section we show how to regularize the initial wave function
$\psi_N$  given in Theorem \ref{thm:main2}.

\begin{proposition}\label{prop:initialdata}
Suppose that (\ref{eq:thm2ass1}) is satisfied. For $\kappa
>0$ we define
\begin{equation}\label{eq:wtpsi}
\wt \psi_N: = \frac{\chi ( \kappa H_N/N ) \psi_N}{\| \chi (\kappa
H_N/N) \psi_N \|} \, .
\end{equation}
Here $\chi \in C^{\infty}_0 ( \bR )$ is a cutoff function such that
$0\leq \chi\leq 1$, $\chi (s) =1$ for $0 \leq s \leq 1$ and $\chi
(s) =0$ for $s \geq 2$. We denote by $\wt \gamma^{(k)}_{N}$, for $k
=1, \dots, N$, the marginal densities associated with $\wt \psi_N$.
\begin{enumerate}
\item[i)] For every integer $k \geq 1$ we have
\begin{equation}
\langle \wt \psi_N , H_N^k \, \wt \psi_N \rangle \leq \frac{2^k
N^k}{\kappa^k}\,.
\end{equation}
\item[ii)] We have
\[ \sup_N \| \psi_N - \wt\psi_N \| \leq C \kappa^{1/2} \; .
\]
\item[iii)] Suppose, moreover, that the assumption (\ref{eq:thm2ass2}) is
satisfied, that is,  suppose that there exists $\ph \in L^2 (\bR^3)$
and, for every $N \in \bN$ and $k =1, \dots ,N$, there exists
$\xi^{(N-k)}_N \in L^2_s (\bR^{3(N-k)})$ with $\|\xi^{(N-k)}_N\| =1$
such that
\begin{equation}\label{eq:fact}
\lim_{N\to\infty}\| \psi_N - \ph^{\otimes k} \otimes \xi^{(N-k)}_N \| = 0\; .
\end{equation}
Then, for $\kappa >0$ small enough, and for every
fixed $k \geq 1$ and $J^{(k)} \in \cK_k$, we have
\begin{equation}\label{eq:init-3}
\lim_{N\to\infty}
\tr \; J^{(k)} \left( \wt \gamma^{(k)}_{N} - |\ph \rangle \langle
\ph|^{\otimes k} \right) = 0\; .
\end{equation}
\end{enumerate}
\end{proposition}

\begin{proof}
The proof of part i) and ii) is analogous to the proof of part i)
and ii) of Proposition 5.1 in \cite{ESY2}. Introduce the shorthand
notation $\Xi := \chi (\kappa H_N /N)$.
In order to prove i), we note that $ {\bf 1}
(H_N \leq 2 N/\kappa) \Xi = \Xi$, where ${\bf 1} (s
\leq \lambda) $ is the characteristic function of $[0, \lambda]$.
Therefore
\begin{equation}
\begin{split}
\langle \wt \psi_N , H_N^k \, \wt \psi_N \rangle &= \left\langle
\frac{\Xi \psi_N}{\| \Xi \psi_N \|}, H^k_N \frac{\Xi \psi_N}{\| \Xi
\psi_N \|} \right\rangle = \left\langle \frac{\Xi \psi_N}{\| \Xi
\psi_N \|}, {\bf 1} (H_N \leq 2 N /\kappa) H^k_N \frac{\Xi
\psi_N}{\| \Xi \psi_N \|} \right\rangle \\ &\leq \|{\bf 1} (H_N \leq
2N /\kappa) H^k_N\| \leq \frac{2^k N^k}{\kappa^k} \; .
\end{split}
\end{equation}

To prove ii), we compute
\begin{equation}
\| \Xi \psi_N - \psi_N \|^2 =  \Big\langle\psi_N , ( 1 - \Xi )^2
\psi_N\Big\rangle \leq \Big\langle\psi_N, {\bf 1} ( \kappa H_N \geq
N) \psi_N\Big\rangle\;.
 \end{equation}
Next we use that ${\bf 1} (s \geq 1) \leq s$, for all $s \geq 0$.
Therefore
\begin{equation}
\| \Xi \psi_N - \psi_N \|^2 \leq \frac{\kappa}{N} \langle\psi_N ,
H_N \psi_N\rangle \leq C \kappa
\end{equation}
by the assumption (\ref{eq:thm2ass1}). Hence
\begin{equation}\label{eq:eps1}
\| \Xi \psi_N - \psi_N \| \leq C \kappa^{1/2}.
\end{equation}
Since  $\| \psi_N \|=1$, part ii) follows by (\ref{eq:eps1}),
because
\begin{equation}
\begin{split}
\Big\| \psi_N - \frac{ \Xi \psi_N}{\| \Xi \psi_N \|} \Big\| &\leq \|
\psi_N - \Xi \psi_N \| + \Big\| \Xi \psi_N -\frac{ \Xi \psi_N}{\|
\Xi \psi_N \|} \Big\| = \| \psi_N - \Xi \psi_N \| + | 1 - \| \Xi
\psi_N \| | \\ &\leq 2 \| \psi_N - \Xi \psi_N \| \, .
\end{split}
\end{equation}

\medskip

Finally, we prove iii). For any sufficiently small $\kappa$ we will
prove that for any fixed $k \geq 1$, $J^{(k)} \in \cK_k$ and $\e >0$
(small enough)
\begin{equation}\label{eq:cutoff}
\left|\tr \; J^{(k)} \left( \wt \gamma^{(k)}_N - |\ph\rangle \langle
\ph|^{\otimes k} \right) \right| \leq \e
\end{equation}
holds if $N\ge N_0(k,\e)$ is large enough. To this end, we choose
$\ph_* \in H^2 (\bR^3)$ with $\| \ph_* \| =1$, such that $\| \ph -
\ph_* \| \leq \e/ (32 k \| J^{(k)} \|)$. Then we have
\begin{equation}\label{eq:ph-ph*} \| \ph^{\otimes k} \otimes
\xi_N^{(N-k)} - \ph_*^{\otimes k} \otimes \xi_N^{(N-k)} \| \leq k \|
\ph - \ph_* \| \leq \frac{\e}{32 \| J^{(k)} \|} \,.
\end{equation} Therefore
\begin{equation}
\Big\| \frac{\Xi \psi_N}{\| \Xi \psi_N \|} - \frac{\Xi \left(
\ph_*^{\otimes k} \otimes \xi_N^{(N-k)} \right)}{\|\Xi \left(
\ph_*^{\otimes k} \otimes \xi_N^{(N-k)} \right)\|} \Big\| \leq
\frac{2}{\| \Xi \psi_N \|} \, \Big\| \Xi \left(\psi_N -
\ph_*^{\otimes k} \otimes \xi_N^{(N-k)} \right) \Big\| \leq 4 \big\|
\psi_N - \ph_*^{\otimes k} \otimes \xi_N^{(N-k)} \big\|
\end{equation}
for $\kappa >0$ small enough (by (\ref{eq:eps1}) and because $\|
\Xi\| \leq 1$). Hence
\begin{equation}\label{eq:Xipsi-ph}
\begin{split}
\Big\| \frac{\Xi \psi_N}{\| \Xi \psi_N \|} - \frac{\Xi \left(
\ph_*^{\otimes k} \otimes \xi_N^{(N-k)} \right)}{\|\Xi \left(
\ph_*^{\otimes k} \otimes \xi_N^{(N-k)} \right)\|} \Big\| &\leq 4 \big\|
\psi_N - \ph^{\otimes k} \otimes \xi_N^{(N-k)} \big\| + 4\big\| \ph^{\otimes
k} \otimes \xi_N^{(N-k)} - \ph_*^{\otimes k} \otimes \xi_N^{(N-k)}\big\|
 \\ &\leq \frac{\e}{6 \| J^{(k)}\|}
\end{split}
\end{equation}
for $N$ large enough. Here we used (\ref{eq:ph-ph*}) and the
assumption (\ref{eq:fact}). Next we define the Hamiltonian
\begin{equation}
\wh H_N := -\sum_{j=k+1}^N \Delta_j + \sum_{k<i<j}^N V_N
(x_i-x_j)\,.
\end{equation}
Note that $\wh H_N$ acts only   on the last $N-k$ variables. We set
$\wh\Xi := \chi (\kappa \wh H_N /N)$. Then, from (\ref{eq:Xipsi-ph}),
we will obtain
\begin{equation}\label{eq:hatXipsi-ph}
\begin{split}
\Big\| \frac{\Xi \psi_N}{\| \Xi \psi_N \|} - \frac{\wh\Xi \left(
\ph_*^{\otimes k} \otimes \xi_N^{(N-k)} \right)}{\|\wh\Xi \left(
\ph_*^{\otimes k} \otimes \xi_N^{(N-k)} \right)\|} \Big\| \leq
\frac{\e}{3 \| J^{(k)}\|}
\end{split}
\end{equation}
for $N$ sufficiently large (if $\kappa >0$ and $\eps >0$ are small
enough).

Before proving (\ref{eq:hatXipsi-ph}), let us show how
(\ref{eq:cutoff}) follows from it. Let
\[ \wh\psi_{N} := \frac{\wh\Xi
\left( \ph_*^{\otimes k} \otimes \xi_N^{(N-k)} \right)}{\|\wh\Xi
\left( \ph_*^{\otimes k} \otimes \xi_N^{(N-k)} \right)\|} =
\ph_*^{\otimes k} \otimes \frac{\wh\Xi \xi_N^{(N-k)}}{\|\wh\Xi
\xi_N^{(N-k)}\|} \] since $\wh \Xi$ acts only on the last $N-k$
variables and since $\| \ph_*\|=1$. Moreover, we define
\[ \wh \gamma_N^{(k)} (\bx_k;\bx'_k) := \int \rd \bx_{N-k} \, \wh
\psi_{N} (\bx_k, \bx_{N-k}) \overline{\wh\psi}_{N}
(\bx'_k,\bx_{N-k})\,. \] Note that $\wh \psi_N$ is not symmetric in
all variables, but it is symmetric in the first $k$ and the last
$N-k$ variables. In particular, $\wh\gamma_N^{(k)}$ is a density
matrix and clearly
\[ \wh \gamma_N^{(k)} =
|\ph_* \rangle \langle \ph_*|^{\otimes k} \quad \text{i.e.} \quad
\wh\gamma_N^{(k)} (\bx_k;\bx'_k) = \prod_{j=1}^k \ph_*(x_j)
\overline{\ph}_* (x'_j) \,.\] Therefore, since
 $\| \wt \psi_N - \wh \psi_{N} \| \leq \e
/(3\| J^{(k)}\|)$ by (\ref{eq:hatXipsi-ph}) and since $\| \ph -
\ph_*\| \leq \e/(32k \|J^{(k)}\|)$, we have
\begin{equation}
\begin{split}
\Big|\tr \; J^{(k)} \left( \wt\gamma^{(k)}_{N} - |\ph \rangle
\langle \ph|^{\otimes k} \right) \Big| \leq \; &\Big|\tr \; J^{(k)}
\left( \wt\gamma^{(k)}_{N} - |\ph_* \rangle \langle \ph_*|^{\otimes
k} \right) \Big| + \Big| \tr \; J^{(k)} \left( |\ph_* \rangle
\langle \ph_*|^{\otimes k} - |\ph\rangle \langle \ph|^{\otimes k}
\right)\Big| \\ \leq \; &2 \| J^{(k)} \| \, \| \wt \psi_N - \wh
\psi_{N} \| + 2k\| J^{(k)}\| \, \| \ph - \ph_* \| \leq \e
\end{split}
\end{equation}
for $N$ sufficiently large (for arbitrary $\kappa, \e
>0$ small enough). This proves (\ref{eq:cutoff}).

\medskip

It remains to prove (\ref{eq:hatXipsi-ph}). To this end, we set
$\psi_{N,*}:=\ph_*^{\otimes k} \otimes \xi_N^{(N-k)}$, and we expand
the operator $\Xi-\wh\Xi=\chi(\kappa H_N/N) - \chi (\kappa \wh
H_N/N)$ using the Helffer-Sj\"ostrand functional calculus (see, for
example, \cite{Dav}). Let $\wt \chi$ be an almost analytic extension
of the smooth function $\chi$ of order three (that is
$|\partial_{\overline z} \wt \chi (z)| \leq C |y|^3$, for $y
=\text{Im} z$ near zero): for example we can take $\wt\chi (z=x+iy):
= [\chi (x) + i y \chi' (x) + \chi'' (x) (iy)^2 /2 + \chi''' (x)
(iy)^3 /6] \theta (x,y)$, where $\theta \in C_0^{\infty} (\bR^2)$
and $\theta (x,y) = 1$ for $z =x +iy$ in some complex neighborhood
of the support of $\chi$. Then
\begin{equation}
\begin{split}
(\Xi - \wh \Xi) \psi_{N,*} &= -\frac{1}{\pi} \int \rd x \, \rd y \,
\partial_{\bar z} \wt \chi (z) \, \left( \frac{1}{z - (\kappa
H_N/N)} - \frac{1}{z - (\kappa \wh H_N/N)} \right) \psi_{N,*} \\
&= - \frac{\kappa}{N \pi} \int \rd x \, \rd y \,
\partial_{\bar z} \wt \chi (z) \, \frac{1}{z - (\kappa
H_N/N)} (H_N - \wh H_N) \frac{1}{z - (\kappa \wh H_N/N)} \psi_{N,*}
\, .
\end{split}
\end{equation}
Taking the norm we obtain
\begin{equation}\label{eq:Helffer}
\| (\Xi - \wh \Xi) \psi_{N,*} \| \leq \frac{C\kappa}{N} \int \rd x
\, \rd y \, \frac{|\partial_{\bar z } \wt \chi (z)|}{|y|} \, \Big\|
 \frac{1}{z- (\kappa H_N/N)} (H_N - \wh H_N ) \frac{1}{z - (\kappa \wh H_N/N)}\psi_{N,*} \Big\|
\, .
\end{equation}
Notice that the operator
\begin{equation}
\label{eq:diff}
H_N -\wh H_N = -\sum_{j=1}^k \Delta_j + \sum_{i\leq
k, i < j \leq N} V_N (x_i -x_j) \,
\end{equation}
is positive hence $(H_N-\wh H_N)^{1/2}$ exists. By using
$\|AB\psi\|^2 \leq \|A\|^2 \langle\psi, B^*B\psi\rangle$, we obtain
\begin{equation}\label{eq:norm2}
\begin{split}
\Big\| \frac{1}{z- (\kappa H_N/N)} & (H_N - \wh H_N ) \frac{1}{z -
(\kappa \wh H_N/N)}\psi_{N,*}
 \Big\|^2 \\
& \leq \Big\| (H_N - \wh H_N)^{1/2}\frac{1}{|z - (\kappa
H_N/N)|^2} (H_N - \wh H_N)^{1/2} \Big\| \, \\ & \hspace{3cm} \times
\left\langle \psi_{N,*} ,
 \frac{1}{\bar{z} - (\kappa \wh H_N/N)} (H_N - \wh H_N)
 \frac{1}{z - (\kappa \wh H_N/N)}\psi_{N,*} \right\rangle\,.
\end{split}
\end{equation}
Moreover (since $\| BA^2B\| = \| AB^2A\| \leq \| AC^2A\|$ for
positive operators $A,B,C$ with $B^2\leq C^2$),
\begin{equation}\label{eq:norm3}
\begin{split}
\Big\| (H_N - \wh H_N)^{1/2}\frac{1}{|z - (\kappa H_N/N)|^2} (H_N -
\wh H_N)^{1/2} \Big\| & = \Big\| \frac{1}{|z - (\kappa H_N/N)|} (H_N
- \wh H_N) \frac{1}{|z - (\kappa H_N/N)|} \Big\|
 \\ &\leq \Big\| \frac{1}{|z - (\kappa H_N/N)|} H_N \frac{1}{|z - (\kappa H_N/N)|} \Big\|
 \\ &\leq \frac{C N}{ |y|^2 \kappa}
\end{split}
\end{equation}
for $z$ in the support of $\wt \chi$, where we used the spectral
theorem in the last step. On the other hand, the second factor on
the r.h.s. of (\ref{eq:norm2}) can be bounded by
\begin{equation}
\begin{split}
\Big\langle \psi_{N,*}, &\frac{1}{\bar{z} - (\kappa \wh H_N/N)} (H_N
- \wh H_N) \frac{1}{z - (\kappa \wh H_N/N)}\psi_{N,*} \Big\rangle \\
&\leq k \Big\langle \psi_{N,*}, \frac{1}{\bar{z} - (\kappa \wh
H_N/N)} \left( -\Delta_1 + k V_N (x_1 -x_2) + N V_N (x_1 -x_{k+1})
\right) \frac{1}{z - (\kappa \wh H_N/N)} \, \psi_{N,*} \Big\rangle
\,. \nonumber
\end{split}
\end{equation}
 Here we used the
fact that $\psi_{N,*}$ is symmetric w.r.t. permutations of the first
$k$ and the last $N-k$ variables, and that the operator $\wh H_N$
preserves this property. Since $N V_N (x_1 -x_{k+1}) \leq C \| V
\|_{L^1} (1-\Delta_1)^2$, and $k V_N (x_1 -x_2) \leq C \| V \|_{L^1}
(1-\Delta_1)^2$ (see (\ref{eq:sobolev2})) we find
\begin{equation}\label{eq:norm5}
\begin{split}
\Big\langle \psi_{N,*}, &\frac{1}{\bar{z} - (\kappa \wh H_N/N)} (H_N
- \wh H_N) \frac{1}{z - (\kappa \wh H_N/N)}\psi_{N,*} \Big\rangle \\
&\leq k \Big\langle \psi_{N,*} \frac{1}{\bar{z} - (\kappa H_N/N)}
\left( -\Delta_1 + (1-\Delta_1)^2 \right)\frac{1}{z - (\kappa
H_N/N)} \psi_{N,*} \Big\rangle \leq C\,  k \, |y|^{-2} \| \ph_*
\|^2_{H_2}
\end{split}
\end{equation}
because $\Delta_1$ commutes with $\wh H_N$ (recall that $\psi_{N,*}=
\ph_*^{\otimes k} \otimes \xi_N^{(N-k)}$). {F}rom
(\ref{eq:Helffer}), (\ref{eq:norm2}), (\ref{eq:norm3}) and
(\ref{eq:norm5}) we find that
\[ \| (\Xi - \wh \Xi ) \psi_{N,*} \| \leq  C_{k,\e} N^{-1/2}\]
for a constant  $C_{k,\e}$ depending on $k$ and $\eps$ (through the
norm $\| \ph^* \|_{H^2}$) but independent of $\kappa$, for $\kappa$
small enough. This implies that
\begin{equation}\label{eq:Xi-hatXi}
\begin{split}
\Big\| \frac{\Xi \left( \ph_*^{\otimes k} \otimes \xi_N^{(N-k)}
\right)}{\| \Xi \left( \ph_*^{\otimes k} \otimes \xi_N^{(N-k)}
\right) \|} - \frac{\wh\Xi \left( \ph_*^{\otimes k} \otimes
\xi_N^{(N-k)} \right)}{\|\wh\Xi \left( \ph_*^{\otimes k} \otimes
\xi_N^{(N-k)} \right)\|} \Big\| &\leq \frac{2}{\|\Xi \left(
\ph_*^{\otimes k} \otimes \xi_N^{(N-k)} \right) \|} \, \|(\Xi - \wh
\Xi ) \psi_{N,*} \|  \\ &\leq 4 \, \| (\Xi - \wh \Xi ) \psi_{N,*} \|
\leq \frac{\e}{6\|J^{(k)}\|}
\end{split}
\end{equation}
for $N$ large enough (and assuming that $\e >0$ and $\kappa >0$ are
small enough, independently of $N$). Here we used that (by
(\ref{eq:fact}), (\ref{eq:eps1}), and (\ref{eq:ph-ph*}))
\begin{equation}
\begin{split}
\|\Xi \left( \ph_*^{\otimes k} \otimes \xi_N^{(N-k)} \right) \| \geq
\; &\| \psi_N \| - \| \Xi \psi_N - \psi_N \| - \| \Xi \left(\psi_N
-\ph^{\otimes k} \otimes \xi_N^{(N-k)} \right)\| \\ &- \| \Xi
\left(\ph^{\otimes k} \otimes \xi_N^{(N-k)} - \ph_*^{\otimes k}
\otimes \xi_N^{(N-k)} \right)\| \\ \geq \; &1 -C \kappa^{1/2} - o(1)
- \frac{\e}{32 \| J^{(k)} \|} \geq 1/2
\end{split}
\end{equation}
for $\kappa,\eps$ small enough and for $N$ large enough. {F}rom
(\ref{eq:Xi-hatXi}) and (\ref{eq:Xipsi-ph}) we obtain
(\ref{eq:hatXipsi-ph}). This completes the proof of part iii).
\end{proof}

\section{Proof of Proposition \ref{prop:Hk}}\label{sec:proofHk}
\setcounter{equation}{0}

This section is devoted to the proof the Proposition \ref{prop:Hk}.
Let us recall the definition of the cutoff functions
\[ \Theta^{(n)}_k=\Theta^{(n)}_k (\bx) =\exp \left( -\frac{2^n}{\ell^{\eps}}
\sum_{i\leq k}\sum_{j \neq i} h (x_i -x_j) \right) \, \] from
(\ref{eq:thetan}) with the function $h$ defined in \eqref{eq:hdef}.
We introduce the notation $h_{ij} = h (x_i -x_j)$ and we also adopt
the convention that $h_{ii} =0$ for any $i \in \bN$. Moreover we
recall that $D_k := \nabla_1 \dots \nabla_k \,.$

\begin{proof}[Proof of Proposition \ref{prop:Hk}]
We prove (\ref{eq:Hk}) by induction over $k$. For $k=1$ we clearly
have
\begin{equation}\label{eq:k=1}
\langle \psi, (H_N +N) \psi \rangle \geq N \int |\nabla_1 \, \psi|^2
+ \frac{N(N-1)}{2} \int V_{N} (x_1 -x_2) |\psi|^2.
\end{equation}
For $k=2$ we have, from (\ref{eq:H2-1}), (\ref{eq:H2-2}) (but
keeping  the term on the sixth line, which was neglected, because of
its positivity, in the last inequality in (\ref{eq:H2-2})),
(\ref{eq:H2-3}), and (\ref{eq:H2-3b}) we find,  for $\rho$ small
enough (recall the definition of $\rho$ in (\ref{eq:defrho})),
\begin{equation}\label{eq:Hk-k=2}
\begin{split}
\langle \psi, (H_N+N)^2 \psi \rangle \geq \; & \langle \psi, H_N^2
\psi \rangle \\ \geq \; &N (N-1) \langle \psi, \fh_1 \fh_2 \psi
\rangle + N \langle \psi, \fh_1^2 \psi \rangle \\ \geq \;& N(N-1)
(1-c\rho) \int (1-w_{12})^2 |\nabla_1 \nabla_2 \phi_{12}|^2 \\ &+
\frac{N(N-1)(N-2)}{2} \int (1-w_{12})^2 \, V_N (x_2-x_3) |\nabla_1
\phi_{12}|^2 + N \int |\fh_1 \psi|^2
\end{split}
\end{equation}
where $\fh_i$, for $i=1, \dots,N$, was defined in (\ref{eq:frakh}).
{F}rom the last term we get
\begin{equation}
\int |\fh_1 \psi|^2 \geq \int \theta_1^{(2)} |\fh_1 \psi|^2 \geq
\int \theta_1^{(2)} \Delta_1 \overline{\psi} \Delta_1 \psi +
\frac{1}{2} \sum_{j \geq 2} \int \theta_1^{(2)} \left( \Delta_1
\overline{\psi} \; V_N(x_1 -x_j) \psi + \text{h.c.} \right)
\end{equation}
where h.c. denotes the hermitian conjugate. The last term is
exponentially small in $N$ because on the support
of the potential $V_N (x_1 -x_j)$ the point
 $x_1$ is close to $x_j$ (on the length scale $N^{-1}$) and
this makes the factor $\theta_1^{(2)}$ exponentially small. Hence we
find  (with the notation
 $\nabla_1^j:=\partial_{x_1^{(j)}}$ where $x_1= (x_1^{(1)},
x_1^{(2)},x_1^{(3)})\in\bR^3$),
\begin{equation}\label{eq:H1^2}
\begin{split}
\int \theta_1^{(2)} |\fh_1 \psi|^2 \geq \; &\int  \theta_1^{(2)}
|\nabla_1^2 \psi|^2 + \int \sum_{i,j=1}^3 \left\{
(\nabla_1^i \theta_1^{(2)})
(\nabla_1^j \overline{\psi})
 \nabla_1^i\nabla_1^j \psi + \text{h.c.} \right\} +
\int \sum_{i,j=1}^3 \nabla_1^i\nabla_1^j \theta_1^{(2)}
\nabla_1^i \overline{\psi}\nabla_1^j\psi\, \\ &- o(N)
\left\{ \int \theta_1^{(1)} |\nabla_1 \psi|^2 + \int |\psi|^2
\right\}
\end{split}
\end{equation}
by using $|\nabla_1\theta_1^{(2)}|\leq C\ell^{-1}\theta_1^{(1)}$
from  Lemma \ref{lm:theta}, part iii).
{F}rom part ii) and iv) of the same lemma we also have
\begin{equation} \begin{split}
\frac{\left|\nabla_1 \theta_1^{(2)} \right|^2}{\theta_1^{(2)}} &\leq
C \ell^{-2} \theta_1^{(1)} \quad \text{and } \quad  \left|
\nabla^2_1 \theta_1^{(2)} \right| \leq C \ell^{-2} \theta_1^{(1)}\; ,
\end{split}
\end{equation}
and therefore we obtain
\begin{equation}
\begin{split}
\sum_{i,j=1}^3 \left|  \int
(\nabla_1^i \theta_1^{(2)})
(\nabla_1^j \overline{\psi})
 \nabla_1^i\nabla_1^j \psi \right|
\leq \a \int \theta_1^{(2)} |\nabla_1^2
\psi|^2 + \a^{-1} \int \frac{|\nabla_1
\theta_1^{(2)}|^2}{\theta_1^{(2)}} |\nabla_1 \psi|^2
\\ \leq o(1) \int \theta_1^{(2)} |\nabla_1^2 \psi|^2 + o(N) \int
\theta_1^{(1)} |\nabla_1 \psi|^2\; ,
\end{split}
\end{equation}
where we used that $N\ell^2 \gg 1$ (and an appropriate choice of the
parameter $\a$). Analogously
\begin{equation}\label{eq:analog}
\sum_{i,j=1}^3 \int \left| \nabla_1^i\nabla_1^j \theta_1^{(2)}
\nabla_1^i \overline{\psi}\nabla_1^j\psi\right|
 \leq o(N) \int
\theta_1^{(1)} |\nabla_1 \psi|^2\,.
\end{equation}
{F}rom (\ref{eq:Hk-k=2})--\eqref{eq:analog},
we find
\begin{equation}
\begin{split}
\langle \psi, (H_N+N)^2 \psi \rangle \geq  \;& N^2 (1-c\rho-o(1))
\int (1-w_{12})^2  |\nabla_1 \nabla_2 \phi_{12}|^2 \\
&+ \frac{N^3}{2} (1-o(1)) \int (1-w_{12})^2  \, V_N (x_2-x_3)
|\nabla_1 \phi_{12}|^2 \\ &+ N (1-o(1)) \int \theta_1^{(2)}
|\nabla_1^2 \psi|^2 - o(N^2) \left\{ \int \theta_1^{(1)} |\nabla_1
\psi|^2 + \int \theta_1^{(1)} |\psi|^2 \right\} \, .
\end{split}
\end{equation}

Next we apply Lemma \ref{lm:phitopsi} (with $n=0$)
to replace, in
the first and second term on the r.h.s. of the last equation,
$\phi_{12}$ by $\psi$. We find
\begin{equation}\label{eq:k=2}
\begin{split}
\langle \psi, (H_N+N)^2 \psi \rangle \geq  \;& N^2 (1-c\rho-o(1))
\int \theta_1^{(2)} |\nabla_1 \nabla_2 \psi|^2
\\
&+ \frac{N^3}{2} (1-o(1)) \int \, \theta_1^{(2)} \, V_N (x_2-x_3)
|\nabla_1 \psi|^2+ N (1-o(1)) \int \theta_1^{(2)} |\nabla_1^2 \psi|^2
\\ & - o(N^2) \int \left\{
\theta_1^{(1)} |\nabla_1 \psi|^2 +  \theta_1^{(1)} |\psi|^2
+  NV_N(x_1-x_2)|\psi|^2
\right\} \, .
\end{split}
\end{equation}
By (\ref{eq:k=1}) we have
\begin{equation}
\begin{split}
o(N^2) \int  \left\{  \theta_1^{(1)} |\nabla_1 \psi|^2  +
\theta^{(1)}_1 |\psi|^2  +
 NV_N(x_1-x_2)|\psi|^2 \right\} &\leq o(N) \langle \psi, (H_N+N)\psi
\rangle \\ &\leq o(1) \langle \psi, (H_N+N)^2 \psi \rangle\,.
\end{split}
\end{equation} Hence, from (\ref{eq:k=2}), we obtain
\begin{equation}
\begin{split}
(1+o(1)) \langle \psi, (H_N+N)^2 \psi \rangle \geq \;& N^2
(1-c\rho-o(1))
\int \theta_1^{(2)} |\nabla_1 \nabla_2 \psi|^2 \\
&+ \frac{N^3}{2} (1-o(1)) \int \, \theta_1^{(2)} \, V_N (x_2-x_3)
|\nabla_1 \psi|^2
\\ &+ N (1-o(1)) \int \theta_1^{(2)} |\nabla_1^2 \psi|^2\,.
\end{split}
\end{equation}
It follows that, for $\rho$  small enough, there exists $C_0>0$
such that we have
\begin{equation}
\begin{split}
\langle \psi, (H_N+N)^2 \psi \rangle \geq & \;C_0^2 N^2 \int
\theta_1^{(2)} |\nabla_1 \nabla_2 \psi|^2 + C_0^2 N^3 \int
\theta_1^{(2)} \, V_N (x_2-x_3) |\nabla_1 \psi|^2 \\ &+ C_0^2 N \int
\theta_1^{(2)} |\nabla_1^2 \psi|^2
\end{split}
\end{equation}
if $N$ is large enough.

\medskip

We assume now that (\ref{eq:Hk}) is correct for all $k\leq n+1$ and
we prove if for $k=n+2$, assuming $n\geq 1$. To this end we note
that, for $N \geq N_0 (n)$, using the induction hypothesis we have
\begin{equation}
\begin{split}
\langle \psi, (H_N +N)^{n+2} \psi \rangle \geq \; & \langle H_N
\psi,
(H_N+N)^n H_N \psi \rangle \\
\geq \; & C_0^n N^n \, \int \Theta^{(n)}_{n-1} \; |D_n H_N \psi|^2
\\ \geq \; & C_0^n N^n
\, \int \Theta^{(n+2)}_n \; |D_n H_N \psi|^2
\end{split}
\end{equation}
where we used that $1 \geq \theta^{(n)}_i \geq \theta^{(n+2)}_i$ for
every $i=1,\dots,n$. We write $H_N = \sum_{j=1}^N \fh^{(n)}_j$, with
\begin{equation}\label{eq:frakhn}
\fh_j^{(n)} = \left\{
\begin{array}{ll}
-\Delta_j + \frac{1}{2} \sum_{i > n, i\neq j} V_N (x_i
-x_j) \qquad &\text{if } j > n \\
-\Delta_j + \frac{1}{2} \sum_{i \leq n} V_N (x_i -x_j) + \sum_{i >
n} V_N (x_i -x_j) \qquad &\text{if } j \leq n\; .
\end{array}\right. \;
\end{equation}
Then we have
\begin{equation}
\begin{split}
\langle \psi, (H_N+N)^{n+2} \psi \rangle \geq \; & C_0^n N^n \,
\sum_{i,j > n} \int \Theta^{(n+2)}_n \; D_n \fh^{(n)}_i \,
\overline{\psi} \, D_n \fh^{(n)}_j \psi
\\ &+ C_0^n N^n \, \left\{ \sum_{i\leq n < j} \int \Theta^{(n+2)}_n
\; D_n \fh^{(n)}_i \, \overline{\psi} \, D_n \fh^{(n)}_j \psi +
\text{h.c.} \right\} \\ &+ C_0^n N^n \, \sum_{i,j \leq n} \int
\Theta^{(n+2)}_n \; D_n \fh^{(n)}_i \, \overline{\psi} \, D_n
\fh^{(n)}_j \psi \, .
\end{split}
\end{equation}

The last term on the r.h.s. (where $i,j \leq n$) is positive and
therefore it can be neglected. In the first term on the r.h.s. we
can neglect all terms where $i=j$ (because they are all positive).
Therefore we obtain
\begin{equation}\label{eq:Hk-1}
\begin{split}
\langle \psi, (H_N+N)^{n+2} \psi \rangle \geq \; & C_0^n N^n \,
\sum_{i,j > n, i\neq j} \int \Theta^{(n+2)}_n \; D_n \, \fh^{(n)}_i
\, \overline{\psi} \; D_n \, \fh^{(n)}_j \, \psi \\ &+ C_0^n N^n \,
\sum_{i\leq n < j} \int \Theta^{(n+2)}_n \; \left\{ D_n \fh^{(n)}_i
\, \overline{\psi} \; D_n \fh^{(n)}_j \, \psi + \text{h.c.} \right\}
\,.
\end{split}
\end{equation}
In Proposition \ref{lm:Hk-1} below we give a lower bound for the
first term in \eqref{eq:Hk-1}, while Proposition \ref{lm:Hk-2}
estimates the second term. Combining these two estimates, we find
that,  for $\rho$ small enough (independently of $N$ and $n$)
and for $N$ large enough,
\begin{equation}\label{eq:negg}
\begin{split}
\langle \psi, (H_N+N)^{n+2} \psi \rangle \geq \; &C_0^n N^{n+2} (1-
c\rho-o(1)) \int \; \Theta^{(n+2)}_{n+1} \; |D_{n+2} \psi|^2
\\ &+ C_0^n N^{n+1} (1-o(1)) \int
\Theta^{(n+2)}_{n+1} \; |\nabla_1 D_{n+1} \,\psi|^2  \\ &+
\frac{C_0^n N^{n+3}}{2} (1-o(1)) \, \int \Theta^{(n+2)}_{n+1} \,
V_N (x_{n+2} -x_{n+3}) \; |D_{n+1} \psi|^2  \\
&-\Omega_n (\psi) \end{split}
\end{equation}
where the error $\Omega_n (\psi)$ is given by
\begin{equation}\label{eq:error}
\begin{split}
\Omega_n (\psi) = \; &o(N^{n+3}) \int \Theta^{(n+1)}_{n} \, V_N
(x_{n+1} -x_{n+2})\, |D_n \, \psi|^2 \\ &+ o(N^{n+1}) \left\{ \int
\Theta^{(n+1)}_{n} |\nabla_1 D_n \psi|^2 + \int \Theta^{(n)}_{n-1}
|\nabla_1 D_{n-1} \psi|^2 \right\}
\\ &+ o(N^{n+2}) \left\{  \int \Theta^{(n+1)}_{n}
|D_{n+1} \psi|^2 + \int \Theta^{(n)}_{n-1} |D_n \psi|^2
\right. \\
&\hspace{2cm} \left. + \int \Theta^{(n-1)}_{n-2} |D_{n-1} \psi|^2
+\int \Theta^{(n-2)}_{n-3} |D_{n-2} \psi|^2\right\}.
\end{split}
\end{equation}
Now we use the induction hypothesis, Eq. (\ref{eq:Hk}), with
$k=n-1,n,n+1$ to bound the negative contributions. For example,
(\ref{eq:Hk}) with $k=n+1$ implies that
\begin{equation*}
\begin{split}
o(N^{n+3}) \int \Theta^{(n+1)}_{n} \, V_N (x_{n+1} -x_{n+2})\, |D_n
\, \psi|^2 &\leq o(N) \langle \psi, (H_N + N)^{n+1} \psi \rangle
\leq o(1) \langle \psi, (H_N+N)^{n+2} \psi \rangle \end{split}
\end{equation*} because $H_N \geq 0$.  The other terms
in  \eqref{eq:error} are treated similarly. It follows that
\begin{equation*}
\begin{split}
(1 +o(1)) \; \langle \psi,\; (H_N+N)^{n+2} \psi \rangle  \geq \;
&C_0^n N^{n+2} (1- c\rho-o(1)) \int \; \Theta^{(n+2)}_{n+1} \;
|D_{n+2} \psi|^2
\\ &+ C_0^n N^{n+1} (1-o(1)) \int
\Theta^{(n+2)}_{n+1} \; |\nabla_1 D_{n+1} \,\psi|^2  \\ &+
\frac{C_0^n N^{n+3}}{2} (1-o(1)) \, \int \Theta^{(n+2)}_{n+1} \, V_N
(x_{n+2} -x_{n+3}) \; |D_{n+1} \psi|^2 \, .
\end{split}
\end{equation*}
Thus, if $\rho$ and $C_0$ are small enough (independently of $n$),
we can find $N_0 (n+2,C_0) > N_0 (n,C_0)$ such that
\begin{equation}
\begin{split}
\langle \psi,\; (H_N+N)^{n+2} \psi \rangle \geq \;& C_0^{n+2}
N^{n+2} \int \; \Theta^{(n+2)}_{n+1} \; |D_{n+2} \psi|^2 + C_0^{n+2}
N^{n+1} \int \Theta^{(n+2)}_{n+1} \; |\nabla_1 D_{n+1} \,\psi|^2  \\
&+ C_0^{n+2} N^{n+3} \, \int \Theta^{(n+2)}_{n+1} \, V_N (x_{n+2}
-x_{n+3}) \; |D_{n+1} \psi|^2.
\end{split}
\end{equation}
\end{proof}

\medskip

 In the rest of this section we will state and prove
Propositions  \ref{lm:Hk-1} and  \ref{lm:Hk-2} used in
\eqref{eq:Hk-1}. Both proofs will be divided into several Lemmas.

\medskip

Similarly to the $H_N^2$-energy estimate from Proposition
\ref{prop:H2}, the key idea in Proposition \ref{lm:Hk-1} is that
$\fh^{(n)}_i\psi$ can be conveniently estimated by the derivatives
of $\phi_{ij}$, where $\phi_{ij}$ is given by the relation $\psi=
(1-w_{ij})\phi_{ij}$. The estimates of  all errors are done in terms
of $\phi_{ij}$ and its derivatives. Finally, Lemma \ref{lm:phitopsi}
will show how to go back from the estimates on $\phi_{ij}$ to
estimates involving $\psi$ with a cutoff supported on a bigger set.

\begin{proposition}\label{lm:Hk-1}
Suppose $\rho$ is small enough and $\ell \gg N^{-1/2}$.
 For $i=1,
\dots,N$, let $\fh^{(n)}_i$ be defined as in (\ref{eq:frakhn}). Then
\begin{equation}\label{eq:lmHk-1}
\begin{split}
C_0^n N^n \, \sum_{i,j > n, i\neq j} &\int \Theta^{(n+2)}_n \; D_n
\, \fh^{(n)}_i \, \overline{\psi} \; D_n \, \fh^{(n)}_j \, \psi
\\ \geq \; &C_0^n N^{n+2} (1- c\rho-o(1)) \int \;
\Theta^{(n+2)}_{n+1} \; |D_{n+2} \psi|^2
\\ &+ \frac{C_0^n N^{n+3}}{2} (1-o(1)) \,
\int \Theta^{(n+2)}_{n+1} \, V_N (x_{n+2} -x_{n+3}) \; |D_{n+1}
\psi|^2 -\Omega_n (\psi)
\end{split}
\end{equation}
where the error term $\Omega_n (\psi)$ has been defined in
(\ref{eq:error}).
\end{proposition}

\begin{proof}
For any $i \neq j$, $i,j
>n$, we write $\psi = (1- w_{ij}) \phi_{ij}$. Then we have,
 similarly to \eqref{eq:h1},
\begin{equation} \begin{split}
(1-w_{ij})^{-1} \fh^{(n)}_i \big[(1-w_{ij}) \phi_{ij}\big] &=
-\Delta_i \phi_{ij} + 2 \frac{ \nabla w_{ij}}{1-w_{ij}} \nabla_i
\phi_{ij} + \frac{1}{2} \sum_{m >n , \, m \neq i,j} V_N (x_i - x_m)
\phi_{ij}
\\ &= L_i \phi_{ij} + \frac{1}{2} \sum_{m >n, \, m \neq i,j} V_N (x_i - x_m)
\phi_{ij} \end{split}
\end{equation}
where the differential operator $L_i := -\Delta_i + 2 \frac{\nabla
w_{ij}}{1-w_{ij}} \nabla_i$ is such that
\be\label{eq:symmm}
 \int (1-w_{ij})^2 (L_i \overline{\phi})
 \chi = \int (1-w_{ij})^2 \overline{\phi} (L_i
\chi) = \int (1-w_{ij})^2 \nabla_i \overline{\phi} \nabla_i \chi.
\ee
Note that the operator $L_i$ also depends on the choice of the index
$j$. Analogously, we have
\[
(1-w_{ij})^{-1} \fh^{(n)}_j \big[ (1-w_{ij})\phi_{ij}\big]
 = L_j \phi_{ij} + \frac{1}{2}
\sum_{m>n, \, m \neq i,j} V_N (x_j - x_m) \phi_{ij} \] with $L_j
=-\Delta_j + 2 \frac{\nabla w_{ji}}{1-w_{ij}} \nabla_j$. Note that
$D_n$ commutes with $L_i$, $L_j$ and $1-w_{ij}$ if $i,j>n$. The
l.h.s of (\ref{eq:lmHk-1}) is thus given by
\begin{equation*}
\begin{split}
C_0^n N^n \, &\sum_{i,j > n, i\neq j} \; \int (1-w_{ij})^2 \;
\Theta^{(n+2)}_n \; D_n \, \left[ \left( L_i + \frac{1}{2} \sum_{m
>n, m \neq i,j} V_N (x_m -x_i) \right) \, \overline{\phi}_{ij}\right]
\\ & \hspace{6cm} \times D_n \, \left[ \left( L_j + \frac{1}{2}
 \sum_{r >n, r \neq i,j} V_N (x_j -x_r) \right) \, \phi_{ij}\right] \\
\geq \; & C_0^n N^n \, \sum_{i,j > n, i\neq j} \; \int (1-w_{ij})^2
\; \Theta^{(n+2)}_n \; L_i D_n \,\overline{\phi}_{ij} \; L_j D_n \,
\phi_{ij}
\\ &+ \frac{C_0^n N^n}{2} \, \sum_{i,j > n, i\neq j} \; \sum_{r>n, \, r \neq i,j} \; \int
(1-w_{ij})^2 \; \Theta^{(n+2)}_n \; V_N (x_j -x_r) \; L_i D_n \,
\overline{\phi}_{ij} \; D_n  \phi_{ij} + \text{h.c.}\, ,
\end{split}
\end{equation*}
because of the positivity of the potential. Proposition
\ref{lm:Hk-1} now follows from Lemma \ref{lm:Hk-1-1} and Lemma
\ref{lm:Hk-1-2}, where we consider separately the two terms on the
r.h.s. of the last equation.
\end{proof}

\begin{lemma}\label{lm:Hk-1-1}
Suppose the assumptions of Lemma \ref{lm:Hk-1} are satisfied. Then
we have
\begin{equation}
\begin{split}
C_0^n N^n \, \sum_{i,j > n, i\neq j} \; \int (1-&w_{ij})^2 \;
\Theta^{(n+2)}_n \; L_i \, D_n \,\overline{\phi}_{ij} \; L_j \, D_n
\, \phi_{ij}
\\ \geq & \; C_0^n N^{n+2} \, (1- c\rho -o(1)) \int
\Theta^{(n+2)}_{n+1} \; |D_{n+2} \psi|^2 \\ &-o(N^{n+2}) \left( \int
\Theta^{(n+1)}_{n} \; |D_{n+1} \, \psi|^2 + \int \Theta_{n-1}^{(n)}
\; |D_n \psi|^2 \right)\,.
\end{split}
\end{equation}
\end{lemma}

\begin{proof}
 By the symmetry \eqref{eq:symmm} we have
\begin{equation}\label{eq:Hk-3}
\begin{split}
C_0^n N^n \, &\sum_{i,j > n, i\neq j} \; \int (1-w_{ij})^2 \;
\Theta^{(n+2)}_n \; L_i D_n \,\overline{\phi}_{ij} \; L_j D_n \,
\phi_{ij}
\\ = & \; C_0^n N^n \, \sum_{i,j > n, i\neq j} \; \int (1-w_{ij})^2
\; \left\{ \Theta^{(n+2)}_n \; \nabla_i D_n \,\overline{\phi}_{ij}
\; \nabla_i L_j D_n \, \phi_{ij} + \nabla_i \Theta^{(n+2)}_n \;
\nabla_i D_n \,\overline{\phi}_{ij} \; L_j D_n \, \phi_{ij} \right\}
\\ = & \; C_0^n N^n \, \sum_{i,j > n, i\neq j} \; \int
(1-w_{ij})^2 \left\{ \Theta^{(n+2)}_n \; |\nabla_j \nabla_i D_n
\,\phi_{ij}|^2 + \nabla_j \Theta^{(n+2)}_n  \; \nabla_i D_n
\,\overline{\phi}_{ij} \; \nabla_j \nabla_i D_n \, \phi_{ij} \right.
\\&\hspace{3cm}+ \nabla_i \Theta^{(n+2)}_n  \; \nabla_j \nabla_i D_n \,
\overline{\phi}_{ij} \; \nabla_j D_n \, \phi_{ij} + \nabla_i
\nabla_j \Theta^{(n+2)}_n \; \nabla_i D_n \,\overline{\phi}_{ij} \;
\nabla_j D_n \, \phi_{ij}
\\ &\hspace{3cm}\left.
+ \; \Theta^{(n+2)}_n \; \nabla_i D_n \,\overline{\phi}_{ij} \;
[\nabla_i, L_j ] D_n \, \phi_{ij} \right\}
\end{split}
\end{equation}
To bound the second and third term on the r.h.s. of (\ref{eq:Hk-3}),
we note that, by part iii) of Lemma \ref{lm:theta},
\begin{equation}\label{eq:nablatheta}
\left| \nabla_j \Theta^{(n+2)}_n  \right| \leq C \ell^{-1} \left(
\frac{2^{n+2}}{\ell^{\eps}} \sum_{m=1}^n h_{mj} \right)
\Theta^{(n+2)}_n \, .
\end{equation}
Therefore the second term on the r.h.s. of (\ref{eq:Hk-3}) can be
bounded by
\begin{equation}\label{eq:Hk-4}
\begin{split}
\sum_{i,j > n, i\neq j} &\left| \int (1-w_{ij})^2 \; \nabla_j
\Theta^{(n+2)}_n  \; \nabla_i D_n \,\overline{\phi}_{ij} \; \nabla_j
\nabla_i D_n \, \phi_{ij}\right| \\ \leq \; & \a \sum_{i,j >n, i
\neq j} \int (1-w_{ij})^2 \; \Theta^{(n+2)}_n |\nabla_j \nabla_i D_n
\, \phi_{ij}|^2 \\ &+ C \ell^{-2} \a^{-1} \sum_{i,j >n, i \neq j}
\int (1-w_{ij})^2 \; \left( \frac{2^{n+2}}{\ell^{\eps}} \sum_{m=1}^n
h_{mj} \right)^2 \Theta^{(n+2)}_n \; |\nabla_i D_n \, \phi_{ij}|^2
\end{split}
\end{equation}
for some $\a >0$. Next we use that $\phi_{ij} = \psi
(1-w_{ij})^{-1}$. Since $i,j >n$, we have \[ \nabla_i D_n \left(
\psi (1-w_{ij})^{-1}\right) = \nabla_i w_{ij} (1-w_{ij})^{-2} D_n
\psi + (1-w_{ij})^{-1} \nabla_i D_n \psi \] and thus
\begin{equation}\label{eq:nablaphi}
\begin{split}
(1-w_{ij})^2 \, |\nabla_i D_n \phi_{ij}|^2 &\leq 2\left( \frac{\nabla
w_{ij}}{1-w_{ij}}\right)^2 |D_n \psi|^2 + 2|\nabla_i D_n \psi|^2 \leq
 \frac{C}{|x_i -x_j|^2} |D_n \psi|^2 +2 |\nabla_i D_n \psi|^2\,.
\end{split}
\end{equation}
Therefore the second term on the r.h.s. of (\ref{eq:Hk-4}) is
bounded by
\begin{equation}
\begin{split}
\sum_{i,j > n, i \neq j} \int (1-&w_{ij})^2 \; \left(
\frac{2^{n+2}}{\ell^{\eps}} \sum_{m=1}^n h_{mj} \right)^2
\Theta^{(n+2)}_n \; |\nabla_i D_n \, \phi_{ij}|^2 \\  \leq \;
&C\sum_{i,j>n, i\neq j} \int \left( \frac{2^{n+2}}{\ell^{\eps}}
\sum_{m=1}^n h_{mj} \right)^2 \Theta^{(n+2)}_n \left\{ |\nabla_i D_n
\psi|^2 + \frac{1}{|x_i - x_j|^2} |D_n \psi|^2 \right\}
 \\ \leq \; & C\sum_{i,j>n, i\neq j} \int
\left( \frac{2^{n+2}}{\ell^{\eps}} \sum_{m=1}^n h_{mj} \right)^2
\left\{ \Theta^{(n+2)}_n \; |\nabla_i D_n \psi|^2 + \left|\nabla_i
\left( \Theta^{(n+2)}_n \right)^{\frac{1}{2}} \right|^2 \; |D_n
\psi|^2\right\}
\end{split}
\end{equation}
where we used Hardy inequality and the fact that $i \neq j$
and $i>n$. Using a
bound similar to (\ref{eq:nablatheta}), and part ii) of Lemma
\ref{lm:theta}, we can continue this estimate
\begin{equation}\label{eq:Hk-5}
\begin{split}
\sum_{i,j > n, i \neq j} \int &(1-w_{ij})^2 \; \left(
\frac{2^{n+2}}{\ell^{\eps}} \sum_{m=1}^n h_{mj} \right)^2
\Theta^{(n+2)}_n \; |\nabla_i D_n \, \phi_{ij}|^2 \\ \leq \; &
C\sum_{i,j>n, i\neq j} \int \left( \frac{2^{n+2}}{\ell^{\eps}}
\sum_{m=1}^n h_{mj} \right)^2 \Theta^{(n+2)}_n \left\{  |\nabla_i
D_n \psi|^2 + \ell^{-2} \left( \frac{2^{n+1}}{\ell^{\eps}}
\sum_{m=1}^n h_{mi} \right)^2 \; |D_n \psi|^2 \right\}
\\ \leq \; & C\sum_{i>n} \int \left( \frac{2^{n+2}}{\ell^{\eps}}
\sum_{j >n} \sum_{m=1}^n h_{mj} \right)^2 \Theta^{(n+2)}_n \;
|\nabla_i D_n \psi|^2 \\ &+ C\ell^{-2} \int \left(
\frac{2^{n+2}}{\ell^{\eps}} \sum_{j >n} \sum_{m=1}^n h_{mj}
\right)^2 \left( \frac{2^{n+1}}{\ell^{\eps}} \sum_{i >n}
\sum_{m=1}^n h_{mi} \right)^2 \Theta_n^{(n+2)} \; |D_n \psi|^2
\\ \leq \; & C\sum_{i>n} \int \Theta^{(n+1)}_{n} \; |\nabla_i
D_n \psi|^2 + C \ell^{-2} \int \Theta_n^{(n)} \; |D_n \psi|^2
\\ \leq \; &C N \int \Theta^{(n+1)}_{n} \; |
D_{n+1} \psi|^2 + C \ell^{-2} \int \Theta_n^{(n)} \; |D_n \psi|^2 \,
,
\end{split}
\end{equation}
because of the permutation symmetry of $\psi$. {F}rom
(\ref{eq:Hk-4}) we find
\begin{equation}\label{eq:Hk-6}
\begin{split}
\sum_{i,j > n, i\neq j} &\left| \int (1-w_{ij})^2 \; \nabla_j
\Theta^{(n+2)}_n \; \nabla_i D_n \,\overline{\phi}_{ij} \;
\nabla_j \nabla_i D_n \, \phi_{ij}\right| \\ \leq \; &\a N^2 \int
(1-w_{n+1,n+2})^2 \; \Theta^{(n+2)}_n |D_{n+2} \, \phi_{n+1,n+2}|^2
\\ &+\a^{-1} C \ell^{-2} N \int \Theta^{(n+1)}_n \; | D_{n+1} \psi|^2 +
\a^{-1} C \ell^{-4} \int \Theta_n^{(n)} \; |D_n
\psi|^2 \, \\
\leq \; & o(N^2) \left( \int (1-w_{n+1,n+2})^2 \; \Theta^{(n+2)}_n
|D_{n+2} \, \phi_{n+1,n+2}|^2 +  \int \Theta^{(n+1)}_n \; | D_{n+1}
\psi|^2 + \int \Theta_{n-1}^{(n)} \; |D_n \psi|^2 \right)
\end{split}
\end{equation}
for an appropriate choice of $\a$ (using that $N\ell^2 \gg 1$). In
the last term we also used that $\theta_n^{(n)} \leq 1$.

\medskip

The third term on the r.h.s. of (\ref{eq:Hk-3}), being the hermitian
conjugate of the second term can be bounded exactly in the same way.

\medskip

Now we consider the fourth term on the r.h.s. of (\ref{eq:Hk-3}). To
this end we use that, since $i \neq j$, and $i,j
>n$, we have, by Lemma \ref{lm:theta}, part v),
\begin{equation}
\left| \nabla_i \nabla_j \; \Theta_n^{(n+2)} \right| \leq C
\ell^{-2} \left( \frac{2^{n+2}}{\ell^{\eps}} \sum_{m=1}^n h_{mj}
\right) \left( \frac{2^{n+2}}{\ell^{\eps}} \sum_{m=1}^n h_{mi}
\right) \Theta_n^{(n+2)}\,.
\end{equation}
Therefore
\begin{equation}\label{eq:Hk-7}
\begin{split}
\sum_{i,j > n, i\neq j} \; \Big| \int (1-w_{ij})^2 \; &\nabla_i
\nabla_j  \Theta^{(n+2)}_n  \; \nabla_i D_n
\,\overline{\phi}_{ij} \; \nabla_j D_n \, \phi_{ij}\Big| \\ \leq \;
&C \ell^{-2} \sum_{i,j > n, i\neq j} \; \int (1-w_{ij})^2 \; \left(
\frac{2^{n+2}}{\ell^{\eps}} \sum_{m=1}^n h_{mj} \right)^2 \,
\Theta^{(n+2)}_n \, |\nabla_i D_n \phi_{ij}|^2
\\ \leq \; &C \ell^{-2} N \int \Theta^{(n+1)}_n \; | D_{n+1} \psi|^2 + C
\ell^{-4} \int \Theta_n^{(n)} \; |D_n \psi|^2 \\
\leq \; & o(N^2) \left( \int \Theta^{(n+1)}_{n} \; | D_{n+1} \psi|^2
+ \int \Theta_{n-1}^{(n)} \; |D_n \psi|^2 \right)\; ,
\end{split}
\end{equation}
where in the second line we used (2.51) and a Schwarz inequality, in
the third line we used the bound (\ref{eq:Hk-5}), while in the last
line we used $N\ell^2 \gg 1$.

\medskip

Next we consider the last term on the r.h.s. of (\ref{eq:Hk-3}). To
this end we note that,  by (\ref{eq:w>0}) and (\ref{eq:w-2}),
\begin{equation*}
\begin{split}
\left| \left[\nabla_i, \frac{\nabla w_{ji}}{1-w_{ij}} \right]
\right| &\leq \frac{\left| \nabla^2 w_{ji} \right|}{1-w_{ij}} +
\left( \frac{ \nabla w_{ji}}{1-w_{ij}} \right)^2 \leq c \rho
\frac{1}{|x_i - x_j|^2}
\end{split}
\end{equation*}
assuming that $\rho$ is small enough. Therefore, the terms in
the sum on the last line of (\ref{eq:Hk-3}) can be bounded by using
Hardy inequality as
\begin{equation}
\begin{split}
&\left|\int (1-w_{ij})^2 \; \Theta^{(n+2)}_n \; \nabla_i D_n
\,\overline{\phi}_{ij} \; [\nabla_i, L_j ] D_n \, \phi_{ij} \right|
\\ &\hspace{2cm}\leq c \rho \int (1-w_{ij})^2
\Theta^{(n+2)}_n \, \frac{1}{|x_i - x_j|^2} |\nabla_i D_n \phi_{ij}|^2\\
&\hspace{2cm}\leq c \rho \int \Theta^{(n+2)}_n \, \frac{1}{|x_i
- x_j|^2} |\nabla_i D_n \phi_{ij}|^2 \\
&\hspace{2cm}\leq c \rho \int  \Theta^{(n+2)}_n \,|\nabla_j \nabla_i
D_n \phi_{ij}|^2 +C \int \left|\nabla_j
\left(\Theta^{(n+2)}_n\right)^{\frac{1}{2}} \right|^2 \,|\nabla_i
D_n \phi_{ij}|^2
\\&\hspace{1.8cm}  \stackrel{\eqref{eq:nablatheta}} {\leq}
 c \rho \int (1-w_{ij})^2 \,
\Theta^{(n+2)}_n \,|\nabla_j \nabla_i D_n \phi_{ij}|^2
\\ &\hspace{2.5cm}
+C \ell^{-2} \int (1-w_{ij})^2 \, \left(
\frac{2^{n+1}}{\ell^{\eps}} \sum_{m=1}^n h_{jm}\right)^2
\Theta^{(n+2)}_n \, |\nabla_i D_n \phi_{ij}|^2 \, .
\end{split}
\end{equation}
Next we sum over $i,j >n$ ($i \neq j$); to control the contribution
originating from the second term on the r.h.s. of the last equation
we use (\ref{eq:Hk-5}). We obtain
\begin{equation}\label{eq:Hk-8}
\begin{split}
\sum_{i,j >n, i \neq j} \Big|\int (1-w_{ij})^2 \; \Theta^{(n+2)}_n
&\; \nabla_i D_n \,\overline{\phi}_{ij} \;
[\nabla_i, L_j ] D_n \, \phi_{ij} \Big| \\
\leq \; &c \rho \sum_{i,j >n, i \neq j} \int (1-w_{ij})^2 \,
\Theta^{(n+2)}_n \, |\nabla_j \nabla_i D_n \phi_{ij}|^2 \\ &+ C
\ell^{-2} N \int \Theta^{(n+1)}_n \; | D_{n+1} \psi|^2 + C \ell^{-4}
\int \Theta_n^{(n)} \; |D_n \psi|^2
\\ \leq \; &c \rho \sum_{i,j >n, i \neq j} \int (1-w_{ij})^2
\, \Theta^{(n+2)}_n \, |\nabla_j \nabla_i D_n \phi_{ij}|^2 \\ &+
o(N^2) \left( \int \Theta^{(n+1)}_n \; | D_{n+1} \psi|^2 + \int
\Theta_{n-1}^{(n)} \; | D_n \psi|^2 \right)\,.
\end{split}
\end{equation}

\medskip

Inserting (\ref{eq:Hk-6}), (\ref{eq:Hk-7}), and (\ref{eq:Hk-8}) into
the right side of (\ref{eq:Hk-3}) it follows that
\begin{equation}\label{eq:Hk-9}
\begin{split}
C_0^n N^n \, &\sum_{i,j > n, i\neq j} \; \int (1-w_{ij})^2 \;
\Theta^{(n+2)}_n \; L_i D_n \,\overline{\phi}_{ij} \; L_j D_n \,
\phi_{ij}
\\ \geq & \; C_0^n N^{n+2}(1- c\rho - o(1)) \int (1-w_{n+1,n+2})^2
\; \Theta^{(n+2)}_n \;
|D_{n+2} \phi_{n+1,n+2}|^2 \\
&-o(N^{n+2}) \left( \int \Theta^{(n+1)}_{n} \; |D_{n+1} \, \psi|^2 +
\int \Theta_{n-1}^{(n)} \; |D_n \psi|^2 \right)\,.
\end{split}
\end{equation}
Lemma \ref{lm:Hk-1-1} now follows from (\ref{eq:phitopsi1}) in Lemma
\ref{lm:phitopsi} below that shows how to replace estimates
involving the function $\phi_{ij}=(1-w_{ij})^{-1}\psi$ with
estimates on $\psi$.
\end{proof}

\begin{lemma}\label{lm:Hk-1-2}
Suppose the assumptions of Lemma \ref{lm:Hk-1} are satisfied. Then
we have
\begin{equation}\label{eq:lmHk-1-2}
\begin{split}
\frac{C_0^n N^n}{2} \, &\sum_{i,j
> n, i\neq j} \; \sum_{r>n, \, r \neq i,j} \; \int (1-w_{ij})^2 \;
\Theta^{(n+2)}_n \; V_N (x_j -x_r) \; L_i \, D_n \,
\overline{\phi}_{ij}\, D_n \, \phi_{ij} + \text{h.c.}
\\ \geq \; &\frac{C_0^n N^{n+3}}{2} (1-o(1)) \,
\int \Theta^{(n+2)}_{n+1} \,
V_N (x_{n+2} -x_{n+3}) \; |D_{n+1} \psi|^2 \\
&- o(N^{n+3})\int \Theta^{(n+1)}_{n} \, V_N (x_{n+1} - x_{n+2}) |D_n
\psi|^2 \, .
\end{split}
\end{equation}
\end{lemma}
\begin{proof}
Using (\ref{eq:symmm}), we find
\begin{equation}\label{eq:Hk-2-1}
\begin{split}
&\frac{C_0^n N^n}{2} \, \sum_{i,j > n, i\neq j} \; \sum_{r>n, \, r
\neq i,j} \; \int (1-w_{ij})^2 \; \Theta^{(n+2)}_n \; V_N (x_j
-x_r)\; L_i D_n \, \overline{\phi}_{ij} D_n \, \phi_{ij}
\\
&\hspace{.2cm} = \frac{C_0^n N^n}{2} \, \sum_{i,j > n, i\neq j}
\sum_{r >n, \, r \neq i,j} \int (1- w_{ij})^2 \, V_N (x_j -x_r) \\
&\hspace{5cm} \times \left\{ \Theta^{(n+2)}_n \; |\nabla_i D_n
\phi_{ij}|^2 + \nabla_i \Theta^{(n+2)}_n \; \nabla_i D_n \,
\overline{\phi}_{ij} \, D_n \phi_{ij} \right\} \,.
\end{split}
\end{equation}

Using (\ref{eq:nablatheta}) (with $j$ replaced by $i$) the second
term in the curly bracket can be bounded by
\begin{equation}\label{eq:Hk-2-2}
\begin{split}
&\sum_{i,j > n, i\neq j} \sum_{r>n , \, r \neq i,j} \left|\int (1-
w_{ij})^2 \, V_N (x_j -x_r)\, \nabla_i \Theta^{(n+2)}_n \,  \nabla_i
D_n \, \overline{\phi}_{ij} D_n \phi_{ij}
\right| \\
&\hspace{1cm}\leq \; C \a \sum_{i,j > n, i\neq j} \sum^N_{r \neq
i,j} \int (1- w_{ij})^2 \, \Theta^{(n+2)}_n \, V_N (x_j -x_r) \;
|\nabla_i D_n \phi_{ij}|^2 \\ &\hspace{1.2cm} +C \ell^{-2} \a^{-1}
\sum_{i,j
> n, i\neq j} \sum^N_{r \neq i,j} \int (1- w_{ij})^2 \left(
\frac{2^{n+2}}{\ell^{\eps}} \sum_{m=1}^n h_{im} \right)^2 \,
\Theta^{(n+2)}_n \, V_N (x_j -x_r) \,  |D_n \phi_{ij}|^2.
\end{split}
\end{equation}
Since $i,j >n$,  and $\psi = (1-w_{ij})\phi_{ij}$, the second term
can be estimated as
\begin{equation}\label{eq:Hk-2-3}
\begin{split}
C \ell^{-2} \a^{-1} &\sum_{i,j > n, i\neq j} \sum_{r>n, r \neq i,j}
\int \left( \frac{2^{n+2}}{\ell^{\eps}} \sum_{m=1}^n h_{im}
\right)^2 \, \Theta^{(n+2)}_n \, V_N (x_j -x_r) \; | D_n \psi|^2 \\
\leq \;& C \ell^{-2} \a^{-1} \sum_{j
> n} \sum_{r>n, r \neq j} \int \left(
\frac{2^{n+2}}{\ell^{\eps}} \sum_{i >n} \sum_{m=1}^n h_{im}
\right)^2 \, \Theta^{(n+2)}_n \, V_N (x_j -x_r) \; |D_n \psi|^2 \\
\leq \; &C \ell^{-2} \a^{-1} \sum_{j
> n} \sum_{r>n, \, r \neq j} \int \Theta^{(n+1)}_{n} \, V_N (x_j
-x_r) \; |D_n \psi|^2
\\ = \; &C \ell^{-2} \a^{-1} (N-n) (N-n-1) \int
\Theta^{(n+1)}_n \, V_N (x_{n+1} - x_{n+2}) |D_n \psi|^2 \, ,
\end{split}
\end{equation}
because of the permutation symmetry of $\psi$ and
$\Theta_n^{(n+1)}$. {F}rom (\ref{eq:Hk-2-3}) and (\ref{eq:Hk-2-2}),
it follows that
\begin{equation}\label{eq:Hk-2-4}
\begin{split}
\sum_{i,j > n, i\neq j} \sum_{r>n, r \neq i,j} &\left|\int (1-
w_{ij})^2 \, \nabla_i \Theta^{(n+2)}_n \, V_N (x_j -x_r) \; \nabla_i
D_n \, \overline{\phi}_{ij} D_n \phi_{ij} \right|
\\ \leq &\;
o(1) \sum_{i,j > n, i\neq j} \sum^N_{r \neq i,j} \int (1- w_{ij})^2
\, \Theta^{(n+2)}_n \, V_N (x_j -x_r) \; |\nabla_i D_n \phi_{ij}|^2
\\ &+ o(N^3)\int \Theta^{(n+1)}_{n} \, V_N
(x_{n+1} - x_{n+2}) |D_n \psi|^2
\end{split}
\end{equation}
where we used that $N \ell^2 \gg 1$ and we made a suitable choice of
the parameter $\a$. Inserting this bound into (\ref{eq:Hk-2-1}),
using the permutation symmetry, and (\ref{eq:phitopsi2}) from Lemma
\ref{lm:phitopsi}, the lemma follows easily.
\end{proof}

The next lemma, showing how to replace estimates on $\phi_{ij}$ with
estimates on $\psi$, has already been used in the previous proofs.

\begin{lemma}\label{lm:phitopsi}
Suppose the assumptions of Proposition \ref{prop:Hk} are satisfied.
Recall that $\phi_{ij}$ is defined by $\psi = (1- w_{ij})
\phi_{ij}$.
\begin{itemize}
\item[i)] For $n\geq 0$, we have
\begin{equation}\label{eq:phitopsi1}
\begin{split}
\int (1-&w_{n+1,n+2})^2 \; \Theta^{(n+2)}_n \; |D_{n+2}
\phi_{n+1,n+2}|^2 \\ \geq \; & (1- o(1)) \int \Theta^{(n+2)}_{n+1}
\; |D_{n+2} \psi|^2 - o(1) \left\{ \int \Theta^{(n+1)}_{n} \,
|D_{n+1}\psi|^2 + \int \Theta^{(n)}_{n-1} \, |D_{n}\psi|^2
\right\}\,.
\end{split}
\end{equation}
\item[ii)] For $n \geq 0$, we have
\begin{equation}\label{eq:phitopsi2}
\begin{split}
\int (&1- w_{n+1,n+2})^2 \; \Theta^{(n+2)}_n \, V_N (x_{n+2}
-x_{n+3}) \; |D_{n+1} \phi_{n+1,n+2}|^2
\\ \geq \; & (1-o(1)) \int \Theta^{(n+2)}_{n+1} \, V_N
(x_{n+2} -x_{n+3}) \; |D_{n+1} \psi|^2 -o(1)\int \Theta^{(n+1)}_{n}
\, V_N (x_{n+1} -x_{n+2}) \; |D_{n} \psi|^2\,.
\end{split}
\end{equation}
\end{itemize}
\end{lemma}

\begin{proof}
In order to prove part i) we start by noticing that
\begin{equation}
\begin{split}
\int (1-w_{n+1,n+2})^2 \; &\Theta^{(n+2)}_n \; |D_{n+2}
\phi_{n+1,n+2}|^2 \geq \int (1-w_{n+1,n+2})^2 \;
\Theta^{(n+2)}_{n+1} \; |D_{n+2} \phi_{n+1,n+2}|^2.
\end{split}
\end{equation}
Using that $\phi_{n+1, n+2}=(1-w_{n+1, n+2})^{-1} \, \psi$ we find
\begin{equation*} \begin{split} D_{n+2}
\phi_{n+1,n+2} = \; & \frac{1}{1-w_{n+1,n+2}} D_{n+2}
\, \psi + \frac{\nabla w_{n+1,n+2}}{(1-w_{n+1,n+2})^2}
D_n\nabla_{n+2} \, \psi  \\ & + \frac{\nabla
w_{n+2,n+1}}{(1-w_{n+1,n+2})^2} D_{n+1} \, \psi +
\left(\frac{\nabla^2 w_{n+1,n+2}}{(1-w_{n+1,n+2})^2} + 2
\frac{(\nabla w_{n+1,n+2})^2}{(1-w_{n+1,n+2})^3} \right) D_n \, \psi
\end{split}
\end{equation*}
and thus,  from (\ref{eq:w>0})
bounds]
\begin{equation}\label{eq:Hk-10}
\begin{split}
\int (1-w_{n+1,n+2})^2 \; &\Theta^{(n+2)}_n \; | D_{n+2}
\phi_{n+1,n+2}|^2 \\ \geq \; &\int \Theta^{(n+2)}_{n+1} \;
|D_{n+2}\psi|^2 - C \int \Theta^{(n+2)}_{n+1} \; |\nabla
w_{n+1,n+2}| \, |D_{n+2}\psi| \; |D_{n+1} \psi|
\\ &-C \int \Theta^{(n+2)}_{n+1}
\;\left( |\nabla w_{n+1,n+2}|^2 + |\nabla^2 w_{n+1,n+2}|\right) \,
|D_{n+2}\psi| \; |D_n \psi| \,.
\end{split}
\end{equation}
The second term can be bounded by
\begin{equation}
\begin{split}
\int \Theta^{(n+2)}_{n+1} \;& |\nabla w_{n+1,n+2}| \, |D_{n+2}\psi|
\; |D_{n+1} \psi|
\\ \leq \; & \a \int \Theta^{(n+2)}_{n+1} \, |D_{n+2}\psi|^2 +
\a^{-1} \int \Theta^{(n+2)}_{n+1} |\nabla w_{n+1,n+2}|^2 \, |D_{n+1}
\psi|^2
\\ \leq \; & \a \int \Theta^{(n+2)}_{n+1} \, |D_{n+2}\psi|^2 +
\a^{-1} \int \Theta^{(n+2)}_{n+1} \; \chi (|x_{n+1} - x_{n+2}| \geq
\ell) \, |\nabla w_{n+1,n+2}|^2 \, |D_{n+1} \psi|^2 \\
&+ \a^{-1} \int \Theta^{(n+2)}_{n+1} \; \chi (|x_{n+1} - x_{n+2}|
\leq \ell) |\nabla w_{n+1,n+2}|^2 \, |D_{n+1}\psi|^2 \\
\leq \; & \a \int \Theta^{(n+2)}_{n+1} \, |D_{n+2}\psi|^2 + C
\a^{-1} a^2 \int \Theta^{(n+2)}_{n+1} \; \frac{\chi (|x_{n+1} -
x_{n+2}|
\geq \ell)}{|x_{n+1} -x_{n+2}|^4} \, |D_{n+1} \psi|^2 \\
&+ C \a^{-1} N^2 \int \Theta^{(n+2)}_{n+1} \; \chi (|x_{n+1} -
x_{n+2}|\leq \ell) \, |D_{n+1}\psi|^2 ,
\end{split}
\end{equation}
where in the last inequality we used that, by Lemma \ref{lm:w},
$|\nabla w_{n+1,n+2}| \leq C N$. Moreover we used that $\nabla w (x)
= a/|x|$ for $|x|
>R/N$ (with $R$ such that $\supp V \subset \{ x \in \bR^3: |x| \leq
R \}$), and that $R/N \leq \ell $ for $N$ large enough. Using that
\begin{equation} \label{eq:cut} \theta^{(n+2)}_{n+1} \chi (|x_{n+1}
- x_{n+2}| \leq \ell) \leq C e^{-C\ell^{-\eps}} \end{equation} we
have (recall that $a=a_0/N$)
\begin{equation*}
\begin{split}
\int &\Theta^{(n+2)}_{n+1} \; |\nabla w_{n+1,n+2}| \, |D_{n+2}\psi|
\; |D_{n+1} \psi|
\\ &\leq \;
\a \int \Theta^{(n+2)}_{n+1} \, |D_{n+2}\psi|^2 +C \a^{-1} a^2
\ell^{-4} \int \Theta^{(n+2)}_{n} \, |D_{n+1}\psi|^2 + C \a^{-1} N^2
e^{-C\ell^{-\eps}}  \int \Theta^{(n+2)}_{n} \, |D_{n+1} \psi|^2.
\end{split}
\end{equation*}
Since $N\ell^{2} \gg 1$, we find
\begin{equation}\label{eq:Hk-11}
\begin{split}
\int &\Theta^{(n+2)}_{n+1} |\nabla w_{n+1,n+2}| \, |D_{n+2}\psi|
|D_{n+1} \psi| \leq o(1) \left\{ \int \Theta^{(n+2)}_{n+1} \,
|D_{n+2}\psi|^2 + \int \Theta^{(n+1)}_{n} \, |D_{n+1}\psi|^2
\right\}.
\end{split}
\end{equation}
As for the third term on the r.h.s. of (\ref{eq:Hk-10}), we proceed
as follows.
\begin{equation}
\begin{split}
 \int \Theta^{(n+2)}_{n+1}
\;&\left( |\nabla w_{n+1,n+2}|^2 + |\nabla^2 w_{n+1,n+2}|\right) \,
|D_{n+2}\psi| \; |D_{n} \psi|
\\ \leq \; & \a \int \Theta^{(n+2)}_{n+1} \, |D_{n+2}\psi|^2 \; +
\a^{-1}  \int \Theta^{(n+2)}_{n+1} \; \left( |\nabla w_{n+1,n+2}|^2
+ |\nabla^2 w_{n+1,n+2}|\right)^2 \; |D_{n} \psi|^2
\\ \leq \; &\a \int \Theta^{(n+2)}_{n+1} \, |D_{n+2}\psi|^2 \; +
C \a^{-1} a^2 \int \Theta^{(n+2)}_{n+1} \; \frac{\chi
(|x_{n+1}-x_{n+2}| \geq \ell)}{|x_{n+1}-x_{n+2}|^6} \; |D_{n}
\psi|^2
\\ &+C \a^{-1}N^4 \int \Theta^{(n+2)}_{n+1} \; \chi (|x_{n+1}-x_{n+2}|
\leq \ell) \; |D_{n} \psi|^2
\end{split}
\end{equation}
where we used the bounds for $|\nabla w|$ and $|\nabla^2 w|$ from
(\ref{eq:w-1}) and that $w(x) = a/|x|$ for $|x| \geq\ell$ since
$\ell\gg R/N$. Using (\ref{eq:cut}) to bound the last term, we
obtain
\begin{equation}\label{eq:Hk-12}
\begin{split}
 \int \Theta^{(n+2)}_{n+1}
\;&\left( |\nabla w_{n+1,n+2}|^2 + |\nabla^2 w_{n+1,n+2}|\right) \,
|D_{n+2}\psi| \; |D_{n} \psi|
\\ \leq \; &\a  \int \Theta^{(n+2)}_{n+1} \,
|D_{n+2}\psi|^2 \; + C \a^{-1} a^2 \ell^{-4} \int \Theta^{(n+2)}_{n}
\;\frac{1}{|x_{n+1} -x_{n+2}|^2} \, |D_n \psi|^2\,
\\ &
 + C \a^{-1} N^4 e^{-C\ell^{-\eps}}  \int
\Theta^{(n)}_{n-1} \; |D_n \psi|^2 .
\end{split}
\end{equation}
To bound the second term on the r.h.s., we apply Hardy inequality.
We have
\begin{equation}
\begin{split}
\int \Theta^{(n+2)}_{n} \; &\frac{1}{|x_{n+1} -x_{n+2}|^2} \, |D_n
\psi|^2\, \\ \leq \; & C \int \Theta^{(n+2)}_{n} \, |D_{n+1} \psi|^2
+ C \int \left|\nabla_{n+1}
\left(\Theta^{(n+2)}_{n}\right)^{\frac{1}{2}}\right|^2 \, |D_n
\psi|^2 \, \\ \leq \; & C \int \Theta^{(n+2)}_{n} \, |D_{n+1}
\psi|^2 + C \ell^{-2} \int \left(\frac{2^{n+1}}{\ell^{\eps}}
\sum_{i=1}^n h_{i,n+1}\right)^2 \Theta^{(n+2)}_{n} \, |D_n \psi|^2
\\ \leq \; & C \int \Theta^{(n+2)}_{n} \, |D_{n+1} \psi|^2 + C
(N-n)^{-1} \ell^{-2} \int \left(\frac{2^{n+1}}{\ell^{\eps}} \sum_{j
\geq n+1} \sum_{i=1}^n h_{ij}\right)^2 \Theta^{(n+2)}_{n} \, |D_n
\psi|^2
\\ \leq \; & C \int
\Theta^{(n+1)}_{n} \, |D_{n+1} \psi|^2 + C (N-n)^{-1} \ell^{-2} \int
\Theta^{(n)}_{n-1} \, |D_n \psi|^2\,.
\end{split}
\end{equation}
Since $N\ell^{2} \gg 1$, it follows from (\ref{eq:Hk-12}) that
\begin{equation*}
\begin{split}
 \int \Theta^{(n+2)}_{n+1} \, &\left( |\nabla w_{n+1,n+2}|^2 + |\nabla^2 w_{n+1,n+2}|\right)
\; \, |D_{n+2}\psi| \; |D_n \psi| \\ \leq \; &o(1) \left\{ \int
\Theta^{(n+2)}_{n+1} \, |D_{n+2}\psi|^2 + \int \Theta^{(n+1)}_{n} \,
|D_{n+1}\psi|^2 + \int \Theta^{(n)}_{n-1} \, |D_n\psi|^2  \right\}
\,.
\end{split}
\end{equation*}
Part i) of Lemma \ref{lm:phitopsi} follows now from
(\ref{eq:Hk-10}), (\ref{eq:Hk-11}) and from last equation.

\medskip

In order to prove part ii) we rewrite the l.h.s. of
(\ref{eq:phitopsi2}) as follows.
\begin{equation}\label{eq:Hk-2-9}
\begin{split}
\int (1- w_{n+1,n+2})^2 &\Theta^{(n+2)}_n \, V_N (x_{n+2} -x_{n+3})
\; |D_{n+1} \phi_{n+1,n+2}|^2
\\ \geq \; &\int (1 - w_{n+1,n+2})^2 \Theta^{(n+2)}_{n+1} \,
V_N (x_{n+2} -x_{n+3}) \; |D_{n+1} \phi_{n+1,n+2}|^2 \,.
\end{split}
\end{equation}
Using
\[D_{n+1} \phi_{n+1,n+2} = \frac{1}{1-
w_{n+1,n+2}} D_{n+1} \psi + \frac{\nabla w_{n+1,n+2}}{(1-
w_{n+1,n+2})^2} D_n \psi \] we find
\begin{equation} \begin{split} \int (1-
w_{n+1,n+2})^2 &\Theta^{(n+2)}_n \, V_N (x_{n+2} -x_{n+3}) \;
|D_{n+1} \phi_{n+1,n+2}|^2
\\ \geq \; &(1-\a) \int \Theta^{(n+2)}_{n+1} \,
V_N (x_{n+2} -x_{n+3}) \; |D_{n+1} \psi|^2
\\ &- C \a^{-1} \int \Theta^{(n+2)}_{n+1}
\, |\nabla w_{n+1,n+2}|^2 \, V_N (x_{n+2} -x_{n+3}) \; |D_n
\psi|^2\,.
\end{split}
\end{equation}
The last term can be controlled by using (\ref{eq:w-1}) and that
$w(x)=a/|x|$ for $|x| >\ell\gg R/N$ by
\begin{equation}
\begin{split}
\int  &\Theta^{(n+2)}_{n+1} |\nabla w_{n+1,n+2}|^2 \, V_N (x_{n+2}
-x_{n+3}) \; |D_n \psi|^2 \\ \leq \; & C N^2 \int
\Theta^{(n+2)}_{n+1} \chi (|x_{n+1} - x_{n+2}| \leq \ell) \,
V_N (x_{n+2} -x_{n+3}) \; |D_n \psi|^2 \\
&+ C a^2 \int \Theta^{(n+2)}_{n+1} \frac{\chi (|x_{n+1} - x_{n+2}|
\geq \ell)}{|x_{n+1} - x_{n+2}|^4} \, V_N (x_{n+2} -x_{n+3}) \; |D_n
\psi|^2 \\ \leq \; &C N^2 e^{-C\ell^{-\eps}} \int
\Theta^{(n+1)}_{n+1} \, V_N (x_{n+2} -x_{n+3}) \; |D_n \psi|^2 + C
a^2 \ell^{-4} \int \Theta^{(n+2)}_{n+1} \,
V_N (x_{n+2} -x_{n+3}) \; |D_n \psi|^2 \, \\
\leq &\; o(1) \int \Theta^{(n+1)}_{n+1} \, V_N (x_{n+2} -x_{n+3}) \;
|D_n \psi|^2\,.
\end{split}
\end{equation}
{F}rom (\ref{eq:Hk-2-9}), we have
\begin{equation}\label{eq:Hk-2-13}
\begin{split}
\int (1- w_{n+1,n+2})^2 \Theta^{(n+2)}_n \, &V_N (x_{n+2} -x_{n+3})
\; |D_{n+1} \phi_{n+1,n+2}|^2 \\ \geq \; &(1-o(1))\int
\Theta^{(n+2)}_{n+1} \, V_N (x_{n+2} -x_{n+3}) \; |D_{n+1} \psi|^2
\\ &-o(1)\int \Theta^{(n+1)}_{n} \, V_N (x_{n+1}
-x_{n+2}) \; |D_n \psi|^2\,.
\end{split}
\end{equation}
In the last term we used $\Theta_{n+1}^{(n+1)}\leq \Theta_n^{(n+1)}$,
the permutation
symmetry of $\psi$ and we shifted the indices  $n+2, n+3 \to n+1,
n+2$
\end{proof}

\begin{proposition}\label{lm:Hk-2}
Suppose $N \ell^2 \gg 1$. Let $\fh^{(n)}_i$ be defined as in
(\ref{eq:frakhn}). Then, if $N$ is large enough (depending on $n$),
\begin{equation}\label{eq:lmHk-2}
\begin{split}
C_0^n N^n \, \sum_{i\leq n < j} \int \Theta^{(n+2)}_n \; D_n
\fh^{(n)}_i \, \overline{\psi} \; D_n \fh^{(n)}_j \, \psi
+\text{h.c.} \geq  \; & C_0^n N^{n+1} (1-o(1)) \int
\Theta^{(n+2)}_{n+1} \; |\nabla_1 \, D_{n+1} \,\psi|^2  \\ & -
\Omega_n (\psi)
\end{split}
\end{equation}
where the error term $\Omega_n (\psi)$ has been defined in
(\ref{eq:error}).
\end{proposition}

\begin{proof}
We rewrite the l.h.s. of (\ref{eq:lmHk-2}) as
\begin{equation}
\begin{split}
C_0^n N^n \, &\sum_{i\leq n < j} \int \Theta^{(n+2)}_n \; D_n
\fh^{(n)}_i \, \overline{\psi} \; D_n \fh^{(n)}_j \, \psi +\text{h.c.} \\
= \; &C_0^n N^n \sum_{i\leq n < j} \int \Theta^{(n+2)}_n \; D_n
\Delta_i \, \overline{\psi} \; D_n \Delta_j \, \psi  \\ &-
\frac{C_0^n N^n}{2} \sum_{i\leq n < j} \sum_{m > n, \, m \neq j}
\int \Theta^{(n+2)}_n \; V_N (x_j -x_m) \; D_n \Delta_i \,
\overline{\psi} \; D_n \, \psi \\
 &- C_0^n N^n \sum_{i\leq n < j} \sum_{r \neq i} \lambda_r
 \int \Theta^{(n+2)}_n
\; D_n (V_N (x_i-x_r) \, \overline{\psi}) \; D_n \Delta_j \, \psi \\
&+ \frac{C_0^n N^n}{4} \sum_{i\leq n < j} \sum_{r \neq i} \lambda_r
\sum_{m>n,\, m \neq j} \int \Theta^{(n+2)}_n \; D_n \, (V_N (x_i
-x_r) \, \overline{\psi}) \; D_n (V_N (x_j-x_m) \, \psi)
+\text{h.c.}
\end{split}
\end{equation}
with $\lambda_r =1$ if $r > n$, and $\lambda_r =1/2$ if $r \leq n$
(recall the definition of $\fh_i^{(n)}$, for $i \leq n$, in
(\ref{eq:frakhn})). The terms on the last two lines are easy to
bound because the potential $V_N (x_i -x_r)$ forces the particle $i$
to be close (on the length scale $N^{-1}$) to the particle $r$. But
then the factor $\theta_i^{(n+2)}$ in $\Theta_n^{(n+2)}$ makes this
contribution exponentially small. More precisely, for $i \leq n$, we
have the bound
\begin{equation}
\left(\nabla^{\alpha} \Theta^{(n+2)}_n \right) \; |\nabla^{\beta}
V_N (x_i - x_r)| \leq e^{-C\ell^{-\eps}} \Theta^{(n+1)}_{n}
\end{equation}
for $\alpha =0,1$, $\beta=0,1,2$, and for all $N$ large enough. It
is therefore easy to prove that
\begin{equation*}
\begin{split}
C_0^n &N^n \, \sum_{i\leq n < j} \int \Theta^{(n+2)}_n \; D_n
\fh^{(n)}_i \, \overline{\psi} \; D_n \fh^{(n)}_j \, \psi +\text{h.c.} \\
= \; &C_0^n N^n \sum_{i\leq n < j} \int \Theta^{(n+2)}_n \; D_n
\Delta_i \, \overline{\psi} \; D_n \Delta_j \, \psi  \\ &-
\frac{C_0^n N^n}{2} \sum_{i\leq n < j} \sum_{m >n, \, m \neq j} \int
\Theta^{(n+2)}_n \; V_N (x_j -x_m)\; D_n \Delta_i \,
\overline{\psi} \; D_n  \, \psi +\text{h.c.}\\
&- O \left( e^{-C\ell^{-\eps}} \right)  \int \left\{
\Theta^{(n+1)}_{n} |D_{n+1} \psi|^2 + \Theta^{(n)}_{n-1} |D_n
\psi|^2 + \Theta^{(n-1)}_{n-2} |D_{n-1} \psi|^2 +
\Theta^{(n-2)}_{n-3} |D_{n-2} \psi|^2 \right\}\,.
\end{split}
\end{equation*}
Lemma \ref{lm:Hk-2} now follows from Lemma \ref{lm:Hk-2-1} and Lemma
\ref{lm:Hk-2-2} below, where we handle the first and, respectively,
the second term on the r.h.s. of the last equation.
\end{proof}

\begin{lemma}\label{lm:Hk-2-1}
Suppose the assumptions of Lemma \ref{lm:Hk-2} are satisfied. Then
we have
\begin{equation}\label{eq:lmHk-2-1}
\begin{split}
C_0^n N^n \sum_{i \leq n < j} \int \Theta^{(n+2)}_n \; &D_n \Delta_i
\, \overline{\psi} \; D_n \Delta_j \, \psi +\text{h.c.}
\\ \geq \; & C_0^n N^{n+1} (1-o(1)) \int \Theta^{(n+2)}_{n+1} \;
|\nabla_1 D_{n+1} \,\psi|^2 \\ &-o (N^{n+2}) \int \Theta^{(n+1)}_{n}
|D_{n+1} \psi|^2 -o (N^{n+1})  \int \Theta^{(n+1)}_n |\nabla_1 D_n
\psi|^2 \, .
\end{split}
\end{equation}
\end{lemma}

\begin{proof}
Integration by parts leads to
\begin{equation}\label{eq:Hk-4-2}
\begin{split}
\sum_{i\leq n < j} \int &\Theta^{(n+2)}_n \; D_n \Delta_i \,
\overline{\psi} \; D_n \Delta_j \, \psi +\text{h.c.} \\ = \;
&\sum_{i\leq n < j} \int \Theta^{(n+2)}_n \; |\nabla_i \nabla_j D_n
\,\psi|^2 + \sum_{i\leq n < j} \int \nabla_i \Theta^{(n+2)}_n \;
\nabla_i \nabla_j D_n  \, \overline{\psi} \; \nabla_j D_n \, \psi
\\ &+ \sum_{i\leq n < j} \int
\nabla_j \Theta^{(n+2)}_n \; \nabla_i D_n  \, \overline{\psi} \;
\nabla_i \nabla_j D_n \, \psi
\\ &+ \sum_{i\leq n < j} \int
\nabla_i \nabla_j \Theta^{(n+2)}_n  \; \nabla_i D_n \,
\overline{\psi} \; \nabla_j D_n \, \psi +\text{h.c.}
\end{split}
\end{equation}
The second term on the r.h.s. of the last equation can be bounded by
\begin{equation}\begin{split}
\sum_{i\leq n < j} \Big| \int \nabla_i & \Theta^{(n+2)}_n \;
\nabla_i \nabla_j D_n \, \overline{\psi} \; \nabla_j D_n \, \psi
\Big| \\ \leq \; &\a \sum_{i\leq n < j} \int \frac{|\nabla_i
\Theta^{(n+2)}_n|^2}{\Theta^{(n+2)}_n} \; |\nabla_j D_n \, \psi|^2 +
\a^{-1} \sum_{i\leq n < j} \int \Theta^{(n+2)}_n \; |\nabla_i
\nabla_j D_n \, \psi|^2
\end{split}
\end{equation}
for some $\a >0$. Next we use that, by Lemma \ref{lm:theta}, part
iv),
\[ \sum_{i\leq n} \frac{|\nabla_i
\Theta^{(n+2)}_n|^2}{\Theta^{(n+2)}_n} \leq C \ell^{-2}
\Theta^{(n+1)}_{n} \] and therefore, since $N\ell^2 \gg 1$,
\begin{equation}\begin{split}\label{eq:Hk-4-3}
\sum_{i\leq n < j} \Big| \int \nabla_i & \Theta^{(n+2)}_n \;
\nabla_i \nabla_j D_n \, \overline{\psi} \; \nabla_j D_n \, \psi
\Big| \\ \leq \; &\a C \ell^{-2} \sum_{j >n} \int \Theta^{(n+1)}_{n}
\; |\nabla_j D_n \, \psi|^2 + \a^{-1} \sum_{i\leq n < j} \int
\Theta^{(n+2)}_n \; |\nabla_i \nabla_j D_n \, \psi|^2 \\ \leq \; & o
(N^2) \int \Theta^{(n+1)}_n \; |D_{n+1} \, \psi|^2 + o(1) \sum_{i
\leq n < j} \int \Theta^{(n+2)}_n \; |\nabla_i \nabla_j D_n \,
\psi|^2\,.
\end{split}
\end{equation}
The estimate of the third term on the r.h.s. of (\ref{eq:Hk-4-2}) is
almost identical to the second term;
\begin{equation}\label{eq:Hk-4-4}
\begin{split}
\sum_{i\leq n < j} \Big|\int \nabla_j &\; \Theta^{(n+2)}_n \;
\nabla_i D_n \, \overline{\psi} \; \nabla_i \nabla_j D_n \, \psi
\Big| \\ \leq \; & \a\sum_{i\leq n < j}  \int \frac{|\nabla_j \,
\Theta^{(n+2)}_n|^2}{\Theta^{(n+2)}_n}  |\nabla_i D_n \psi|^2
+\a^{-1} \sum_{i\leq n < j} \int \Theta^{(n+2)}_n \; |\nabla_i
\nabla_j D_n \, \psi|^2
\\ \leq \; &C \a \ell^{-2} \sum_{i \leq n} \int
\Theta^{(n+1)}_{n} |\nabla_i D_n \psi|^2 + \a^{-1} \sum_{i\leq n <
j}\int \Theta^{(n+2)}_n \; |\nabla_i \nabla_j D_n \, \psi|^2 \\ \leq
\; & o(N) \int \Theta^{(n+1)}_n \; |\nabla_1 D_n \psi|^2 + o(1)
\sum_{i \leq n < j}\int \Theta^{(n+2)}_n \; |\nabla_i \nabla_j D_n
\, \psi|^2\,.
\end{split}
\end{equation}
Finally, to bound the fourth term on the r.h.s. of
(\ref{eq:Hk-4-2}), we use that, by Lemma \ref{lm:theta}, part vi),
\begin{equation}
\sum_{j >n} |\nabla_j \nabla_i \, \Theta^{(n+2)}_n| \leq C \ell^{-2}
\Theta^{(n+1)}_n \quad \text{and} \qquad \sum_{i \leq n} |\nabla_j
\nabla_i \, \Theta^{(n+2)}_n| \leq C \ell^{-2} \Theta^{(n+1)}_n .
\end{equation}
This implies that
\begin{equation}\label{eq:Hk-4-5}
\begin{split}
\sum_{i\leq n < j} \Big|\int \nabla_j \nabla_i &\, \Theta^{(n+2)}_n
\; \nabla_i D_n \, \overline{\psi} \; \nabla_j D_n \, \psi \Big|
\\ \leq \; & C \sum_{i \leq n} \int \sum_{j >n} | \nabla_i \nabla_j
\Theta^{(n+2)}_n|  |\nabla_i D_n \psi|^2 + \sum_{j >n} \int \sum_{i
\leq n} |\nabla_i \nabla_j
(\Theta^{(n+2)}_n)| \; |\nabla_j \, D_n \, \psi|^2 \\
\leq \; &o(N) \int  \Theta^{(n+1)}_{n} |\nabla_1 D_n \psi|^2 +
o(N^2) \int \Theta^{(n+1)}_n |D_{n+1} \psi|^2 \,.
\end{split}
\end{equation}
Lemma \ref{lm:Hk-2-1} now follows from (\ref{eq:Hk-4-2}),
(\ref{eq:Hk-4-3}), (\ref{eq:Hk-4-4}) and (\ref{eq:Hk-4-5}).
\end{proof}

\begin{lemma}\label{lm:Hk-2-2}
Suppose the assumptions of Lemma \ref{lm:Hk-2} are satisfied. Then
we have, for $N$ large enough (depending on $n$),
\begin{equation}\label{eq:lmHk-2-2}
\begin{split}
- \frac{C_0^n N^n}{2} \sum_{i\leq n < j} \sum_{m >n, \, m \neq j}
\int \Theta^{(n+2)}_n &\; V_N (x_j -x_m)\; D_n \Delta_i \,
\overline{\psi}
\; D_n \, \psi + \text{h.c.} \\
\geq \; &-o(N^{n+3}) \int \Theta^{(n+1)}_n \, V_N (x_{n+1}
-x_{n+2})\, |D_n \, \psi|^2\,.
\end{split}
\end{equation}
\end{lemma}

\begin{proof}
We have
\begin{equation}\label{eq:Hk-5-1a}
\begin{split}
-  \sum_{i\leq n < j} \sum_{m >n, \, m \neq j} \int \Theta^{(n+2)}_n
\; & V_N (x_j -x_m) \; D_n \Delta_i \,
\overline{\psi} \; D_n \, \psi +\text{h.c.} \\
= \; &\sum_{i\leq n < j} \sum_{m >n, \, m \neq j} \int
\Theta^{(n+2)}_n \;
V_N (x_m - x_j) |\nabla_i D_n \, \psi|^2\\
&+  \sum_{i\leq n < j} \sum_{m >n, \, m \neq j} \int \nabla_i \,
\Theta^{(n+2)}_n \; V_N (x_j -x_m)\, \nabla_i D_n \, \overline{\psi}
\; D_n \, \psi +\text{h.c.} \,.
\end{split}
\end{equation}
The second term can be bounded by
\begin{equation}\label{eq:Hk-5-2}
\begin{split}
\Big|\sum_{i\leq n < j} \sum_{m >n, \, m \neq j} \int \nabla_i \,
\Theta^{(n+2)}_n \; &V_N (x_j -x_m)\, \nabla_i D_n \,
\overline{\psi} \; D_n \, \psi +\text{h.c.} \Big| \\  \leq \;& \a
\sum_{i\leq n < j} \sum_{m >n , \, m\neq j} \int \frac{|\nabla_i
\Theta^{(n+2)}_n|^2}{\Theta^{(n+2)}_n}  \; V_N (x_j -x_m)\, |D_n \,
\psi|^2  \\ &+ \a^{-1} \sum_{i\leq n < j} \sum_{m >n, \, m \neq j}
\int \Theta^{(n+2)}_n \; V_N (x_j -x_m)\, |\nabla_i D_n \,\psi|^2
\,.
\end{split}
\end{equation}
Since, by Lemma \ref{lm:theta}, part iv),
\begin{equation}
\sum_{i \leq n} \frac{|\nabla_i \, \Theta^{(n+2)}_n
|^2}{\Theta_n^{(n+2)}} \leq C \ell^{-2} \Theta^{(n+1)}_{n}\,,
\end{equation}
using the permutation symmetry and optimizing $\alpha$, we obtain
\begin{equation}
\begin{split}
\Big|\sum_{i\leq n < j} \sum_{m>n, m \neq j} \int \nabla_i \,
&\Theta^{(n+2)}_n \; V_N (x_j -x_m)\, \nabla_i D_n \,
\overline{\psi} \; D_n \, \psi +\text{h.c.} \, \Big| \\ \leq \;&
o(N^3) \int
\Theta^{(n+1)}_n \, V_N (x_{n+1} -x_{n+2})\, |D_n \, \psi|^2 \\
&+ o(1) \sum_{i\leq n < j} \sum_{m >n, \, m \neq j} \int
\Theta^{(n+2)}_n \; V_N (x_j -x_m)\, |\nabla_i D_n \,\psi|^2\,.
\end{split}
\end{equation}
Inserting the last bound in (\ref{eq:Hk-5-1a}), we conclude the
proof of Lemma \ref{lm:Hk-2-2}.
\end{proof}

\appendix

\section{Properties of the cutoff function $\theta_i^{(n)}$}
\setcounter{equation}{0}

Recall that the cutoff functions
$\Theta_k^{(n)}=\Theta_k^{(n)}(\bx)$ defined for $k=1,\dots,N$ and
$n \in \bN$, in Eq. (\ref{eq:thetan}). In the following lemma we
collect some of their important properties which were used in the
energy estimate, Proposition \ref{prop:Hk}.

\begin{lemma}\label{lm:theta}
\begin{itemize}
\item[i)] The functions $\Theta_k^{(n)}$ are monotonic in both
parameters, that is for any $n, k \in \bN$,
\[ \Theta_{k+1}^{(n)}  \leq \Theta_k^{(n)}  \leq 1\; ,\qquad
 \Theta_k^{(n+1)}  \leq \Theta_k^{(n)}  \leq 1 \; .
 \]
Moreover,  $\Theta^{(n)}_k$ is permutation symmetric  in the first
$k$ and the last $N-k$ variables.
\item[ii)] We have, for any $n\in \bN$, $k=1,\dots,N$,
\begin{equation}\label{eq:lmthetaii}
\left( \frac{2^n}{\ell^{\eps}} \sum_{i=1}^k \sum_{j\neq i}^N h_{ij}
\right)^m \Theta_k^{(n)} \leq C_m \; \Theta_k^{(n-1)}\,.
\end{equation}
\item[iii)] For every $k = 1, \dots , N$, $n \in \bN$, we have
\begin{equation}\label{eq:lmthetaiiia}
\begin{split}
|\nabla_i & \Theta_k^{(n)}| \leq C \ell^{-1} \left(
\frac{2^{n}}{\ell^{\eps}} \sum_{r=1}^N h_{ri} \right) \Theta_k^{(n)}
\leq C \ell^{-1}
\Theta_k^{(n-1)} \qquad \text{if } i \leq k \\
|\nabla_i &\Theta_k^{(n)}| \leq C \ell^{-1} \left(
\frac{2^{n}}{\ell^{\eps}} \sum_{r=1}^k h_{ri} \right) \Theta_k^{(n)}
\leq C \ell^{-1} \Theta_k^{(n-1)} \qquad \text{if } i
> k
\end{split}
\end{equation}
\item[iv)] For every $k =1,\dots ,N$, $n \in\bN$ we have
\begin{equation}\label{eq:lmthetaiv}
\begin{split}
\sum_{j=1}^N &\frac{\left| \nabla_j \Theta_k^{(n)}
\right|^2}{\Theta_k^{(n)}} \leq C \ell^{-2} \Theta_k^{(n-1)}
\end{split}
\end{equation}
\item[v)]
For every fixed $k =1,\dots,N$ and $n \in \bN$ we have
\begin{equation}\label{eq:lmthetav}
\begin{split}
\left| \nabla_i \nabla_j \Theta_k^{(n)} \right| &\leq C \ell^{-2}
\left( \frac{2^{n}}{\ell^{\eps}} \sum_{m=1}^k h_{mj} \right) \left(
\frac{2^{n}}{\ell^{\eps}} \sum_{r=1}^k h_{ri} \right) \Theta_k^{(n)}
\leq C \ell^{-2} \; \Theta_k^{(n-1)},
\quad  \text{if} \quad i \neq j \text{ and } i,j >k \\
\left| \nabla_i \nabla_j \Theta_k^{(n)} \right| & \leq C \ell^{-2}
\; \Theta_k^{(n-1)}, \quad \text{ for any } \quad i, j
\end{split}
\end{equation}
\item[vi)] For every fixed $k =1,\dots,N$ and $n \in \bN$ we have
\begin{equation}\label{eq:lmthetavi}
\begin{split}
\sum_{i,j} \left| \nabla_i \nabla_j \Theta_k^{(n)} \right| \leq C
\ell^{-2} \Theta_k^{(n-1)}\,.
\end{split}
\end{equation}
\end{itemize}
\end{lemma}
\begin{proof}
Part i) follows trivially from the definition of $\theta_i^{(n)}$.
Part ii) follows from $x^{m} e^{-x} \leq C_{m} e^{-x/2}$ for every
real $x$. To prove part iii), we observe that, for $i >k$
\begin{equation}
\nabla_i \Theta_k^{(n)} = -\Big(\frac{2^n}{\ell^{\eps}} \sum_{r=1}^k
\nabla h_{ir} \Big)\exp \left( -\frac{2^n}{\ell^{\eps}} \sum_{r=1}^k
\sum_{j \neq r} h_{jr} \right)\,.
\end{equation}
Since $|\nabla h (x)| \leq C \ell^{-1} h(x)$, we obtain
\begin{equation}\label{eq:nablai}
\left|\nabla_i \Theta_k^{(n)} \right| \leq C \ell^{-1} \left(
\frac{2^{n}}{\ell^{\eps}} \sum_{r=1}^k h_{ir} \right) \; \exp \left(
-\frac{2^n}{\ell^{\eps}} \sum_{r=1}^k \sum_{j \neq r} h_{jr}
\right)\, .
\end{equation}
Similarly, for $i \leq k$, we have
\begin{equation}
\nabla_i \Theta_k^{(n)} = - \Bigg(\frac{2^n}{\ell^{\eps}} \sum_{r=1}^N
\nabla h_{ir} (1+\eta_r) \Bigg)\exp \left( -\frac{2^n}{\ell^{\eps}}
\sum_{r=1}^k \sum_{j \neq r} h_{jr} \right)
\end{equation}
with $\eta_r =0$ if $r >k$ and $\eta_r =1$ if $r \leq k$. Therefore,
in this case
\begin{equation}\label{eq:nablai3}
\left|\nabla_i \Theta_k^{(n)} \right| \leq C \ell^{-1} \left(
\frac{2^{n}}{\ell^{\eps}} \sum_{r=1}^N h_{ir} \right) \; \exp \left(
-\frac{2^n}{\ell^{\eps}} \sum_{r=1}^N \sum_{j \neq r} h_{jr}
\right)\,.
\end{equation}
Eqs. (\ref{eq:nablai}) and (\ref{eq:nablai3}), together with part
ii), prove (\ref{eq:lmthetaiiia}).

As for part iv), we have, from (\ref{eq:nablai}),
\begin{equation}
\begin{split}
\sum_{j=k+1}^N \frac{\left|\nabla_j \Theta_k^{(n)}
\right|^2}{\Theta_k^{(n)}} \leq \; &C \ell^{-2} \sum_{j=k+1}^N
\left( \frac{2^{n}}{\ell^{\eps}} \sum_{r=1}^k h_{jr} \right)^2 \exp
\left( -\frac{2^n}{\ell^{\eps}} \sum_{r=1}^k \sum_{j \neq r} h_{jr}
\right)
\\ \leq \; & C \ell^{-2} \left( \frac{2^{n}}{\ell^{\eps}} \sum_{j=k+1}^N
\sum_{r=1}^k h_{jr} \right)^2 \exp \left( -\frac{2^n}{\ell^{\eps}}
\sum_{r=1}^k \sum_{j \neq r} h_{jr} \right)
\\ \leq \; &C \ell^{-2} \Theta_k^{(n-1)}
\end{split}
\end{equation}
by part ii) of this lemma. The contribution to (\ref{eq:lmthetaiv})
from terms with $j \leq k$ can be controlled similarly, using
(\ref{eq:nablai3}). The proof of part v) and vi) is based on simple
explicit computations and the same bounds used for part iii) and
iv).
\end{proof}

\section{Example of an Initial Data}\label{sec:W}
\setcounter{equation}{0}

In this section, we denote by $(1-\omega (x))$ the ground state
solution of the Neumann problem
\[ \left( - \Delta + \frac{1}{2} V_N (x)\right) (1-\omega(x)) =
e_{\ell} (1-\omega (x)) \] on the ball $\{ x : |x| \leq \ell \}$
with the normalization condition $\omega (x) =0$ if $|x| = \ell$. We
extend $\omega (x) =0$ for all $x \in\bR^3$ with $|x| > \ell$. We
will choose $\ell$ such that $a \ll \ell \ll 1$. Recall that $a=a_0
/N$ is the scattering length of the potential $V_N (x) = N^2 V
(Nx)$. Assuming that $V \geq 0$ is smooth spherical symmetric and
compactly supported, we have, from Lemma A.2 in \cite{ESY}, the
following properties of $e_{\ell}$ and $\omega (x)$.
\begin{itemize}
\item[i)] If $a/\ell$ is small enough, then
\begin{equation}\label{eq:eell}
e_{\ell} = 3a \ell^{-3} (1+ o (a/\ell)) \end{equation}
\item[ii)] There exists $c_0 >0$ such that
\[ c_0 \leq 1-\omega (x) \leq 1\] for all $x \in \bR^3$. Moreover
\begin{equation}\label{eq:ome} |\omega (x)| \leq Ca \frac{{\bf 1}(|x| \leq \ell)}{|x|+ a}
\qquad \text{and } \quad |\nabla \omega (x)| \leq C a \frac{{\bf 1}
(|x| \leq \ell)}{(|x| + a)^2} \,. \end{equation}
\end{itemize}
We define the $N$-body wave function \[ W_N (\bx) := \prod_{i<j}^N
(1-\omega (x_i -x_j))\,. \] For $m =1,\dots ,N$, we also define
\[ W_N^{[m]} (x_{m+1}, \dots , x_N) := \prod_{m<i<j}^N (1-\omega (x_i
-x_j)). \]

\begin{lemma}\label{lm:W}
Define \[ \psi_N (\bx) = \frac{W_N (\bx) \prod_{j=1}^N \ph (x_j)}{\|
W_N(\bx)\prod_{j=1}^N \ph (x_j)\|} \]  for any $\ph \in H^1 (\bR^3)$
with $\| \ph \|_{L^2} =1$. Then, if $a \ll \ell \ll 1$, we have
\begin{equation}\label{eq:en} \langle \psi_N, H_N \psi_N \rangle \leq C N
\end{equation}
and, for any fixed $k$,
 \begin{equation}\label{eq:fa}
\lim_{N\to\infty}\| \psi_N - \ph^{\otimes k}
\otimes \xi^{(N-k)}_N \| = 0  \; ,\end{equation}
where
\[ \xi^{(N-k)}_N (x_{k+1}, \dots , x_N) := \frac{\prod_{k<i<j} (1-\omega (x_i -x_j)) \prod_{j=k+1}^N
\ph (x_j)}{\|\prod_{k<i<j} (1-\omega (x_i -x_j)) \prod_{j=k+1}^N \ph
(x_j)\|} \,. \]
\end{lemma}
\begin{proof}
Let $\phi_N (\bx) := \prod_{j=1}^N \ph (x_j)$, and, for $m=1,
\dots,N$, $\phi_N^{[m]} (x_{m+1}, \dots, x_N) := \prod_{j
>m}^N \ph (x_j)$. We start by noticing that
\begin{equation}\label{eq:nor}
(1-o(1)) \Big\| W_N^{[1]}\,\phi_N^{[1]} \Big\|^2 \leq \Big\| W_N \,
\phi_N \Big\|^2 \leq \Big\| W_N^{[1]}\, \phi_N^{[1]}\Big\|^2\,.
\end{equation}
Here $\| W_N^{[1]} \phi_N^{[1]} \|$ is the norm on
$L^2(\bR^{3(N-1)})$. The upper bound in \eqref{eq:nor} is clear
since $1-\omega \leq 1$ and $\| \ph \|=1$. To prove the lower bound,
we note that, by (\ref{eq:ome}), and using the notation $\omega_{ij}
= \omega (x_i -x_j)$,
\begin{equation*}
\begin{split}
\|W_N \, \phi_N \|^2 = \; &\int \rd \bx \, \prod_{i<j}^N
(1-\omega_{ij})^2 \, |\phi_N (\bx)|^2  \\ = \; &\int \rd \bx \,
\prod_{1<i<j}^N (1-\omega_{ij})^2 \, |\phi_N (\bx)|^2 - \int \rd\bx
\, \left(1 - \prod_{j=2}^N (1-\omega_{1,j})^2 \right)
\prod_{1<i<j}^N (1-\omega_{ij})^2 \, |\phi_N (\bx)|^2 \\
\geq \; & \|\varphi\|^2 \| W_N^{[1]} \phi^{[1]}_N \|^2 -
2\sum_{j=1}^N \int \rd\bx \, \omega_{1,j}
 \Big[ W_N^{[1]} (x_2, \dots , x_N)\Big]^2  \, |\phi_N (\bx)|^2 \\
\geq \; & \| W_N^{[1]} \phi^{[1]}_N \|^2 - C N a
 \int \rd\bx \, \frac{{\bf 1} (|x_1 -x_j| \leq \ell)}{|x_1 -x_j|}
\Big[ W_N^{[1]} (x_2, \dots , x_N)  \Big]^2\, |\phi_N (\bx)|^2 \\
\geq \; & (1 - C N a \ell \| \ph \|_{H^1}^2) \| W_N^{[1]} \phi^{[1]}_N
\|^2
\end{split}
\end{equation*}
using  that ${\bf 1} (|x_1 -x_j| \leq \ell) \leq \ell |x_1
-x_j|^{-1}$, and then applying a Hardy inequality in the variable
$x_1$. This proves (\ref{eq:nor}), because $\ell \ll 1$.
Analogously, we can prove that
\begin{equation}\label{eq:nor-k}
(1-o_k(1)) \Big\| W_N^{[k]}\,\phi_N^{[k]} \Big\|^2 \leq \Big\| W_N
\, \phi_N \Big\|^2 \leq \Big\| W_N^{[k]}\, \phi_N^{[k]}\Big\|^2
\end{equation}
where $o_k(1) \to 0$ as $N\to\infty$, for every fixed $k \geq 1$, and
where $\| W_N^{[k]} \phi_N^{[k]} \|$ is the norm on
$L^2(\bR^{3(N-k)})$.

Next we prove (\ref{eq:fa}). To this end we
remark that, by (\ref{eq:nor-k}),
\begin{equation}\label{eq:norm-diff1}
\Big\| \frac{W_N \phi_N}{\| W_N \phi_N\|} - \frac{W_N \phi_N}{\|
W_N^{[k]} \phi^{[k]}\|} \Big\| \leq \Big| \frac{\| W_N
\phi_N\|}{\|W_N^{[k]} \phi^{[k]}\|} -1 \Big| \to 0
\end{equation}
as $N \to \infty$. Moreover, since
\[ \ph^{\otimes k} \otimes \xi^{(N-k)}_N = \frac{W_N^{[k]}
\phi_N}{\| W_N^{[k]} \phi_N^{[k]} \|}\; , \]
we observe from \eqref{eq:norm-diff1} and \eqref{eq:nor-k} that
\begin{equation}\label{eq:norm-diff}
\limsup_{N\to\infty} \|\psi_N -  \ph^{\otimes k} \otimes
\xi^{(N-k)}_N \|^2 \leq \limsup_{N\to\infty} \frac{\Big\| (W_N -
W_N^{[k]})\phi_N \Big\|^2}{  \| W_N^{[k]} \phi_N^{[k]} \|^2}
\end{equation}
Now we have
\begin{equation*}
\begin{split}
\Big\| (W_N - W_N^{[k]})\phi_N \Big\|^2 = \; & \int \rd \bx \,
\left(1- \prod_{i<j<k, i<k<j} (1-\omega_{ij})^2 \right) \big[ W_N^{[k]}
(x_{k+1}, \dots ,x_N) \big]^2\prod_{j=1}^N |\ph (x_j)|^2
\\ \leq \; &C \sum_{i<k} \sum_{j=1}^N \int \rd \bx \,
\omega_{ij}\big[ W_N^{[k]} (x_{k+1}, \dots,x_N) \big]^2\prod_{j=1}^N |\ph (x_j)|^2 \\
\leq &C N k a \ell \| \ph \|_{H^1}^2 \| W_N^{[k]} \phi^{[k]}_N \|^2
\end{split}
\end{equation*}
by using (\ref{eq:ome}) and Sobolev inequality in $x_i$ (see Lemma
\ref{lm:sobolev}, part i)). By \eqref{eq:norm-diff} and $\ell\ll 1$,
this proves (\ref{eq:fa}).

\medskip

Finally, we prove (\ref{eq:en}). To this end we observe that
\begin{equation}\label{eq:en-1}
\frac{1}{W_N} H_N (W_N \phi_N) = \sum_{j=1}^N L_j \phi_N + e_{\ell}
\sum_{j \neq m}^N  {\bf 1} (|x_m -x_j| \leq \ell) \phi_N -
\sum_{i=1}^N \sum_{j,m \neq i, j\neq m}^N
\frac{\nabla\omega_{ij}}{1-\omega_{ij}} \cdot
\frac{\nabla\omega_{im}}{1-\omega_{im}} \phi_N \end{equation}
 where
\[ L_j = -\Delta_j + 2\sum_{m \neq j}
\frac{\nabla\omega_{jm}}{1-\omega_{jm}} \cdot \nabla_j \; .\] Note
that
\[
\int \, W_N^2 \, \overline{\phi}_N \, L_j \psi_N  = \int  W_N^2 \,
L_j \overline{\phi}_N \, \psi_N = \int \, W_N^2  \, \nabla_j
\overline{\phi}_N \, \nabla_j \psi_N \,.
\]
{F}rom (\ref{eq:en-1}) we find, by using (\ref{eq:eell}), $W_N\leq
W_N^{[k]}$ and by applying the Sobolev type inequalities of Lemma
\ref{lm:sobolev} and the permutational symmetries,
\begin{equation}
\begin{split}
\langle W_N &\phi_N, H_N W_N \phi_N \rangle  \\ = \; &\sum_{j=1}^N
\int  \, W_N^2  |\nabla_j \phi_N |^2 + e_{\ell}
\sum_{j\neq m}^N \int \rd \bx \, W_N^2 (\bx) {\bf 1} (|x_j -x_m|
\leq \ell) |\phi_N(\bx)|^2 \\& - \sum_{i=1}^N \sum_{j,m \neq i,
j\neq m}^N \int  \, W_N^2  \,
\frac{\nabla\omega_{ij}}{1-\omega_{ij}} \cdot
\frac{\nabla\omega_{im}}{1-\omega_{im}} |\phi_N |^2 \\
\leq \;&N \| \ph \|_{H^1}^2 \Big\| W_N^{[1]} \phi_N^{[1]} \Big\|^2 +
C N(N-1) a \| \ph \|_{H^1}^4 \Big\| W^{[2]}_N\phi^{[2]}\Big\|^2 \\
&+ CN(N-1)(N-2) a^2 \, \| \ph \|_{H^1}^4 \Big\|
W^{[3]}_N\phi^{[3]}\Big\|^2
\end{split}
\end{equation}
for any $\e >0$. {F}rom (\ref{eq:nor-k}), and since $\ell \ll 1$, we
have
\begin{equation}
\Big \langle \frac{W_N \phi_N}{\| W_N \phi_N\|}, H_N \frac{W_N
\phi_N}{\| W_N \phi_N\|} \Big\rangle \leq C N
\end{equation}
which completes the proof of (\ref{eq:en}).
\end{proof}

\section{Trapped condensates}
\label{sec:trap}

In this Appendix we show that Theorem \ref{thm:main2}
can be applied to the ground state of interacting Bose Hamiltonians
with a trap.
Recall the definition of the
Hamiltonian $H_N$ without a trap from (\ref{eq:ham1}),
and define
\[ \Htrap = H_N + \sum_{j=1}^N \Vtrap (x_j) = \sum_{j=1}^N \left(-\Delta_j +
\Vtrap (x_j) \right) + \sum_{i<j}^N V_N (x_i -x_j) \] with a smooth
trapping potential $\Vtrap \geq 0$ satisfying \( \lim_{|x| \to
\infty} \Vtrap (x) = \infty \,. \) Denote by $\psitrap$ the positive
normalized ground state vector of $\Htrap$. The corresponding
Gross-Pitaevskii energy functional is given by
\[ \cE_{\text{GP}}^{\text{trap}} (\phi) = \int \rd x \,
\left( |\nabla \phi (x)|^2
+ \Vtrap (x) |\phi (x)|^2 + 4\pi a_0 |\phi (x)|^4 \right) \] and we
denote by $\phitrap$ the $L^2$-normalized, positive minimizer of
$\cE_{\text{GP}}^{\text{trap}}$. As proven in \cite{LS}, the
ground state energy per particle is given by minimum value of
$ \cE_{\text{GP}}^{\text{trap}} $ as $N\to\infty$,
\be
     \frac{1}{N} \langle \psitrap, \Htrap\psitrap\rangle
\to \cE_{\text{GP}}^{\text{trap}} (\phitrap),
\label{energy}
\ee
and the one-particle marginal density $\gamma^{(1)}_{N, \text{trap}}$
associated with $\psitrap$ satisfies $\gamma^{(1)}_{N, \text{trap}}
\to |\phitrap \rangle \langle \phitrap|$ (with
convergence in the trace-norm).
{F}rom \eqref{energy}, $\langle
\psitrap, \Htrap \psitrap \rangle \leq C N$ and
since $H_N \leq \Htrap$, we  obtain that
$\psitrap$ satisfies (\ref{eq:thm2ass1}).
The goal of this section is to prove in Proposition
\ref{prop:C1} below
that $\psitrap$  satisfies the asymptotic factorization
property (\ref{eq:thm2ass2}). From Theorem \ref{thm:main2}
we therefore immediately  obtain  the
following corollary:

\begin{corollary}\label{cor:main3}
Suppose $V$ satisfies the same conditions as in Theorem
\ref{thm:main2}. Let $\psi_{N,t}$ be the solution of the
Schr\"odinger equation without a trap, $i\partial_t \psi_{N,t} =
H_N\psi_{N,t}$, but with initial data given by the trapped ground
state, $ \psi_{N,0}:= \psitrap$. For $k =1, \dots, N$, let
$\gamma_{N,t}^{(k)}$ be the one-particle marginal density associated
with $\psi_{N,t}$. Then, for every $t \in \bR$, and $k \geq 1$,
\begin{equation}\label{eq:restrap}
\gamma^{(k)}_{N,t} \to |\ph_t \rangle \langle \ph_t|^{\otimes k}
\qquad \text{as } N\to\infty
\end{equation}
in the weak* topology of $\cL^1 (L^2 (\bR^{3k}))$. Here $\ph_t$ is
the solution to the Gross-Pitaevskii equation
\[ i\partial_t \ph_t = -\Delta \ph_t + 8 \pi a_0 |\ph_t|^2 \ph_t \]
with initial data $\ph_{t=0} = \phitrap$. $\;\;\Box$
\end{corollary}

\begin{proposition}\label{prop:C1} For any fixed $k=1,2,\ldots $,
there exists a sequence of normalized wave functions,
$\xi_N^{(N-k)} \in L^2 (\bR^{3(N-k)})$,  $N> k$, such that
\[ \big\| \psitrap - [\phitrap]^{\otimes k} \otimes \xi_N^{(N-k)}
 \big\| \to 0 \] as $N \to
\infty$.
\end{proposition}

We will prove this proposition only for $k=1$, the proof
for arbitrary $k\ge 1$ can be obtained similarly. For brevity,
we set $\xi_N= \xi_N^{N-1}$.
For the proof, we make use of the following three lemmas.

\begin{lemma}\label{lm:C1}
There exists a constant $C >0$ independent of $R,N$ such that
\begin{equation}
\| {\bf 1}(|x_1| > R ) \psitrap \| \leq C e^{-R}
\end{equation}
where ${\bf 1} (s > \lambda)$ denotes the characteristic function of
the interval $[\lambda , \infty)$.
\end{lemma}

\begin{lemma}\label{lm:C2}
We have $\phitrap (x) > 0$ for all $x \in \bR^3$. Moreover
\[ \| (1 -\Delta) \phitrap \| < \infty , \qquad  \langle \phitrap, \Vtrap (x) \phitrap \rangle
< \infty \] and there exists a constant $C>0$ such that
\[ \| {\bf 1} (|x| > R ) \phitrap \| \leq C e^{-R} \] for all $R >0$.
\end{lemma}

\begin{lemma}\label{lm:C3}
For fixed $R>0$, $N \in \bN$ define $\wt\xi_{R,N} \in L^2
(\bR^{3(N-1)})$ by
\[ \wt\xi_{R,N} (\bx_{N-1}) = \frac{1}{\int_{|x_1| < R} \rd
x_1 \, |\phitrap (x_1)|^2} \int_{|x_1| < R} \rd x_1 \, |\phitrap
(x_1)|^2 \, \frac{\psitrap (x_1, \bx_{N-1})}{\phitrap (x_1)} \]
where $\bx_{N-1} = (x_2, \dots ,x_N)$. Then we have
\begin{equation}
\int \rd \bx_{N-1} \int_{|x_1| <R} \rd x_1 \, \left| \psitrap (x_1,
\bx_{N-1}) - \phitrap (x_1)\,  \wt \xi_{R,N} (\bx_{N-1}) \right|^2
\leq c_R \, d_N
\end{equation}
where $c_R < \infty$ is independent of $N$ and $d_N$ is independent
of $R$ and satisfies $d_N \to 0$ as $N \to \infty$.
\end{lemma}

Using these three lemmas we can prove Proposition \ref{prop:C1}.

\begin{proof}[Proof of Proposition \ref{prop:C1} for $k=1$.]
 Using the notation
introduced in Lemma \ref{lm:C3} we have
\begin{equation}\label{eq:C1}
\begin{split}
\| \psitrap - \phitrap \otimes \wt \xi_{R,N} \|^2 = \; &\int \rd
\bx_{N-1} \int \rd x_1 \, |\psitrap (x_1, \bx_{N-1}) - \phitrap
(x_1) \wt \xi_{R,N} (\bx_{N-1}) |^2 \\ = \; & \int \rd \bx_{N-1}
\int_{|x_1| < R} \rd x_1 \, |\psitrap (\bx) - \phitrap (x_1) \wt
\xi_{R,N} (\bx_{N-1}) |^2 \\ & + \int \rd \bx_{N-1} \int_{|x_1| \geq
R} \rd x_1 \, |\psitrap (\bx) - \phitrap (x_1) \wt \xi_{R,N}
(\bx_{N-1}) |^2 \\  \leq \; & c_R \, d_N + C e^{-R}
\end{split}
\end{equation}
where we used Lemma \ref{lm:C3} to bound the term on the second
line, and Lemmas \ref{lm:C1} and \ref{lm:C2} to bound the term on
the third line. Eq. (\ref{eq:C1}) implies that
\begin{equation}\label{eq:C2}
\begin{split}
\left\| \psitrap - \phitrap \otimes \frac{ \wt \xi_{R,N}}{\| \wt
\xi_{R,N} \|} \right\| &= \left\| \psitrap - \frac{ \phitrap \otimes
\wt \xi_{R,N}}{\| \phitrap \otimes \wt \xi_{R,N} \|} \right\|
\\ &\leq \frac{2\, \| \psitrap - \phitrap \otimes \wt \xi_{R,N} \|}{1-\|
\psitrap - \phitrap \otimes \wt \xi_{R,N} \|}\,.
\end{split}
\end{equation}
Now choose a sequence $R_N$ such that $R_N \to \infty$ and $c_{R_N}
d_N \to 0$ as $N \to\infty$. Then, taking $\xi_N = \wt \xi_{R_N,N} /
\| \wt \xi_{R_N,N} \|$, we clearly have $\| \xi_N \| =1$ for all
$N$, and, by (\ref{eq:C1}) and (\ref{eq:C2}),
\[ \| \psitrap - \phitrap \otimes \xi_N \| \to 0 \qquad \text{as } N
\to \infty \,.\]
\end{proof}

We still have to prove Lemmas \ref{lm:C1}, \ref{lm:C2}  and
\ref{lm:C3}. Lemma \ref{lm:C2} is a standard result which follows
from the fact that $\phitrap$ is the solution of the
elliptic non-linear eigenvalue equation
\begin{equation}
-\Delta \phitrap + \Vtrap \phitrap + 8 \pi a_0 |\phitrap|^2 \phitrap
= \mu \phitrap
\end{equation}
with some constant $\mu$.
Lemma \ref{lm:C3} has been proven in \cite{LS}, more precisely,
it follows from  Eq. (13) of \cite{LS} by noticing that
the two terms in the parenthesis in
this equation converge to zero, uniformly in $R$, because of Eq.
(7) and Lemma 1 in \cite{LS}.
It only remains to prove Lemma \ref{lm:C1}.
To this end we use
the following two lemmas.

\begin{lemma}\label{lm:C4}
Let $\chi \in C^{\infty} (\bR)$ with $\chi (s) = 0$ if $s <1$ and
$\chi (s) =1$ if $s >2$, and let $f \in C^1 (\bR)$ be a
monotonically increasing function with $\sup_x |f' (x)| <
\infty$. Then we have, for $R >0$ large enough,
\[ \chi (|x_1|/R) \left( \Htrap - |f' (|x_1|)|^2 - E_N \right) \chi
(|x_1|/R) \geq \chi (|x_1|/R)^2, \]
where $E_N$ denotes the ground state energy of $\Htrap$.
\end{lemma}
\begin{proof}
Define \[ \wt H^{\text{trap}}_{N-1} = \sum_{j=2}^N \left(-\Delta_j +
\Vtrap (x_j) \right) + \sum_{2 \leq i < j}^N V_N (x_i -x_j) \] and
let $\wt E_{N-1} = \inf \sigma (\wt H^{\text{trap}}_{N-1})$.
Moreover, we define $\wt \psi^{\text{trap}}_{N-1} \in L^2
(\bR^{3(N-1)})$ to be the positive normalized ground state of $\wt
H^{\text{trap}}_{N-1}$. Then we have, since $-\Delta_1 \geq 0$ and
$V_N(x) \geq 0$,
\begin{equation}\label{eq:lm1-1}
\begin{split}
\chi (|x_1|/R) \Big( \Htrap - &|f' (|x_1|)|^2 - E_N \Big) \chi
(|x_1|/R) \\ &\geq \chi (|x_1|/R) \left( \wt H^{\text{trap}}_{N-1} +
\Vtrap (x_1) - |f' (|x_1|)|^2 - E_N \right) \chi (|x_1|/R) \\ & \geq
\chi (|x_1|/R)^2 \left( \Vtrap (x_1) - C - (E_N - \wt E_{N-1})
\right)
\end{split}
\end{equation}
where we used the assumption $|f'|\leq C$. Next we remark that there
exists a constant $C>0$ such that \[ E_N \leq \wt E_{N-1} + C \qquad
\text{for all } N \,. \] In fact (using the symmetry of the wave
function)
\begin{equation}
\begin{split}
E_N \leq \langle \phitrap \otimes \wt \psi_{N-1}^{\text{trap}},
H_N\, \phitrap \otimes \wt \psi_{N-1}^{\text{trap}} \rangle = \;
&\wt E_{N-1} + \langle \phitrap, \left( -\Delta_1 + \Vtrap (x_1)
\right) \phitrap \rangle \\ &+ \langle \phitrap \otimes \wt
\psi_{N-1}^{\text{trap}}, (N-1) N^2 V (N (x_1 - x_2)) \phitrap
\otimes \wt \psi_{N-1}^{\text{trap}} \rangle \\ \leq \; &\wt E_{N-1}
+ C \; \| (1-\Delta)
\phitrap \|^2 + C \langle \phitrap, \Vtrap (x_1) \phitrap\rangle \\
\leq \; &\wt E_{N-1} + C
\end{split}
\end{equation}
where we used the operator inequality $W (x_1 -x_2) \leq C \| W
\|_{L^1} (1-\Delta_1)^2$ and Lemma \ref{lm:C2}. Since $\lim_{|x| \to
\infty} \Vtrap (x) = \infty$, the lemma now follows from
(\ref{eq:lm1-1}).
\end{proof}

\begin{lemma}\label{lm:C5}
Suppose that $f,\chi$ are as in Lemma \ref{lm:C4}. Then we have, for
$R$ large enough,
\begin{equation}
\| e^{f(|x_1|)} \chi (|x_1|/R) \psitrap \| \leq C_R
\end{equation}
for some constant $C_R$ depending on $R$ but not on $N$.
\end{lemma}
\begin{proof}
We compute
\[ e^{f(|x_1|}( \Htrap - E_N) e^{-f(|x_1|)} = \Htrap - |f'
(|x_1|)|^2 - E_N + i \left(p_1 \cdot \frac{x_1}{|x_1|} f' (|x_1|) +
f' (|x_1|) \frac{x_1}{|x_1|} \cdot p_1 \right)\,, \] with $p_1 = -i
\nabla_1$. Therefore, for $R$ large enough,
\begin{equation}\label{eq:lm2-1}
\begin{split}
\text{Re} \, \Big\langle e^{f(|x_1|)} \chi (|x_1|/R) &\psitrap,
e^{f(|x_1|)} \left( \Htrap - E_N \right) e^{-f(|x_1|)} e^{f(|x_1|)}
\chi (|x_1|/R) \psitrap \Big\rangle  \\ = \; & \Big\langle e^{f(|x_1|)}
\psitrap \chi (|x_1|/R),  \left( \Htrap - |f' (|x_1|)|^2 - E_N \right)
\chi (|x_1|/R)e^{f(|x_1|)} \psitrap \Big\rangle \\ \geq \; &\|
e^{f(|x_1|)} \chi (|x_1|/R) \psitrap \|^2
\end{split}
\end{equation}
where we used Lemma \ref{lm:C4}. On the other hand
\begin{equation}\label{eq:lm2-2}
\begin{split}
\text{Re} \, \Big\langle e^{f(|x_1|)} \chi (|x_1|/R) \psitrap,
& \;  e^{f(|x_1|)} \left( \Htrap - E_N \right) e^{-f(|x_1|)} e^{f(|x_1|)}
\chi (|x_1|/R) \psitrap \Big\rangle  \\ \leq \; &\|  e^{f(|x_1|)} \chi
(|x_1|/R) \psitrap \| \Big\| e^{f(|x_1|)} \left( \Htrap - E_N \right)
\chi (|x_1|/R) \psitrap \Big\| \\ \leq \; & \| e^{f(|x_1|)} \chi
(|x_1|/R) \psitrap \| \, \Big\| e^{f(|x_1|)} \left[ \Htrap , \chi
(|x_1|/R) \right]  \psitrap \Big\|
\end{split}
\end{equation}
because $(\Htrap - E_N) \psitrap = 0$. Combining (\ref{eq:lm2-1})
and (\ref{eq:lm2-2}) we obtain that, for $R$ large enough,
\[ \|
e^{f(|x_1|)} \chi (|x_1|/R) \psitrap \| \leq \Big\| e^{f(|x_1|)} \Big[
H_N , \chi (|x_1|/R)\Big] \psitrap \Big\| \,. \] Next we note that
\[\left[ H_N , \chi (|x_1|/R) \right] =   -2i R^{-1} \chi' (|x_1|/R)
\frac{x_1}{|x_1|} \cdot \nabla_1 + R^{-2} \chi'' (|x_1|/R) + R^{-1}
\frac{\chi' (|x_1|/R)}{|x_1|} \,. \] Since $f$ is monotone
increasing, we see that \begin{equation} \begin{split} \left\| e^{f
(|x_1|)} \chi' (|x_1|/R) \frac{x_1}{|x_1|} \right\| &\leq C
e^{f(2R)},  \quad \left\|e^{f (|x_1|)} \frac{\chi' (|x_1|/R)}{|x_1|}
\right\| \leq C R^{-1} e^{f(2R)} \quad \text{and} \\ \left\| e^{f
(|x_1|)} \chi'' (|x_1|/R) \right\| &\leq C e^{f(2R)} .
\end{split}
\end{equation}
The energy estimate \eqref{energy} and $V_N\ge0$ imply that $\|
\nabla_1 \psitrap \| \leq C$ uniformly in $N$. From these estimates
the lemma follows.
\end{proof}

\begin{proof}[Proof of Lemma \ref{lm:C1}]
Suppose $\chi$ is as in Lemmas \ref{lm:C4} and \ref{lm:C5}. For a
fixed $R_0$ large enough, we have, by Lemma \ref{lm:C5},
\[ \| e^{|x_1|} \psitrap \| \leq \|e^{|x_1|} \chi (|x_1|/R_0)\psitrap
\| + \| e^{|x_1|} \left( 1 - \chi (|x_1|/R_0)\right) \psitrap \|
\leq C .\] Therefore
\[ \| {\bf 1}(|x_1| >R) \psitrap \| \leq \| e^{-|x_1|} {\bf 1} (|x_1| >
R) e^{|x_1|} \psitrap \| \leq C e^{-R} .\]
\end{proof}

\section{Properties of the one-body scattering solution
$1-w(x)$}\label{app:w} \setcounter{equation}{0}

In this section we prove part i) and iii) of Lemma \ref{lm:w}.

\begin{lemma}\label{lm:ph0}
Suppose that $V \geq 0$ is smooth, spherical symmetric with compact
support and with scattering length $a_0$. Let
\begin{equation}\label{eq:defrho2}
\rho = \sup_{r \geq 0} r^2 V(r)+\int_0^{\infty} \rd r \, r \,
V(r),\end{equation} and suppose $\ph_0 (x)$ is the solution of
\begin{equation}\label{eq:scatte}
\left(-\Delta + \frac{1}{2} V\right) \ph_0 = 0 \qquad \text{with }
\quad \ph_0 \to 1 \quad \text{as } |x| \to \infty \,.\end{equation}
\begin{itemize}
\item[i)] There exists $C_0>0$, depending on $V$, such that $C_0 \leq \ph_0
(x) \leq 1$ for all $x \in \bR^3$. Moreover there exists a universal
constant $c$ such that
\begin{equation}\label{eq:ph0-bou}
1 - c \rho \leq \ph_0 (x) \leq 1 \qquad \text{for all } x \in \bR^3
\, . \end{equation}
\item[ii)] There exists a universal constant $c>0$ such that
\begin{equation}\label{eq:oneder}
|\nabla \ph_0 (x)| \leq c \frac{a_0}{|x|^2}, \qquad |\nabla \ph_0
(x)| \leq c \frac{\rho}{|x|} \qquad \text{and } \quad |\nabla^2
\ph_0 (x)| \leq c \frac{\rho}{|x|^2} \,.
\end{equation}
Moreover there are constant $C_1,C_2$, depending on the potential
$V$, such that
\begin{equation}\label{eq:twoder}
|\nabla \ph_0 (x)| \leq C_1 \qquad |\nabla^2 \ph_0| \leq C_2 .
\end{equation}
\end{itemize}
\end{lemma}

\begin{proof}
Let $R$ be such that $\supp V \subset \{ x \in \bR^3 : |x| \leq
R\}$, and let $a_0$ denote the scattering length of $V$. Then we fix
$\wt R >R$ such that $a_0/\wt R \leq \min \, (\rho, 1/2)$, with
$\rho$ defined in (\ref{eq:defrho2}).

\medskip

In order to prove part i), we observe that, for $|x| \geq \wt R$,
$\ph_0 (x) = 1 -a_0/|x|$. Hence
\begin{equation}\label{eq:ph-bou-1}
\frac{1}{2} \leq \ph_0 (x) \leq 1,  \qquad \text{and } \qquad 1-
\rho \leq \ph_0 (x) \leq 1 \, , \qquad \text{for } |x| \geq  \wt R\,
.
\end{equation}
Next, by Harnack principle the ratio between the supremum and the
infimum of $\ph_0$ in a given ball is bounded: therefore $\ph_0$ is
bounded away from zero in the ball $|x| \leq \wt R$ and thus there
exists $C_0 >0$ such that $\ph_0 (x) \geq C_0$ for all $x \in
\bR^3$. Moreover by the maximum principle, and since, from
(\ref{eq:scatte}), $-\Delta \ph_0 \leq 0$, it follows that $\ph_0
(x) \leq 1$, for all $x \in \bR^3$. To prove (\ref{eq:ph0-bou}) for
$|x| \leq \wt R$, we write $\ph_0 (x) = m(r)/r$, with $r=|x|$. Then
$m'(\wt R)= 1$, and, from (\ref{eq:scatte}),
\begin{equation}\label{eq:m''}
-m'' (r) + \frac{1}{2} V(r) m(r) =0 \,. \end{equation} Since $ 0<
\ph_0 (x) \leq 1$, it follows that $m(0) =0$ and $0< m(r)/r \leq 1$.
Therefore, for $r <\wt R$,
\begin{equation}
m'(r)=m'(\wt R) - \int_r^{\wt R} \rd s \, m'' (s) = 1 -
\frac{1}{2}\int_r^{\wt R} \rd s \, s \, V(s) \frac{m(s)}{s} \geq 1 -
c \int_0^{\infty} \rd s \, s \, V(s) \geq 1 - c \rho \,
\end{equation}
and
\begin{equation}
m(r) = \int_0^r \rd s \, m'(s) \geq r (1- c \rho) \qquad \Rightarrow
\quad  \ph_0 (r) = \frac{m(r)}{r} \geq 1-c \rho \qquad \text{for all
} r < \wt R\,.
\end{equation}
The last equation, together with (\ref{eq:ph-bou-1}), implies
(\ref{eq:ph0-bou}).

\medskip

Next we prove ii). For $|x| \geq \wt R$, we have $\ph_0 (x) = 1
-a_0/|x|$ and thus
\begin{equation}\label{eq:nablaph0}
|\nabla \ph_0 (x)| \leq \frac{a_0}{|x|^2} \leq \frac{a_0}{{\wt
R}|x|} \leq \frac{\rho}{|x|}, \qquad \text{for } |x| \geq \wt R,
\end{equation}
by definition of $\wt R$. Next, for $|x| < \wt R$, we write $\ph_0
(x) = m(r)/r$, with $r = |x|$. Then
\begin{equation}\label{eq:oneder-1}
\begin{split}
|\nabla \ph_0 (x)| &= \Big| \frac{m'(r)r -m(r)}{r^2}\Big| \\ &=
\Big| \frac{1}{r} \int_0^r \rd s \, m'' (s) - \frac{1}{r^2} \int_0^r
\rd s \int_0^s \rd \kappa \, m''(\kappa) \Big| \\ &= \frac{1}{r^2}
\int_0^r \rd \kappa \, \kappa \, m''(\kappa)\,,
\end{split}
\end{equation}
because $m'' (\kappa) \geq 0$. {F}rom (\ref{eq:m''}) we obtain
\begin{equation}
\begin{split}
|\nabla \ph_0 (x)| \leq \frac{1}{2r^2} \int_0^r \rd \kappa \,
\kappa^2 V (\kappa) \, \frac{m (\kappa)}{\kappa} \leq
c\frac{a_0}{|x|^2} \,,
\end{split}
\end{equation}
because $8\pi a_0 = \int V(x) \ph_0(x)$ (see Lemma \ref{lm:w}), part
iv). Moreover, again from (\ref{eq:oneder-1}) and (\ref{eq:m''}), we
have
\begin{equation}\label{eq:oneder-2}
\begin{split}
|\nabla \ph_0 (x)| \leq \frac{1}{2r^2} \int_0^r \rd \kappa \,
\kappa^2 V (\kappa) \, \frac{m (\kappa)}{\kappa} \leq c
\frac{\sup_{\kappa \geq 0} \kappa^2 \, V(\kappa)}{r} \leq c
\frac{\rho}{r}\, .
\end{split}
\end{equation}
Together with (\ref{eq:nablaph0}) we obtain the first two
inequalities in (\ref{eq:oneder}). {F}rom (\ref{eq:nablaph0}) and
from the first inequality in (\ref{eq:oneder-2}), it also follows
that there exists $C_1$, depending on the bounded potential $V$,
such that $|\nabla \ph_0 (x)| \leq C_1$. To prove the second bounds
in (\ref{eq:oneder}) and (\ref{eq:twoder}), we note that
\begin{equation}\label{eq:twoder-2}
|\nabla^2 \ph_0 (x)| \leq \frac{a_0}{|x|^3} \leq \frac{\rho}{|x|^2}
\qquad\text{for } |x| >{\wt R},
\end{equation}
by the definition of $\wt R$. For $|x| \leq \wt R$, we have
(expanding $m(r)$ and $m'(r)$ and using that $m(0)=0$)
\begin{equation}\label{eq:twoder-3}
\begin{split}
|\nabla^2 \ph_0 (x)| &\leq \Big| \frac{m'' (r)}{r} -
2\frac{m'(r)}{r^2} +2 \frac{m(r)}{r^3} \Big| \\
&= \Big| \frac{1}{2} V(r) \frac{m(r)}{r} + \frac{2}{r^3} \int_0^r
\rd s \, s^2 \, V(s) \frac{m(s)}{s} \Big| \\ &\leq c
\frac{\left(\sup_{s \geq 0} s^2 V(s) \right)}{r^2} \leq c
\frac{\rho}{r^2}
\end{split}
\end{equation}
Last equation, together with (\ref{eq:twoder-2}), implies the third
bound in (\ref{eq:oneder}). Moreover, from (\ref{eq:twoder-2}) and
the second line in (\ref{eq:twoder-3}), it also follows that there
exists $C_2$, depending on the bounded potential $V$, such that
$|\nabla^2 \ph_0 (x)| \leq C_2$.
\end{proof}

\begin{proof}[Proof of Lemma \ref{lm:w}, part i) and iii)]
By scaling $1-w(x) = \ph_0 (Nx)$, with $\ph_0$ defined in Lemma
\ref{lm:ph0}. Therefore part i) of Lemma \ref{lm:w} follows
immediately by part i) of Lemma \ref{lm:ph0}, and part iii) of Lemma
\ref{lm:w} follows from (\ref{eq:oneder}) and (\ref{eq:twoder}).
\end{proof}

\end{document}